\documentclass[11pt]{article}
\usepackage{amsmath}
\usepackage{amsfonts}
\usepackage{amssymb}
\usepackage[utf8]{inputenc}
\usepackage{amsmath}
\usepackage{amsfonts}
\usepackage{graphicx,color}
\usepackage{hyperref}
\usepackage{booktabs}
\usepackage[super]{nth}
\hypersetup{
  colorlinks=TRUE,
  linkcolor=blue,
  urlcolor=blue,
  citecolor=blue
}
\usepackage{authblk}
\usepackage{epsfig}
\usepackage{amssymb}
\usepackage{fancyhdr}
\usepackage{amsmath}
\usepackage{multirow}
\usepackage{algorithm}
\usepackage[round,authoryear,sort&compress]{natbib}
\usepackage{geometry}

\usepackage{algorithm}
\usepackage{algorithmic}
\usepackage{float}
\usepackage{rotating}

\title{Variational Inference for Sparse Poisson Regression}
\author[1]{Mitra Kharabati\thanks{kharabatimitra@gmail.com}}
\author[1]{Morteza Amini\thanks{Corresponding author, e-mail: morteza.amini@ut.ac.ir}}
\author[2]{Mohammad Arashi\thanks{arashi@um.ac.ir}}
\affil[1]{Department of Statistics, School of Mathematics, Statistics, and Computer Science, College of Science, University of Tehran, Tehran, Iran}
\affil[2]{Department of Statistics, Faculty of Mathematical Sciences, Ferdowsi University of Mashhad, Mashhad 9177948974, Razavi Khorasan, Iran}
\begin{document}
\maketitle

\begin{abstract}
We have utilized the non-conjugate Variational Bayesian (VB) method for the problem of the sparse Poisson regression model. 
To provide approximate conjugacy in the model, the likelihood is approximated by a quadratic function, 
yielding conjugacy between the approximation component and the Gaussian prior on the regression coefficient. 
Three sparsity-enforcing priors {\color{black}(Laplace, Continuous Spike and Slab, and Bernoulli)} are used for this problem. 
The proposed models are compared with each other, the associated MCMC models, 
and two frequentist sparse Poisson methods (LASSO and SCAD) to evaluate their estimation, prediction, and sparsity performance. 
{In a simulation study, the proposed VB methods closely approximate the posterior parameter distribution while achieving significantly 
faster computation than benchmark MCMC methods.} Using several benchmark count response data sets, the prediction performance 
of the proposed methods is evaluated in real-world applications.
\end{abstract}

\noindent\textbf{Keywords}: Non-conjugate Variational Bayes; Bayesian Variable Selection; Count Regression

\section{Introduction}

Poisson regression and its special cases have been applied to { numerous problems in various scientific fields} \citep[see][among others]{h14,
fa23,
wi21,
cha18,
cox09}. {Many researchers have considered the Bayesian estimation of the parameters of the Poisson regression model}. The MCMC sampling methods, including the Metropolis-Hastings algorithm \citep{ha70} and Gibbs sampler \citep{ge84}{,} are used as common methods for estimation of the posterior distribution in these models. \cite{el73} developed the Bayesian analysis of the Poisson regression model. \cite{ch97} proposed a Poisson model { which allows varying exposures and individual shrinkage factors in a regression setting}. { Hierarchical Bayesian Poisson regression models were examined by \cite{kim13}.} There are many other works, including 
\cite{da22,
t22,
z14,
vi02,
se21}
and
\cite{xi14}, 
where Bayesian inference is applied to the Poisson regression model and its varieties, using MCMC sampling methods.

A challenging problem in all regression models is variable selection; that is, {choosing} a subset of covariates that are the most effective for the response variables. {Sparse} estimation is one of the most common solutions to this problem. In {frequentist} methods, the sparsity of the coefficients of the regression model is obtained using sparsity penalty functions. The least absolute shrinkage and selection operator \citep[LASSO-][]{t96}, elastic net \citep{li10}, smoothly clipped absolute deviation \citep[SCAD-][]{fan01}, {and} mini-max concave penalty \citep[MCP-][]{z10} are the most well-known penalized regression models for sparse estimation of the regression parameters. The sparse Poisson regression model has been considered by many researchers, including 
\cite{ah14,
a15,
f03,
gu19,
jea19,
sa21,
lc15}{,} among others.

{ From a Bayesian perspective, regression parameter sparsity is achieved through the use of sparsity-inducing priors.} Some examples of such priors are Laplace \citep{see08}, horseshoe \citep{cha09}, Student’s t \citep{tip01}, and spike-and-slab \citep{mit88,geo97} priors. A huge amount of research is devoted to sparse Bayesian regression models. Some examples are  
\cite{b12,
car9,
geo97,
gr13,
ji13,
mit88,
ni22,
pol14,
ri10,
see08,
shin22,
xu15}.

The Bayesian sparse regression has also been of interest to many researchers. { The majority of studies in this field utilize MCMC sampling to estimate the parameters of sparse Poisson regression models.} \citep{ga10,d16,po10,ben21,tang23,ji07,shan12,cr12,bar20,ram09,bia20,je21}. Some of these works consider the generalized linear models (GLM), which include Poisson, logistic, and many other non-Gaussian regression models. 

{ Although MCMC is the most widely used approach for approximating the posterior distribution in Bayesian inference, it incurs a high computational cost, particularly when the dimensionality of the regression parameters is large.} There are a variety of methods for approximation of the posterior distribution with a lower computational cost, including integrated nested Laplace approximation \citep[INLA-][]{ru09}, mean-field variational Bayes \citep[VB-][]{b17}, expectation propagation \citep[EP-][]{minka01,minka13}{,} and message passing \citep[MP-][]{win5}. The mean-{field} VB is the most frequently used method in this category. The VB method is applied in many problems, especially {in} Bayesian regression models \citep[see][and the references therein]{sm06,minka00,wand11,b13,dr13,gol11,ph13,tit14,sea24,bea24,aea24}. The VB method is also applied to the sparse linear regression models \citep{o17,Ray22,tit11,h16}.

In the case of the Poisson regression model, the VB method is applied by a few authors. \cite{lu15} developed VB for Bayesian {semiparametric} regression with count data, using Poisson and Negative Binomial regression models.
\cite{c10} applied {VB} for {point-process} generalized linear models in {neural-spike-train} analysis.

The method used in \cite{lu15} was based on the non-conjugate VB technique. In this method, an assumed density filtering technique is used along with the mean-field VB for approximation of the posterior density. Precisely speaking, since the approximated component for the coefficients of the regression model is not conjugate with its Gaussian prior, we assume a Gaussian component and optimize the mean and the covariance function by minimizing the Kullback-Leibler divergence between the unnormalized posterior and its approximation. Another method in non-conjugate VB is the approximation method proposed by \cite{ja00}, which was successfully applied to the logistic regression problem. In this method, the likelihood function is approximated by a quadratic function, which provides the conjugacy of the approximation component with the Gaussian prior on the regression coefficient. The method of \cite{ja00} has been applied to the sparse logistic regression model by \cite{zh19}.

In this paper, we have employed the VB method for the problem of sparse Poisson regression models. A similar approximation method to that of \cite{ja00} is used to provide conjugacy in the model approximation. Three sparsity-enforcing priors are used for this problem, and the VB approximated formulas are calculated for all three models. The proposed models are compared with each other and two frequentist sparse Poisson methods (LASSO and SCAD) to evaluate the estimation, prediction, and sparsity performance of the proposed methods. Throughout a simulated data example, the accuracy of the VB methods is computed with respect to the corresponding benchmark MCMC methods. {The proposed VB methods yield a close approximation to the posterior distribution of the parameters while exhibiting substantially greater computational efficiency than MCMC methods.} Using several benchmark count response data sets, the prediction performance of the proposed methods is evaluated in real-world applications.

The rest of the paper is organized as follows. In Section 2, the mean-field variational Bayes method is introduced. Three variational Bayesian sparse Poisson regression models are then proposed in Section 3. The derivation of the components, as well as the evidence lower bound (ELBO) for these methods, is given in Appendix A. In Section 4, a hard thresholding method is introduced to derive the final sparse estimates of the regression coefficients. {Section 5 introduces the posterior predictive density function for predicting the response variable based on new covariate samples.} Section 6 presents a simulation study for evaluating the estimation, prediction, and sparsity performance of the three proposed methods and comparing them with two frequentist sparse Poisson regression methods (LASSO and SCAD Poisson regression models) and the associated MCMC methods. A single replication of the simulation study is also used for a visual comparison of the MCMC empirical posteriors with their approximated counterparts. Some of {the plots} and additional analysis are given in { supplementary material} for the sake of {brevity}. Finally, in Section 7, the test set prediction performance of the proposed methods is compared with { that of} the frequentist alternatives, using 6 benchmark count response data sets. Some concluding remarks are given in Section 8.

 \section{Mean field variational Bayes}

Let $\mathbf{x}$ be a vector of observed data, and ${ \boldsymbol\theta}$ be a parameter vector with joint distribution $p(\mathbf{x} , { \boldsymbol\theta})$.
In the Bayesian inference framework, the inference about ${ \boldsymbol\theta}$ is done based on the posterior distribution
$p({ \boldsymbol\theta} |\mathbf{x} ) = p(\mathbf{x}, { \boldsymbol\theta})/p(\mathbf{x})$
where 
$p(\mathbf{x}) = \int  p(\mathbf{x} , { \boldsymbol\theta}) d { \boldsymbol\theta}$.

Variational Bayes (VB) is a method for finding an approximate distribution $q({ \boldsymbol\theta})$ of the posterior distribution $p({ \boldsymbol\theta} |x )$, by minimizing the Kullback-Leibler divergence ${\rm KL} \left[ q({ \boldsymbol\theta}) || p({ \boldsymbol\theta}|x) \right]$ as a measure of closeness \citep{tr23}. In the mean-field VB, we assume that {the parameter} vector ${ \boldsymbol\theta}$ is divided into $M$ partitions  
$ \{ { \boldsymbol\theta}_1, \ldots , { \boldsymbol\theta}_M \} $, and we want to approximate
$p({ \boldsymbol\theta} |x ) $
by 
$$q({ \boldsymbol\theta}) = \prod_{j=1}^{M} q({ \boldsymbol\theta}_j),$$
{\color{black} that is we assume ${ \boldsymbol\theta}_1, \ldots , { \boldsymbol\theta}_M$ to be independent, in the mean-field approximation.}

The best VB approximation 
$q^{\ast}(\cdot)$
is then obtained as 
$$q^{\ast} = \arg \min_q   {\rm KL} \left[ q({ \boldsymbol\theta}) || p({ \boldsymbol\theta}|\mathbf{x}) \right],$$
where
\begin{equation}\label{kl}
{\rm KL} \left[ q({ \boldsymbol\theta}) || p({ \boldsymbol\theta}|\mathbf{x}) \right] = \int q({ \boldsymbol\theta}) \log \frac{q({ \boldsymbol\theta})}{p({ \boldsymbol\theta}| \mathbf{x})} d{ \boldsymbol\theta} =   \log p(\mathbf{x}) - \int q({ \boldsymbol\theta}) \log \frac{p(\mathbf{x},{ \boldsymbol\theta})}{q({ \boldsymbol\theta})} d{ \boldsymbol\theta},
\end{equation}
which results in 
\begin{equation}\label{vbf}
\log q( { \boldsymbol\theta}_j ) = {\rm E}_{(-{ \boldsymbol\theta}_j)} \left[ \log {p}(\mathbf{x},{ \boldsymbol\theta}) \right] {+ {\rm Const.}} , j=1,\ldots,M,
\end{equation}
in which the expectation $ {\rm E}_{(-{ \boldsymbol\theta}_j)} $ is the expectation with respect to 
$$q(-{ \boldsymbol\theta}_j) = \prod_{i(\neq j)=1}^{M} q({ \boldsymbol\theta}_i).$$
It is clear from \eqref{kl} that minimizing KL is equivalent to maximizing the evidence lower bound (ELBO), defined as 
$$ {\rm ELBO} =  \int q({ \boldsymbol\theta}) \log \frac{p(\mathbf{x},{ \boldsymbol\theta})}{q({ \boldsymbol\theta})} d{ \boldsymbol\theta}$$
ELBO is equal to $\log p(x)$ when the KL divergence is zero, which means a perfect fit.
When the fit is not perfect, ${\rm ELBO}[q({ \boldsymbol\theta})] < \log p(\mathbf{x})$.

\section{Variational Bayesian sparse Poisson regression models}

In the following subsections, we consider three different {sparsity-enforcing} priors for the coefficients of the Poisson regression model. Then, we apply an approximation method similar to that of \cite{ja00} to the likelihood function, to provide approximated conjugacy for the regression coefficient vector. Finally, we propose the VB components for the parameters of each model. The computation process of these components is given in Appendix A. 

\subsection{Laplace prior}

Suppose that $y_i \mathop{\sim}\limits^{\mathrm{ind}} \text{Poiss} (\lambda_i = \exp(\beta_0 + X_i { \boldsymbol\beta})), \quad i=1, \ldots , n $, where {\color{black}$\mathbf{X}_i = (X_{i1},\ldots,X_{i(p-1)})$} and ${ \boldsymbol\beta = (\beta_1,\ldots,\beta_{p-1})^\top}$. A {sparsity-enforcing} prior for ${ \boldsymbol\beta}$ is the Laplace prior \citep{see08}. Thus, we consider the independent priors 
$\beta_j \mathop{\sim}\limits^{\mathrm{ind}}  \;\text{Laplace} (0 , { \eta}),\; {\color{black}j=1,\ldots , p-1}$. It is well-known that this prior is identical to the hierarchical model 
$$\beta_j|\tau_j \mathop{\sim}\limits^{\mathrm{ind}}  \;{\rm N} (0 , \tau_j) , \quad\tau_j \mathop{\sim}\limits^{\mathrm{iid}} \;\text{ Exp} \left(\frac{{ \eta}}{2}\right),\quad {\color{black}j=1,\ldots , p-1}.$$
Adding the intercept priors and the hyper-priors, the Bayesian sparse Poisson regression model considered here is as follows
\begin{align}
y_i|\beta_0,{ \boldsymbol\beta} \mathop{\sim}\limits^{\mathrm{ind}} & \;\text{Poiss} (\lambda_i = \exp(\beta_0 + \mathbf{X}_i { \boldsymbol\beta})), \quad i=1, \ldots , n \nonumber\\
\beta_j|\tau_j \mathop{\sim}\limits^{\mathrm{ind}} & \;{\rm N} (0 , \tau_j) , \quad\tau_j|{ \eta} \mathop{\sim}\limits^{\mathrm{iid}} \;\text{ Exp} \left(\frac{{ \eta}}{2}\right),\quad {\color{black} j=1,\ldots , p-1},\nonumber\\
\beta_0|\tau_0 \sim & \;{\rm N} (0 , \tau_0), \;\; 
 \;\;
{ \eta} \sim \text{Gamma}(\nu,\delta)\nonumber \\
\;\tau_0|a \sim & \text{Inv-Gamma} \left(\frac{1}{2},\frac{1}{a}\right), \;\;
a \sim \text{Inv-Gamma}\left(\frac{1}{2}, \frac{1}{A}\right)\label{lapm}
\end{align}

\begin{figure}[!ht]
\centerline{\includegraphics[scale = 0.5]{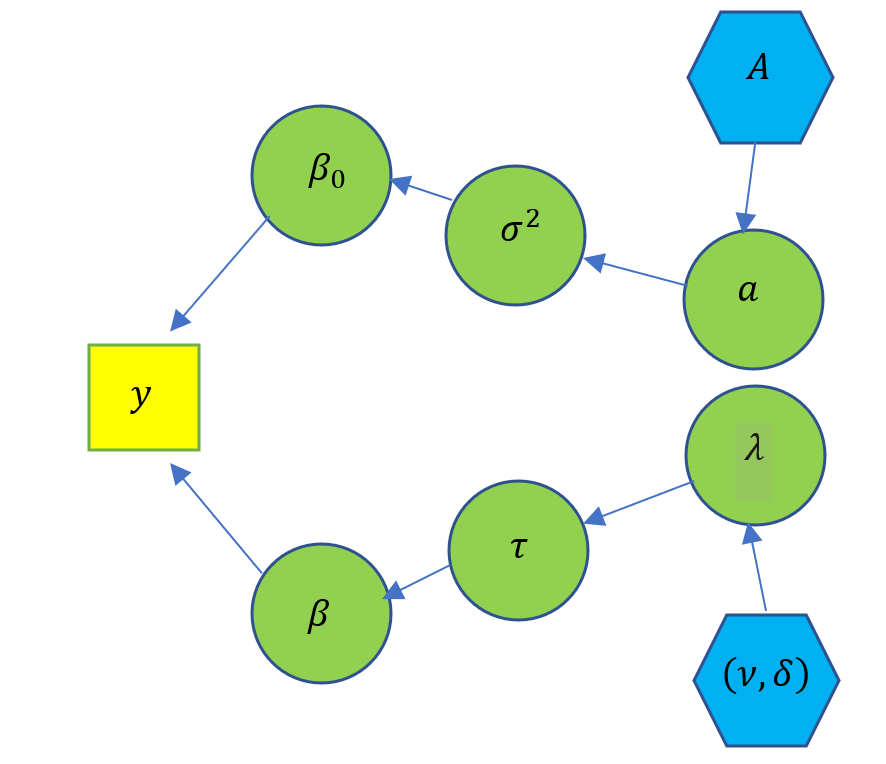}}
\caption{\color{black}A directed acyclic graph (DAG) representing the hierarchical structure of Model \eqref{lapm}. The data hierarchy is shown from top to bottom: observed data (squares) are generated from latent probabilistic processes, which are governed by model parameters (circles) and fixed hyperparameters (polygons). Arrows indicate conditional dependencies; for instance, the response variable $y$ depends on the parameters $\beta$ and $z$, illustrating the model's core data-generating mechanism.}\label{dag1}
\end{figure}

The hierarchical hyper-priors for $\tau_0|a$ and $a$ in \eqref{lapm} {result} in the marginal hyper-prior $\sqrt{\tau_0} \sim {\rm HalfCauchy}(\sqrt{A})$, which is suggested by \cite{lu15}.

{ A graphical representation of the hierarchical model \eqref{lapm} is presented as a directed acyclic graph (DAG)} in Figure \ref{dag1}, which shows the relation between data elements (squares), model parameters (circles), and fixed hyperparameters (polygons). {This figure shows how the priors and hyperpriors of the parameters and hyperparameters are related to each other in a hierarchical structure.}

The mean-field VB approximation density $q$  is assumed to be 
$$q(\beta_0,{ \boldsymbol\beta},{ \boldsymbol\tau},{ \eta},a) = q(\beta_0)q({ \boldsymbol\beta})q({ \boldsymbol\tau})q({ \eta})q(a),$$
where ${ \boldsymbol\tau} = (\tau_0,\tau_1,\ldots,{\color{black}\tau_{p-1}})^\top$.

Because of the $\exp(\cdot)$ in the likelihood function of model \eqref{lapm}, the VB component for ${ \boldsymbol\beta}$ is not conjugate. Similar to the idea proposed by \cite{ja00} (for the logistic regression model), we consider the following { quadratic approximation, for the values of $x$ close to  $\xi,$ 
\begin{equation}\label{inq}
e^x  \approx e^{\xi } \left[ (1-\xi) (1+x)+ \frac{x^2}{2}+ \frac{\xi^2}{2} \right] = g(x,\xi).
\end{equation}
The exact equality holds in \eqref{inq}, when $x = \xi$. Figure \ref{exp} shows this approximation. As one can see from Figure \ref{exp}, we have $e^x > g(x,\xi),$ for $x > \xi$, and $e^x < g(x,\xi),$ for $x < \xi$.

\begin{figure}
\centerline{\includegraphics[scale = 0.5]{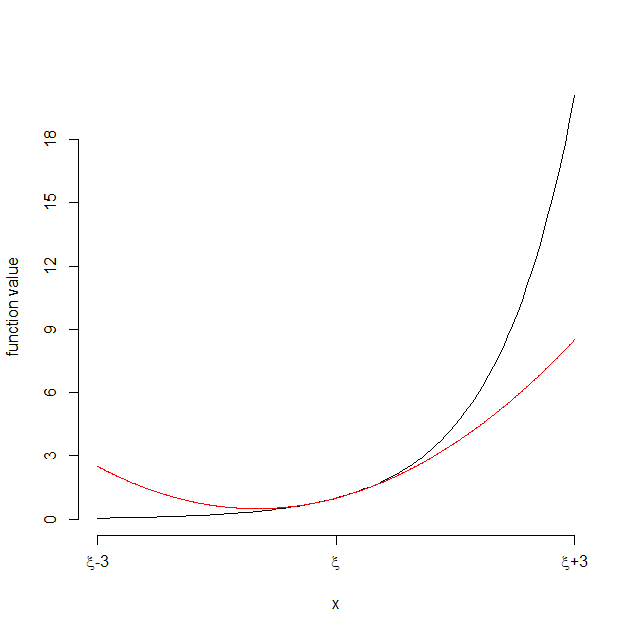}}
\caption{ The exponential function (black line) and its quadratic approximation (red line).}\label{exp}
\end{figure}

\begin{figure}[h]
\centering
\includegraphics[width=0.8\textwidth]{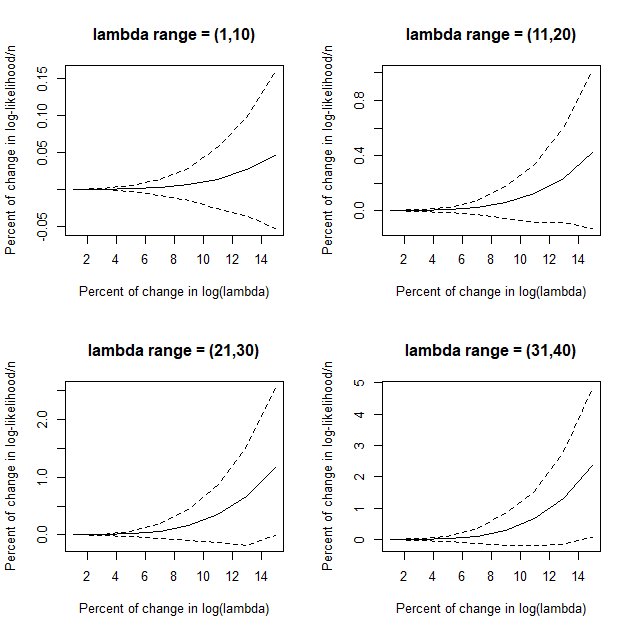}
\caption{\color{black} Approximation error analysis for Taylor-expanded Poisson log-likelihood across different $\lambda$ ranges. The relative error in log-likelihood (as percentage) is shown against the percentage perturbation in $\log(\lambda)$. Solid lines represent mean errors, while dashed lines indicate the 2.5\% and 97.5\% quantiles from 1000 simulations.}
\label{fig:poissonapprox}
\end{figure}

Using the approximation in \eqref{inq}, we have }
\begin{align}
\log p(\mathbf{y} \vert { \boldsymbol\beta}) = & -\sum_{i=1}^{n} \exp(\mathbf{X}_i { \boldsymbol\beta}) + \sum_{i=1}^{n} y_i \mathbf{X}_i { \boldsymbol\beta} - \sum_{i=1}^{n} \log y_i! \nonumber\\
& \approx - \sum_{i=1}^{n} e^{\xi_i} \left[ (1- \xi_i) (1+ \mathbf{X}_i { \boldsymbol\beta})+ \frac{\xi_i^2}{2} + \frac{1}{2} { \boldsymbol\beta}^\top \mathbf{X}_i \mathbf{X}_i^\top { \boldsymbol\beta} \right] \nonumber\\
& + \sum_{i=1}^{n} y_i \mathbf{X}_i { \boldsymbol\beta} - \sum_{i=1}^{n} \log y_i ! \nonumber\\
& := \log \tilde{p}(\mathbf{y}  \vert { \boldsymbol\beta}, \boldsymbol\xi).\label{approx}
\end{align}
{ Since, the equality holds in \eqref{inq} when $\xi= x$, we let} $\xi_i = {\rm E}_q(\mathbf{X}_i{ \boldsymbol\beta}) = \mathbf{X}_i {\color{black}{ \boldsymbol\mu}_{{ \boldsymbol\beta}(L)}}$, for $i=1,\ldots,n$, where ${\color{black}{ \boldsymbol\mu}_{{ \boldsymbol\beta}(L)}}$ is given in \eqref{smub}. { Let $\boldsymbol\xi = (\xi_1,\ldots,\xi_n)^\top$, and} define $M_{\boldsymbol\xi} = e^{\xi} (1 - \xi)$ and  $S_{X}^{\boldsymbol\xi} = \sum_{i=1}^{n}  e^{\xi_i} \mathbf{X}_i \mathbf{X}_i^\top $. 

{\color{black}To evaluate the precision of this approximation, a simulation study was conducted. For each of four different ranges of the Poisson rate parameter $\lambda$—specifically (1,10), (11,20), (21,30), and (31,40)—we generated 1000 datasets. Each dataset contained 100 observations, where the counts $y$ were drawn from a Poisson distribution with rates $\lambda$ drawn uniformly from the respective interval. The approximation error was assessed by introducing controlled perturbations to $\log(\lambda)$, varying the maximum perturbation size from 1\% to 15\%. For each scenario, the relative error between the true and the approximated log-likelihood was computed. The results, visualized in Figure~\ref{fig:poissonapprox}, plot the percentage change in the log-likelihood against the percentage change in $\log(\lambda)$. Each panel displays the mean relative error (solid line) along with the 2.5\% and 97.5\% quantiles (dashed lines) across the simulations. The analysis reveals that the approximation error is generally small (often below 1\%) for modest perturbations and smaller values of $\lambda$. However, both the mean error and its variability tend to increase with larger perturbation sizes and higher $\lambda$ ranges. This confirms that the Taylor approximation remains sufficiently accurate for practical use in VB inference for Poisson regression, provided the variational distribution does not induce excessive dispersion in $\log(\lambda)$, especially when the underlying rates are large.}

To provide conjugacy in the VB component $q({ \boldsymbol\beta})$, we replace $P(\mathbf{y} \vert { \boldsymbol\beta})$ by $\tilde{P}(\mathbf{y} \vert { \boldsymbol\beta},\boldsymbol\xi)$, in the computation of this element. 

We denote this method by Laplace-VB. In Appendix A, we have {shown} that the optimal posterior densities for the parameters are as follows
$$q( { \boldsymbol\beta} ) = {\rm N}({ \boldsymbol\beta}; { \boldsymbol\mu}_{{ \boldsymbol\beta}(L)}, { \boldsymbol\Sigma}_{{ \boldsymbol\beta}(L)})
$$
where ${\rm N}(x; { \boldsymbol\mu}, { \boldsymbol\Sigma})$ stands for the multivariate normal density function and 
\begin{equation}\label{smub}
{ \boldsymbol\Sigma}_{{ \boldsymbol\beta}(L)} =  \left[ S_{X}^{\boldsymbol\xi} + \text{diag}{\rm E}_q({ \boldsymbol\tau}^{-1}) \right] ^{-1}, \quad 
{ \boldsymbol\mu}_{{ \boldsymbol\beta}(L)} =  { \boldsymbol\Sigma}_{{ \boldsymbol\beta}(L)} \mathbf{X}^\top (\mathbf{y} - M_{\boldsymbol\xi}).
\end{equation}
Let $D^{(L)}_{{ \boldsymbol\beta}} = { \boldsymbol\mu}_{{ \boldsymbol\beta}(L)}{ \boldsymbol\mu}_{{ \boldsymbol\beta}(L)}^\top + { \boldsymbol\Sigma}_{{ \boldsymbol\beta}(L)}$. 
Furthermore,  
$$q({\eta}) = \text{Gamma} ({\eta}; { p+ \nu -1} , \delta + \frac{1}{2} \sum_{j=1}^{ p-1} {\rm E}_q(\tau_j)  ),
$$
 ${\rm E}_q({\eta}) = \frac{ { p+ \nu -1}}{\delta + \frac{1}{2} {\sum_{j=1}^{p-1}} {\rm E}_q(\tau_j)}$, 
{\color{black} and for $j = 1,\ldots,p-1$}, 
$$
 q(\tau_j )  = \text{Gen-Inv-Gauss} \Big(\tau_j; \frac{1}{2}, {\rm E}_q({\eta}), {\rm E}_q(\beta_j^2)\Big),
$$
independently, where 
$$ \text{Gen-Inv-Gauss} \Big(x; p_*, a_* , b_* , \Big) = \frac{(a_*/b_*)^{p_*/2}}{2K_p(\sqrt{a_*b_*})}x^{p_*-1}e^{-(a_*x + b_*/x)/2},$$
is the pdf of { the generalized} inverse Gaussian distribution, where $K_p(\cdot)$ is the modified Bessel function of the second kind. Hence, for {$j=1,\ldots,p-1$}, 
$${\rm E}_q(\tau_j) = \frac{\sqrt{{\rm E}_q(\beta_j^2)}K_{3/2}(\sqrt{{\rm E}_q({\eta}){\rm E}_q(\beta_j^2)})}{\sqrt{{\rm E}_q({\eta})}K_{1/2}(\sqrt{{\rm E}_q({\eta}){\rm E}_q(\beta_j^2)})},$$ 
$${\rm E}_q(\tau_j^{-1}) =  \frac{\sqrt{{\rm E}_q({\eta})}K_{3/2}(\sqrt{{\rm E}_q({\eta}){\rm E}_q(\beta_j^2)})}{\sqrt{{\rm E}_q(\beta_j^2)}K_{1/2}(\sqrt{{\rm E}_q({\eta}){\rm E}_q(\beta_j^2)})} - \frac{1}{{\rm E}_q(\beta_j^2)},$$
 and 
$${\rm E}_q(\log \tau_j) = \log\sqrt{\frac{{\rm E}_q(\beta_j^2)}{{\rm E}_q({\eta})}} + \frac{\partial}{\partial t} \log K_t(\sqrt{{\rm E}_q({\eta}){\rm E}_q(\beta_j^2)})|_{t=1/2}.$$
Furthermore,  
$$q(\tau_0) = \text{Inv-Gamma} (\tau_0 ; 1 , \frac{1}{2} d_{00}^{(L)} + {\rm E}_q(a^{-1}) ),
$$
where ${ d_{jj}^{(L)}} = {\rm E}_q(\beta_j^2) = (D^{(L)}_{{ \boldsymbol\beta}})_{jj}$, and thus
${\rm E}_q(\tau_0^{-1}) =  \dfrac{1}{\frac{1}{2} d_{00}^{(L)} + {\rm E}_q(a^{-1})}$ and ${\rm E}_q(\log \tau_0) =   \log (\frac{1}{2}  d_{00}^{(L)} + {\rm E}_q(a^{-1}) )-\psi (1)$, 
where $\psi(\cdot)$ is the digamma function, and  
$$q(a) =  \text{Inv-Gamma} (a ; 1 , {\rm E}_q(\tau_0^{-1}  )+ A^{-1}),
$$
and therefore
${\rm E}_q(a^{-1}) =  \dfrac{1}{{\rm E}_q(\tau_0^{-1}  )+ A^{-1}} $ and 
${\rm E}_q(\log a) = -\psi (1) + \log ({\rm E}_q(\tau_0^{-1}  )+ A^{-1}) )$. 

The approximated ELBO is  calculated as follows {(see Appendix A)}
\begin{align}
\log p(\mathbf{y};q)   & \approx
{\color{black}{\rm E}_q (\log \tilde{p} (\mathbf{y}\vert { \boldsymbol\beta} , \boldsymbol\xi) ) +
{\rm E}_q(\log p({ \boldsymbol\beta} \vert \tau)) +
{\rm E}_q(\log p({ \eta}))  \nonumber\allowdisplaybreaks}\\
& {\color{black} \quad +
{\rm E}_q(\log p(\tau_0 \vert a))+
{\rm E}_q (\log p(a)) 
- {\rm E}_q(\log q({ \boldsymbol\beta}))\nonumber\allowdisplaybreaks}\\
& {\color{black} \quad
- {\rm E}_q (\log q ({ \eta})) 
- {\rm E}_q(\log q (\tau_0) )
- {\rm E}_q(\log q (\tau) ) 
- {\rm E}_q(\log q (a))\nonumber\allowdisplaybreaks}\\
& = - M_{\boldsymbol\xi}^T (1+\mathbf{X} { \boldsymbol\mu}_{{ \boldsymbol\beta}(L)} ) - \frac{1}{2}\sum_{i=1}^{n}  \xi_i^2 e^{\xi_i}  - \frac{1}{2} tr(S_{X}^{\boldsymbol\xi} D_{{ \boldsymbol\beta}}^{(L)}) + \mathbf{y} ^\top \mathbf{X} { \boldsymbol\mu}_{{ \boldsymbol\beta}(L)} \nonumber\allowdisplaybreaks\\
& \quad- \frac{1}{2} {\rm E}_q(\log \tau_0)   - \frac{1}{2} {\rm E}_q(\tau_0^{-1}) {\rm E}_q(\beta_0^2) 
- \dfrac{\delta ({ p+ \nu -1})}{\delta + \frac{1}{2}  {\sum_{j=1}^{p-1}} {\rm E}_q(\tau_j)} \nonumber\allowdisplaybreaks\\
&\quad-  2 \log (A^{-1} + {\rm E}_q(\tau_0^{-1}))- \frac{1}{2} \log (d_{00}^{(L)}/2 + {\rm E}_q(a^{-1}))\nonumber\allowdisplaybreaks \\
& \quad - {\rm E}_q(a^{-1}) {\rm E}_q(\tau_0^{-1}) -  \dfrac{A^{-1}}{A^{-1} + E (\tau_0^{-1})}  +\frac{1}{2} \log \vert { \boldsymbol\Sigma}_{{ \boldsymbol\beta}(L)} \vert\nonumber\allowdisplaybreaks\\
& \quad- \frac{1}{4}  {\sum_{j=1}^{p-1}} \frac{{\rm E}_q({\eta})}{{ d_{jj}^{(L)}}} + {\sum_{j=1}^{p-1}} \log K_{1/2}(\sqrt{{\rm E}_q({\eta}){ d_{jj}^{(L)}}}) + \text{Const.} \label{elbo1}
\end{align} 

Algorithm \ref{al1} presents the algorithm for the Laplace-VB method. 

\begin{algorithm}
    \caption{Laplace-VB method for sparse Poisson regression model.}\label{al1}
    \begin{algorithmic}
        \STATE $\bullet$ Set proper initial values for the hyper-parameters of $q(\cdot)$ functions, {\color{black} using an initial sparse frequentist GLM.}
	\STATE $\bullet$ Set $\epsilon$ equal to an arbitrary small value,
            \WHILE{The absolute relative change in ELBO is greater than $\epsilon$}
                \STATE  $\bullet$ update ${ \boldsymbol\Sigma}_{{ \boldsymbol\beta}(L)} \leftarrow  \left[ S_{X}^{\boldsymbol\xi} + \text{diag}{\rm E}_q({ \boldsymbol\tau}^{-1})\right]^{-1}, $ and  $ { \boldsymbol\mu}_{{ \boldsymbol\beta}(L)} \leftarrow  { \boldsymbol\Sigma}_{{ \boldsymbol\beta}(L)} \mathbf{X}^\top (\mathbf{y} - M_{\boldsymbol\xi}),$
 		\STATE $\bullet$ {update $\xi_i = \mathbf{X}_i { \boldsymbol\mu}_{ \boldsymbol\beta (L)}$, for $i=1,\ldots,n$,}
                \STATE $\bullet$ update ${\rm E}_q(\beta_j^2)= { d_{jj}^{(L)}}  \leftarrow  { \boldsymbol\mu}_{{ \boldsymbol\beta}(L)j}^{2} + { \boldsymbol\Sigma}_{{ \boldsymbol\beta}(L)jj} $, for 
{$j = 0,\ldots, p-1$,}
                \STATE $\bullet$ update ${\rm E}_q({ \eta}) \leftarrow \frac{{ p+ \nu -1}}{\delta + \frac{1}{2} {\sum_{j=1}^{p-1}}{\rm E}_q(\tau_j)}$
		\STATE $\bullet$   update ${\rm E}_q(\tau_j) \leftarrow \frac{\sqrt{{\rm E}_q(\beta_j^2)}K_{3/2}(\sqrt{{\rm E}_q({\eta}){\rm E}_q(\beta_j^2)})}{\sqrt{{\rm E}_q({\eta})}K_{1/2}(\sqrt{{\rm E}_q({\eta}){\rm E}_q(\beta_j^2)})},$ for 
{$j=1,\ldots,p-1$},
		\STATE $\bullet$   update ${\rm E}_q(\tau_j^{-1}) \leftarrow  \frac{\sqrt{{\rm E}_q({\eta})}K_{3/2}(\sqrt{{\rm E}_q({\eta}){\rm E}_q(\beta_j^2)})}{\sqrt{{\rm E}_q(\beta_j^2)}K_{1/2}(\sqrt{{\rm E}_q({\eta}){\rm E}_q(\beta_j^2)})} - \frac{1}{{\rm E}_q(\beta_j^2)},$ for 
{$j=1,\ldots,p-1$},
		\STATE $\bullet$  update  ${\rm E}_q(\tau_0^{-1}) \leftarrow  \dfrac{1}{\frac{1}{2} d_{00}^{(L)} + {\rm E}_q(a^{-1})}$ 
		\STATE $\bullet$ update ${\rm E}_q(a^{-1}) \leftarrow  \dfrac{1}{{\rm E}_q(\tau_0^{-1}  )+ A^{-1}} $,  
		\STATE $\bullet$ calculate ELBO from \eqref{elbo1}
            \ENDWHILE
    \end{algorithmic}
\end{algorithm}

\subsection{Continuous spike and slab prior}

One of the most famous {sparsity-enforcing} priors is the spike and slab prior \citep{mit88,geo97}, which is a mixture of a point mass $\delta(\cdot)$ and a continuous prior, which is usually considered to be Gaussian. Let {$\mathbf{Z} = (Z_1,\ldots,Z_{p-1})^\top$} be a vector of the latent binary {variables}. Then, the spike and slab prior {for} the vector of regression coefficient ${ \boldsymbol\beta}$ {is} defined as
\begin{equation}\label{ssp1}
p({ \boldsymbol\beta} ) = \prod_{j=1}^{ p-1} \left[ Z_j {\rm N} (\beta_j;0, { \tau^2} ) + ( 1 -Z_j ) \delta(\beta_j) \right],
\end{equation}
where $\delta(\beta_j) = 1,$ if $\beta_j = 0$ and $=0$, otherwise. 

Continuous relaxations of \eqref{ssp1}, with $\delta(\cdot)$ replaced by a peaked
continuous density, {are} considered by many authors \citep[see e.g.,][among others]{geo93,ir03,ir05,Ro18}. One of the most famous variations of the continuous spike and slab prior is as follows 
\begin{equation}\label{ssp1}
p({ \boldsymbol\beta} ) = \prod_{j=1}^{ p-1} \left[ Z_j {\rm N} ({ \boldsymbol\beta}_j;0, { \tau^2} ) +
 ( 1 -Z_j ) {\rm N} ({ \boldsymbol\beta}_j; 0 , c { \tau^2} ) \right],
\end{equation}
in which the constant $c$ is small enough to enforce sparsity to ${ \boldsymbol\beta}_j$. {\color{black}Several sensitivity analysis was conducted to assess the influence of the hyperparameter $c$, which governs the variance of the spike component in the continuous spike-and-slab prior. The model's predictive performance was evaluated on some real datasets (see Section 7)  using the test relative prediction error for values of $c$ ranging from $0.001$ to $0.00001$. The results indicated that the predictive error was completely invariant to the choice of $c$ within this interval. This robustness suggests that the variable selection mechanism performs consistently, effectively shrinking irrelevant coefficients to zero provided $c$ is chosen sufficiently small. Based on this observed insensitivity, the value $c = 0.001$ was maintained for all reported analyses.}

\begin{figure}
\centerline{\includegraphics[scale = 0.5]{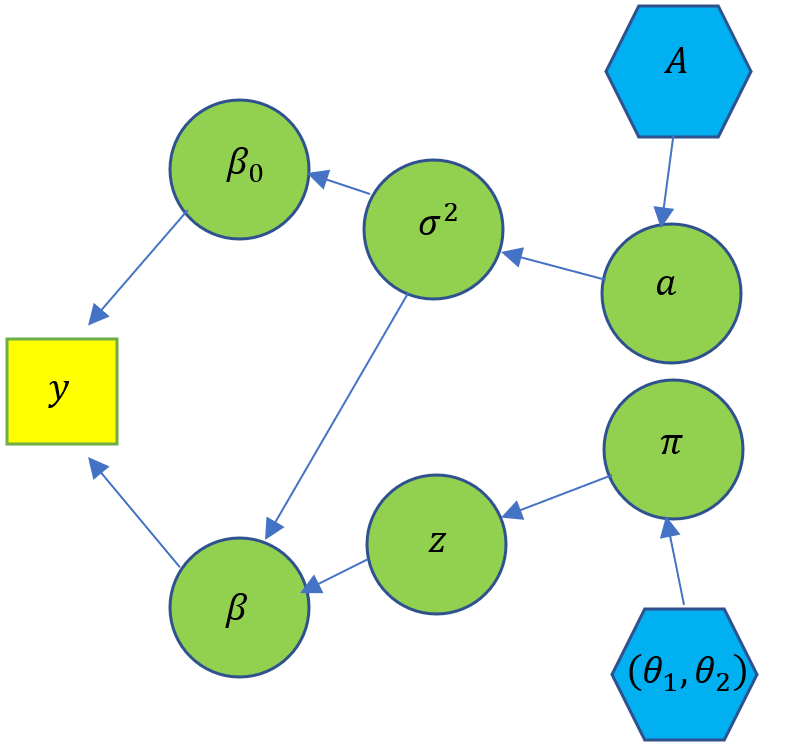}}
\caption{The directed acyclic graph (DAG) for model \eqref{model2}. { Squares denote data elements, circles represent model parameters, and polygons indicate fixed hyperparameters.}}\label{dag2}
\end{figure}

Thus, the model is assumed to be as follows
\begin{align}
y_{i}|{ \boldsymbol\beta} \mathop{\sim}\limits^{\mathrm{ind}} &  \;\text{Poiss} ( \lambda_{i} = \exp (\mathbf{X}_{i} { \boldsymbol\beta} ) ) , \quad  i=1, \ldots , n, \nonumber\\
\mathbf{X}_i = & \;(1,X_{i1},\ldots,{\color{black}X_{i(p-1)}}), \; { \boldsymbol\beta = (\beta_0,\beta_1,\ldots,\beta_{p-1})^\top} \nonumber\\
p( { \boldsymbol\beta} \vert \mathbf{Z}, { \tau^2}) = & \; \prod_{j=0}^{ p-1} \left[ Z_j {\rm N} (0 , { \tau^2} ) + ( 1 - Z_j ) {\rm N} (0 , c { \tau^2} ) \right] , \nonumber\\
Z_0 = 1, & \;\;{\rm w.p.1} \nonumber\\
Z_j|\pi_j  \mathop{\sim}\limits^{\mathrm{ind}} & \;\text{Ber} ( \pi_j ) ,  \;\; 
\pi_j \mathop{\sim}\limits^{\mathrm{iid}}  \text{Beta}  ( {\rho_1 , \rho_2}  )  \quad { j = 1 , \ldots , p-1} ,\nonumber\\
{ \tau^2}|a \sim &\; \text{Inv-Gamma} \left( \dfrac{1}{2} , \dfrac{1}{a} \right), \;\;
a \sim  \;\text{Inv-Gamma} \left( \dfrac{1}{2} , \dfrac{1}{A} \right). \label{model2}
\end{align}
The corresponding DAG for model \eqref{model2} is presented in Figure \ref{dag2}. 
Again, we use the approximation given in \eqref{approx} to provide conjugacy in VB components. Here,
we let $\xi_i = \mathbf{X}_i { \boldsymbol\mu}_{{ \boldsymbol\beta}}$, for $i=1,\ldots,n$, with ${ \boldsymbol\mu}_{{ \boldsymbol\beta}}$ is given in \eqref{smub2}.

This method { is referred to as} CS-VB. The optimal posterior densities for the parameters are computed in Appendix A and are as follows 
$$q( { \boldsymbol\beta} ) = {\rm N}({ \boldsymbol\beta}; { \boldsymbol\mu}_{{ \boldsymbol\beta}(C)}, { \boldsymbol\Sigma}_{{ \boldsymbol\beta}(C)})
$$
where 
$${ \boldsymbol\Sigma}_{{ \boldsymbol\beta}(C)} =  \left[ S_{X}^{\boldsymbol\xi} +  {\rm E}_q (\sigma^{-2} ) \left[\text{diag} ({ P^{(C)}}) + c ^{-1} ( \text{I} - \text{diag} ({ P^{(C)}})) \right] \right] ^{-1}
$$
and 
\begin{equation}\label{smub2}
{ \boldsymbol\mu}_{{ \boldsymbol\beta}(C)} =  { \boldsymbol\Sigma}_{{ \boldsymbol\beta}(C)} \mathbf{X}^\top (\mathbf{y} - M_{\boldsymbol\xi}).
\end{equation}
Let $D^{(C)}_{{ \boldsymbol\beta}} = { \boldsymbol\mu}_{{ \boldsymbol\beta}(C)}{ \boldsymbol\mu}_{{ \boldsymbol\beta}(C)}^\top + { \boldsymbol\Sigma}_{{ \boldsymbol\beta}(C)}$. 
Furthermore,  
$$q({ \tau^2}) = \text{Inv-Gamma}\left({ \tau^2}; \alpha_{{ \tau^2}} ,\beta_{{ \tau^2}} \right),
$$
where $ \alpha_{{ \tau^2}} = \dfrac{p-1}{2} $ and 
$$\beta_{{ \tau^2}} = \frac{1}{2} \text{tr(diag}({ P^{(C)}}) D_{{ \boldsymbol\beta}}^{(C)}) + \dfrac{1}{2c}\text{tr(diag}(1-{ P^{(C)}}) D_{{ \boldsymbol\beta}}^{(C)}) + {\rm E}_q(a^{-1}),$$
and thus, 
${\rm E}_q(\log { \tau^2}) = \log(\beta_{{ \tau^2}}) - \psi(\alpha_{{ \tau^2}}),$
and
${\rm E}_q({\tau^{-2}}) = \frac{\alpha_{{ \tau^2}}}{\beta_{{ \tau^2}}},$
$$q(a) = \text{Inv-Gamma}\left(a ; 1 , {\rm E}_q({\tau^{-2}})+ {A}^{-1}\right),
$$
and thus, 
${\rm E}_q(\log a) = \log\left( {\rm E}_q({\tau^{-2}})+{A}^{-1}\right) - \psi(1),$
and
${\rm E}_q(a^{-1}) = ({{\rm E}_q({\tau^{-2}})+ {A}^{-1}})^{-1},$
$$q(Z_j)  =\text{Ber} (Z_j; { P_j^{(C)}}),\quad { j=1,\ldots,p-1},
$$
where 
\begin{equation}\label{pj}
{ P_j^{(C)}} = \sigma \left({\rm E}_q(\log \pi_j) - {\rm E}_q(\log (1-\pi_j)) - \frac{1}{2} {\rm E}_q({\tau^{-2}}) { d_{jj}^{(C)}} (1-\frac{1}{c})\right), 
\end{equation}
and $ { P^{(C)}} = (1,{ P_1^{(C)}},\ldots,{ P_{p-1}^{(C)}}), $
{ in which $\sigma(v) = (1+\exp(-v))^{-1}$ is the sigmoid function,} ${ d_{jj}^{(C)}} = (D_{{ \boldsymbol\beta}}^{(C)})_{jj}$ ,and
$$q(\pi_j) = \text{Ber} (\pi_j;{ P_j^{(C)}} +{\rho_1} , {\rho_2} - { P_j^{(C)}} +1),
$$
and therefore 
${\rm E}_q(\log \pi_j) = \psi({\rho_1} + { P_j^{(C)}}) - \psi({\rho_1 + \rho_2} +1)$
and
$ {\rm E}_q(\log (1-\pi_j))= \psi({\rho_2} - { P_j^{(C)}} + 1) - \psi({\rho_1 + \rho_2} +1)$.

The approximated ELBO is also obtained as follows {(see Appendix A)}
\begin{align}
\log p(\mathbf{y};q) 
& \approx
{{\rm E}_q (\log \tilde{p} (\mathbf{y}\vert { \boldsymbol\beta} , \boldsymbol\xi) ) +
{\rm E}_q(\log p({ \boldsymbol\beta} \vert \mathbf{Z},\tau^2)) +
{\rm E}_q(\log p({ \mathbf{Z} \vert \pi}))  \nonumber\allowdisplaybreaks}\\
& { \quad +
{\rm E}_q(\log p(\tau^2 \vert a))+
{\rm E}_q (\log p(a)) +
{\rm E}_q (\log p(\pi)) 
- {\rm E}_q(\log q({ \boldsymbol\beta}))\nonumber\allowdisplaybreaks}\\
& { \quad
- {\rm E}_q (\log q ({ \mathbf{Z}})) 
- {\rm E}_q(\log q (\tau^2) )
- {\rm E}_q(\log q (\pi) ) 
- {\rm E}_q(\log q (a))\nonumber\allowdisplaybreaks}\\
= &  - M_{\boldsymbol\xi}^T (1 + \mathbf{X} { \boldsymbol\mu}_{{ \boldsymbol\beta}(C)}) - \frac{1}{2}\sum_{i=1}^{n} e^{\xi_i} \xi_i^2   - \frac{1}{2} tr(S_{X}^{\boldsymbol\xi} D_{{ \boldsymbol\beta}}^{(C)}) +\mathbf{y} ^\top \mathbf{X} { \boldsymbol\mu}_{{ \boldsymbol\beta}(C)}\allowdisplaybreaks \nonumber \\
&  -\frac{1}{2} (1- \frac{1}{c}) {\rm E}_q({ \tau^{-2}}) {\sum_{j=1}^{p-1}} { P_j^{(C)}} d_{jj}^{(C)}  - \frac{1}{2c} {\rm E}_q({ \tau^{-2}}) {\sum_{j=1}^{p-1}} { d_{jj}^{(C)}}\allowdisplaybreaks \nonumber \\ 
& - {\left(1+\frac{p}{2}\right)} {\rm E}_q(\log { \tau^2}) - \log (A^{-1} + {\rm E}_q({ \tau^{-2}}) ) \allowdisplaybreaks \nonumber \\
& - \frac{1}{2}{\rm E}_q({ \tau^{-2}}){\rm E}_q(a^{-1}) + {\sum_{j=1}^{p-1}} { P_j^{(C)}}  E\left(\frac{\log \pi_j}{\log (1-\pi_j)}\right) \allowdisplaybreaks \nonumber \\
& + ({ \rho}_1 -1 )    {\sum_{j=1}^{p-1}} {\rm E}_q(\log \pi_j)+  ({ \rho}_2 -1 )    {\sum_{j=1}^{p-1}}  {\rm E}_q(\log(1- \pi_j)) \allowdisplaybreaks \nonumber\\
& - {{\rm E}_q(a^{-1})}/{A} +\frac{1}{2} \log \vert { \boldsymbol\Sigma}_{{ \boldsymbol\beta}(C)} \vert - \alpha_{{ \tau^2}} \log { \boldsymbol\beta}_{{ \tau^2}} \allowdisplaybreaks \nonumber\\
& + \log \Gamma (\alpha_{{ \tau^2}}) +(\alpha_{{ \tau^2}} -1 ) {\rm E}_q(\log { \tau^2}) + { \boldsymbol\beta}_{{ \tau^2}} {\rm E}_q({ \tau^{-2}})   \allowdisplaybreaks\nonumber \\
& -E \Big[ {\sum_{j=1}^{p-1}} { P_j^{(C)}} \log ({ P_j^{(C)}}) + (1-{ P_j^{(C)}}) \log(1-{ P_j^{(C)}}) \Big] + \text{Const.}\label{elbo2}
\end{align}

Algorithm \ref{al2} presents the algorithm for the CS-VB method. 

\begin{algorithm}
    \caption{CS-VB method for sparse Poisson regression model.}\label{al2}
    \begin{algorithmic}
        \STATE $\bullet$ Set proper initial values for the hyper-parameters of $q(\cdot)$ functions, {\color{black} using an initial sparse frequentist GLM.}
	\STATE $\bullet$ Set $\epsilon$ equal to an arbitrary small value,
            \WHILE{The absolute relative change in ELBO is greater than $\epsilon$}
                \STATE $\bullet$ update 
${ \boldsymbol\Sigma}_{{ \boldsymbol\beta}(C)} \leftarrow  \left[ S_{X}^{\boldsymbol\xi} +  E ({ \tau^{-2}} ) \left[\text{diag} ({ P^{(C)}}) + c ^{-1} ( \text{I} - \text{diag} ({ P^{(C)}})) \right] \right] ^{-1}$ 
                \STATE $\bullet$ update 
${ \boldsymbol\mu}_{{ \boldsymbol\beta}(C)} \leftarrow  { \boldsymbol\Sigma}_{{ \boldsymbol\beta}(C)} \mathbf{X}^\top (\mathbf{y} - M_{\boldsymbol\xi}),$ and ${ d_{jj}^{(C)}} = { \boldsymbol\mu}_{{ \boldsymbol\beta}(C) j}^2 + { \boldsymbol\Sigma}_{{ \boldsymbol\beta}(C) jj}$
  		\STATE $\bullet$ {update $\xi_i = \mathbf{X}_i { \boldsymbol\mu}_{ \boldsymbol\beta (C)}$, for $i=1,\ldots,n$,}
               \STATE $\bullet$ update ${\rm E}_q({ \tau^{-2}}) \leftarrow{\dfrac{p-1}{2}}\left(\frac{1}{2} \text{tr(diag}({ P^{(C)}}) D_{{ \boldsymbol\beta}}^{(C)}) + \dfrac{1}{2c}\text{tr(diag}(1-{ P^{(C)}}) D_{{ \boldsymbol\beta}}^{(C)})+ {\rm E}_q(a^{-1})\right)^{-1},$
               \STATE $\bullet$ update ${\rm E}_q(a^{-1}) = \left({\rm E}_q({ \tau^{-2}} )+A^{-1}\right)^{-1},
$
               \STATE $\bullet$ update ${\rm E}_q(\log \pi_j) = \psi({ \rho}_1 + { P_j^{(C)}}) - \psi({ \rho}_1 + { \rho}_2 +1)$
and
$ {\rm E}_q(\log (1-\pi_j))= \psi({ \rho}_2 - { P_j^{(C)}} + 1) - \psi({ \rho}_1 + { \rho}_2 +1)$,
               \STATE $\bullet$ update ${ P_j^{(C)}} \leftarrow \sigma \left({\rm E}_q(\log\pi_j) - {\rm E}_q(\log (1-\pi_j))- \frac{1}{2} {\rm E}_q({ \tau^{-2}}) { d_{jj}^{(C)}} (1-\frac{1}{c})\right), $ 
		\STATE $\bullet$ calculate ELBO from \eqref{elbo2}
            \ENDWHILE
    \end{algorithmic}
\end{algorithm}

\subsection{Bernoulli sparsity-enforcing prior}

To enforce sparsity {in} the regression coefficient vector ${ \boldsymbol\beta}$, \cite{zh19} and \cite{o17} have used the product of a diagonal matrix $\Gamma$ and the vector ${ \boldsymbol\beta}$, in which the diagonal elements of $\Gamma$ are Bernoulli distributed. When the $j$th diagonal element of $\Gamma$ is zero, the effect of ${ \boldsymbol\beta}_j$ is removed from the linear regression function. \cite{zh19} used this prior for the logistic regression model. { As mentioned by \cite{o17} this model is sometimes called the Bernoulli-Gaussian \citep{sea11} and is closely related to $\ell_0$ regularization and the spike and slab prior \citep{wo11}.}

\begin{figure}
\centerline{\includegraphics[scale = 0.5]{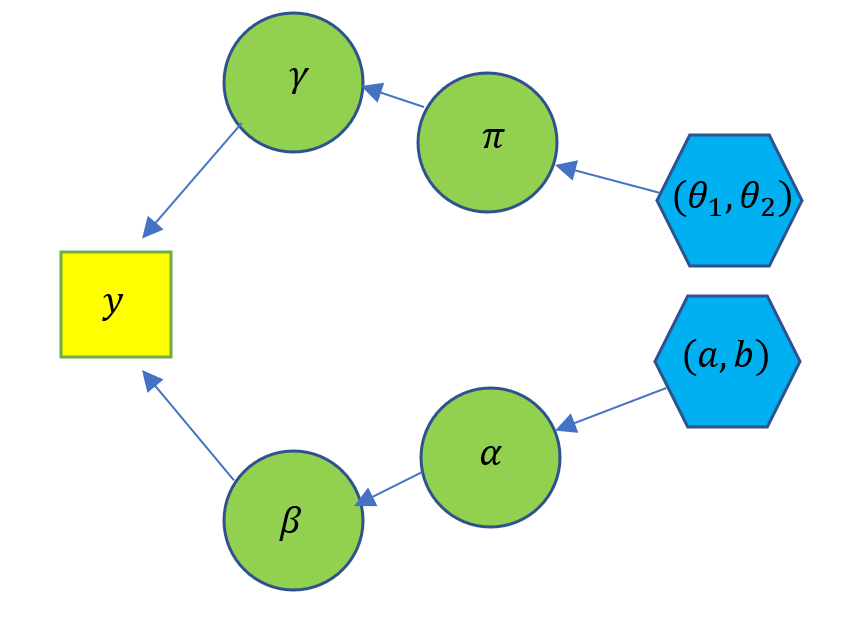}}
\caption{The directed acyclic graph (DAG) for model \eqref{model3}. { Squares denote data elements, circles represent model parameters, and polygons indicate fixed hyperparameters.}}\label{dag3}
\end{figure}

Thus, the Bayesian sparse Poisson model with a Bernoulli prior is as follows
\begin{align}
y_i|{ \boldsymbol\beta} \mathop{\sim}\limits^{\mathrm{ind}} &\text{Poiss} (\lambda_i = \exp(\mathbf{X}_i \Gamma { \boldsymbol\beta})), \quad i=1, \ldots ,n\nonumber\\
\mathbf{X}_i = & (1, X_{i1}, \ldots , X_{i(p-1)}), \quad \Gamma = \text{diag} (1 , \gamma_1 , \ldots , \gamma_{p-1} ), \nonumber\\
 { \boldsymbol\beta} & = (\beta_0,\beta_1,\ldots,\beta_{p-1})^\top,  \quad  {\boldsymbol\gamma} = (\gamma_1 , \ldots , \gamma_{p-1} )^\top \nonumber\\
{ \boldsymbol\beta}| \boldsymbol\alpha \sim & \;{\rm N}_{p+1} (0 , {\rm diag}( \boldsymbol\alpha^{-1})) \nonumber\\
 \boldsymbol\alpha = & \;(\alpha_0 , \alpha_1 , \ldots , \alpha_{p-1})^\top, \;\;
\alpha_j \mathop{\sim}\limits^{\mathrm{ind}}  \text{Gamma} (a_j , b_j) ,\; j= 0, \ldots , {p-1} \nonumber\\
\gamma_j|\pi_j \mathop{\sim}\limits^{\mathrm{ind}} & \text{Ber} (\pi_j) , \;\;
\pi_j \mathop{\sim}\limits^{\mathrm{iid}}  \text{Beta} ({ \rho}_1 , { \rho}_2) \quad j= 1 , \ldots , {p-1}, \label{model3}
\end{align}
The corresponding DAG for model \eqref{model3} is presented in Figure \ref{dag3}. 
Similar to the approximation given in \eqref{approx}, we find an upper bound for the likelihood function as follows, to provide conjugacy in VB components
\begin{align*}
\log p(y_i ;  { \boldsymbol\beta} , { \boldsymbol\gamma})= & - \exp (\mathbf{X}_i \Gamma { \boldsymbol\beta}) + y_i \mathbf{X}_i \Gamma { \boldsymbol\beta} - \log y_i !\\
{ \approx}  & - e^{\xi_i} \Big[ (1 - \xi_i ) (1+\mathbf{X}_i \Gamma { \boldsymbol\beta}) +\dfrac{\xi_i^2}{2} \\
&   + \frac{1}{2} { \boldsymbol\beta}^\top \Gamma \mathbf{X}_i^\top \mathbf{X}_i \Gamma { \boldsymbol\beta} \Big] + y_i \mathbf{X}_i \Gamma { \boldsymbol\beta} + \log y_i ! \\
:= & \log \tilde{p}(y_i \vert  { \boldsymbol\beta}, { \boldsymbol\gamma}, \xi_i).
\end{align*}
Here, we let $\xi_i = \mathbf{X}_i {\rm diag}({ P^{(B)}}) { \boldsymbol\mu}_{{ \boldsymbol\beta}(B)}$, for $i=1,\ldots,n$, with ${ \boldsymbol\mu}_{{ \boldsymbol\beta}(B)}$ and  ${ P^{(B)}}$ given in \eqref{smub3} and \eqref{PP}, respectively.

We denote this method by Bernoulli-VB. In Appendix A, we prove that the optimal posterior densities for the parameters are as follows
$$q({ \boldsymbol\beta}) = {\rm N}({ \boldsymbol\beta}; { \boldsymbol\mu}_{{ \boldsymbol\beta}(B)}, { \boldsymbol\Sigma}_{{ \boldsymbol\beta}(B)}) $$
where
\begin{align*}
{ \boldsymbol\Sigma}_{{ \boldsymbol\beta}(B)} = \left[  S_X^{\boldsymbol\xi}   \odot \Omega + \text{diag} \Big( {\rm E}_q(\alpha) \Big) \right]^{-1},
\end{align*}
\begin{equation}\label{smub3}
{ \boldsymbol\mu}_{{ \boldsymbol\beta}(B)} =  { \boldsymbol\Sigma}_{{ \boldsymbol\beta}(B)} \text{diag} ({ P^{(B)}}) \mathbf{X}^\top ( \mathbf{y} - M_{\boldsymbol\xi})
\end{equation}
in which 
$ \Omega =  ({ P^{(B)}}) ({ P^{(B)}}) ^\top + \text{diag}({ P^{(B)}})  ({ I_{p}} - \text{diag}({ P^{(B)}}) )$. Let $D^{(B)}_{{ \boldsymbol\beta}} = { \boldsymbol\mu}_{{ \boldsymbol\beta}(B)}{ \boldsymbol\mu}_{{ \boldsymbol\beta}(B)}^\top + { \boldsymbol\Sigma}_{{ \boldsymbol\beta}(B)}$. 
Furthermore,  for $j = 1,\ldots, {p-1}$
\begin{align*}
q(\gamma_j) = & \text{Ber} (\gamma_j;{ P_j^{(B)}})
\end{align*}
where 
\begin{equation}\label{PP}
{ P_j^{(B)}} = \sigma\left((\mathbf{y} - M_{\boldsymbol\xi})^T X_j  { \boldsymbol\mu}_{{ \boldsymbol\beta}(B) j} - \frac{1}{2}  S_{X_{jj}}^{\xi} d_{jj}^{(B)}  - \frac{1}{2} \sum_{i\neq j}  P_i^{(B)} S_{X_{ij}}^{\xi} d_{ij}^B + {\rm E}_q\left(\log \frac{\pi_j}{1-\pi_j}\right)\right),  
\end{equation}
and $ { P^{(B)}} = (1,{ P_1^{(B)}},\ldots,{ P_{p-1}^{(B)}}), $ 
$$q(\alpha_j) = \text{Gamma} \left(\alpha_j; a_j+\frac{1}{2} , b_j + \frac{d_{jj}^{(B)}}{2}\right),
$$
thus, ${\rm E}_q(\alpha_j) = {(a_j+{1}/{2})}/{(b_j + {d_{jj}^{(B)}}/{2})},$ 
${\rm E}_q(\log \alpha_j) = \psi(a_j+{1}/{2}) - \log(b_j + {d_{jj}^{(B)}}/{2})$, 
and 
$$q(\pi_j) = \text{Beta} (\pi_j; { P_j^{(B)}} + { \rho}_1  , { \rho}_2 - { P_j^{(B)}}+1 ),
$$
and hence, ${\rm E}_q(\log\pi_j) = \psi({ P_j^{(B)}} + { \rho}_1) - \psi({ \rho}_1 + { \rho}_2 +1 ),$
and ${\rm E}_q(\log(1-\pi_j)) = \psi({ \rho}_2 - { P_j^{(B)}} +1) - \psi({ \rho}_1 + { \rho}_2 +1 ).$

Furthermore, ELBO is approximated as follows {(see Appendix A)}
\begin{align}
\log p(\mathbf{y};q)
& \approx
{{\rm E}_q (\log \tilde{p} (\mathbf{y}\vert { \boldsymbol\beta} , \boldsymbol\gamma,  \boldsymbol\xi) ) +
{\rm E}_q(\log p({ \boldsymbol\beta} \vert \boldsymbol\alpha)) +
{\rm E}_q(\log p({  \boldsymbol\gamma \vert \pi}))  \nonumber\allowdisplaybreaks}\\
& { \quad +
{\rm E}_q(\log p(\boldsymbol\alpha))+
{\rm E}_q (\log p(\pi)) 
- {\rm E}_q(\log q({ \boldsymbol\beta}))\nonumber\allowdisplaybreaks}\\
& { \quad
- {\rm E}_q (\log q ({ \boldsymbol\gamma})) 
- {\rm E}_q(\log q (\boldsymbol\alpha) )
- {\rm E}_q(\log q (\pi) )\nonumber\allowdisplaybreaks}\\
 = & (\mathbf{y}- M_{\boldsymbol\xi})^\top \mathbf{X} { P^{(B)}} { \boldsymbol\mu}_{{ \boldsymbol\beta}(B)} - \frac{1}{2} \text{tr} \left[D^{(B)}_{{ \boldsymbol\beta}} (S_X^{\boldsymbol\xi} \odot \Omega)\right] + \allowdisplaybreaks\nonumber\\
 &- \sum_{i=1}^{n} \big( e^{\xi_i} + \xi_i - \frac{\xi_i^2}{2} \big) + \frac{1}{2} {\sum_{j=1}^{p-1}} {\rm E}_q(\log \alpha_j) \allowdisplaybreaks\nonumber\\
 & - \frac{1}{2}\text{tr} \big[D^{(B)}_{{ \boldsymbol\beta}}  \text{diag} ({\rm E}_q(\alpha))  \big] +\frac{1}{2} \log \big(\vert{ \boldsymbol\Sigma}_{{ \boldsymbol\beta}(B)}\vert \big) \allowdisplaybreaks\nonumber\\
 & -{\sum_{j=1}^{p-1}} [{ P_j^{(B)}} \log { P_j^{(B)}} + (1 - { P_j^{(B)}}) \log (1- { P_j^{(B)}})]\allowdisplaybreaks\nonumber\\
 & +{\sum_{j=1}^{p-1}} \log \Gamma ({ \rho}_1 + { P_j^{(B)}}) + {\sum_{j=1}^{p-1}}\log \Gamma ({ \rho}_2 - { P_j^{(B)}} + 1)- \frac{1}{2} {\sum_{j=1}^{p-1}} {\rm E}_q(\log \alpha_j)\allowdisplaybreaks\nonumber\\ 
 & - {\sum_{j=1}^{p-1}}\big[(a_j + 1/2)\log(b_j + { d_{jj}^{(B)}}/2) -  b_j \frac{a_j + 1/2}{b_j+{{ d_{jj}^{(B)}}}/{2}} \big]+ \text{Const.}\label{elbo3}
\end{align}

Algorithm \ref{al3} presents the algorithm for the Bernoulli-VB method. 

\begin{algorithm}
    \caption{Bernoulli-VB method for sparse Poisson regression model.}\label{al3}
    \begin{algorithmic}
        \STATE $\bullet$ Set proper initial values for the hyper-parameters of $q(\cdot)$ functions, {\color{black} using an initial sparse frequentist GLM.}
	\STATE $\bullet$ Set $\epsilon$ equal to an arbitrary small value,
            \WHILE{The absolute relative change in ELBO is greater than $\epsilon$}
                \STATE $\bullet$ update ${ \boldsymbol\Sigma}_{{ \boldsymbol\beta}(B)} \leftarrow \left[  S_X^{\boldsymbol\xi}   \odot \Omega + \text{diag} \Big( {\rm E}_q(\alpha) \Big) \right]^{-1},$ $ \Omega =  ({ P^{(B)}}) ({ P^{(B)}})^\top + \text{diag} ({ P^{(B)}}) ({ I_{p}} - \text{diag} ({ P^{(B)}})), $
                \STATE $\bullet$ update 
${ \boldsymbol\mu}_{{ \boldsymbol\beta}(B)} \leftarrow  { \boldsymbol\Sigma}_{{ \boldsymbol\beta}(B)} \text{diag} ({ P^{(B)}}) \mathbf{X}^\top ( \mathbf{y} - M_{\boldsymbol\xi}),$ 
                \STATE $\bullet$ update ${\rm E}_q(\alpha_j) = {(a_j+1/2)}/{(b_j + {{ d_{jj}^{(B)}}}/{2})},$ for $j = 1,\dots,{p-1},$
                \STATE $\bullet$ update  
${\rm E}_q(\log\pi_j) = \psi({ P_j^{(B)}} + { \rho}_1) - \psi({ \rho}_1 + { \rho}_2 +1 ),$
and ${\rm E}_q(\log(1-\pi_j)) = \psi({ \rho}_2 - { P_j^{(B)}} +1) - \psi({ \rho}_1 + { \rho}_2 +1 ),$ for $j = 1,\dots,{p-1},$
               \STATE $\bullet$ update ${ P_j^{(B)}} = \sigma\left((\mathbf{y} - M_{\boldsymbol\xi})^T X_j  { \boldsymbol\mu}_{{ \boldsymbol\beta}(B)j} - \frac{1}{2}  S_{X_{jj}}^{\xi} { d_{jj}^{(B)}}  - \frac{1}{2} \sum_{i\neq j}  P_i^{(B)} S_{X_{ij}}^{\xi} d_{ij}^{(B)} + {\rm E}_q\left(\log \frac{\pi_j}{1-\pi_j}\right)\right), $ for $j = 1,\dots,{p-1},$
 		\STATE $\bullet$ {update $\xi_i = \mathbf{X}_i {\rm diag}({ P^{(B)}}){ \boldsymbol\mu}_{ \boldsymbol\beta(B)}$, for $i=1,\ldots,n$,}
		\STATE $\bullet$ calculate ELBO from \eqref{elbo3}
            \ENDWHILE
    \end{algorithmic}
\end{algorithm}

\section{Sparsity-enforcing thresholds}

Due to the structure of the VB methods, the posterior means of the coefficients (${ \boldsymbol\mu}_{ \boldsymbol\beta(M)}$, $M L, C, B$) are not sparse estimators. To obtain a sparse estimator, we propose sparsity-enforcing thresholds.

{For the Bernoulli-VB method, we let 
$$\widehat{\beta}_j^B = \left\{
\begin{array}{lr}
{ \boldsymbol\mu}_{{ \boldsymbol\beta}(B)j}, & { P_j^{(B)}} > 0.5.\\
0, & {\rm o.w.} 
\end{array}
\right.$$
Also, we update ${ P_j^{(B)}} \leftarrow I({ P_j^{(B)}} > 0.5)$. 
 For the Laplace-VB and CS-VB methods, we use the hard threshold (subset selection) strategy \citep[see e.g.,][]{dj94}, that is 
$$\widehat{\beta}_j^M = \left\{
\begin{array}{lr}
{ \boldsymbol\mu}_{{ \boldsymbol\beta}(M)j}, & |{ \boldsymbol\mu}_{{ \boldsymbol\beta}(M)j}|>\hat{\kappa},\\
0, & {\rm o.w.,} 
\end{array}
\right.$$
for $M = L, C$, where $\hat{\kappa}$ is optimally selected over a grid of values by minimizing Akaike Information Criterion 
$${\rm AIC} = -\log \widehat{L}({ \boldsymbol\beta}|x) + 2 {\rm df},$$
where $\widehat{L}({ \boldsymbol\beta}|x)$ is the estimated likelihood of the model and ${\rm df}$ is the { degrees of freedom} of the model, that is, the number of nonzero coefficients in the regression model. 

{\color{black}
The AIC criterion is asymptotically equivalent to leave-one-out cross-validation and is specifically designed to find the model with the best predictive performance. We have preferred AIC rather than BIC, since the BIC criterion is less effective for prediction in finite samples, especially when the true model is complex and may not be in the candidate set. BIC tends to select more parsimonious models than AIC, which can lead to underfitting and poorer predictive performance. For the hard thresholding procedure, when the dimension is high, we need to evaluate a large number of potential models. Using AIC provided a computationally efficient and deterministic criterion without the need for extensive resampling, which would have been computationally prohibitive. 

The above hard-thresholding strategy is a suitable choice for our framework compared to soft-thresholding and posterior inclusion probabilities. The soft thresholding produces sparsity via continuous shrinkage, which biases the non-zero coefficients toward zero. Furthermore, the posterior inclusion probabilities from the VB approximation are poorly calibrated due to the known over-confidence of the mean-field approximation, making them unreliable as a soft selection tool. Therefore, applying a hard threshold to the coefficients is the most coherent way to distill a single, interpretable model from the approximated posterior, providing a clear set of selected variables without additional shrinkage.}
}
\section{Posterior predictive mass function}

The posterior predictive mass function (ppmf) of the response variable associated with the covariates of a new sample $x_0$ given the training data set $(X,y)$ is obtained as follows
\begin{align*}
p(y_0|x_0,X,y) & = \int p(y_0|x_0,{ \boldsymbol\theta}) p({ \boldsymbol\theta} | X,y) d{ \boldsymbol\theta} \\
& \approx \int {p}(y_0|x_0,{ \boldsymbol\theta}) q({ \boldsymbol\theta}) d{ \boldsymbol\theta},
\end{align*}
where ${ \boldsymbol\theta} = { \boldsymbol\beta}$ for Laplace-VB and CS-VB methods, and ${ \boldsymbol\theta} = ({ \boldsymbol\gamma},{ \boldsymbol\beta})^\top$ for Bernoulli-VB method. 

For Laplace-VB and CS-VB methods, we have, for $M = L, C$
$$p(y_0|x_0,X,y) \approx \frac{1}{y_0 !}\int \exp\{-\exp\{x_0{ \boldsymbol\beta}\}\} \exp\{y_0 x_0 { \boldsymbol\beta}\} {\rm N}({ \boldsymbol\beta}; { \boldsymbol\mu}_{{ \boldsymbol\beta}(M)}, { \boldsymbol\Sigma}_{{ \boldsymbol\beta}(M)}) \;d{ \boldsymbol\beta}$$
Using {\color{black} re-parametrization} $\lambda_0 = \exp\{x_0{ \boldsymbol\beta}\}$, we see that 
\begin{align}
p(y_0|x_0,X,y) & \approx \frac{1}{y_0 !}\int e^{-\lambda_0} \lambda_0^{y_0} {\rm LN}(\lambda_0; x_0{ \boldsymbol\mu}_{{ \boldsymbol\beta}(M)}, x_0{ \boldsymbol\Sigma}_{{ \boldsymbol\beta}(M)}x_0^\top) \;d\lambda_0,\label{Lpd}
\end{align}
{ where ${\rm LN}(x; \mu, \sigma^2)$ stands for the pdf of the univariate log-normal distribution. }

The predictive probability mass function in \eqref{Lpd} can be obtained using a {univariate} numerical integration, and might be used for obtaining a point predictor (e.g., most likely predictor) and a prediction interval (e.g., the most likely prediction interval) for $y_0$ given $x_0$ and $(X,y)$. {\color{black} Evaluating this single-variable integral with deterministic numerical methods (e.g., Gaussian quadrature) is computationally very cheap and achieves high precision without the sampling error inherent in Monte Carlo methods, which would require drawing a large number of samples from a high-dimensional multivariate posterior. Consequently, the above method is not only more accurate but also faster than a sampling-based alternative for calculating prediction intervals.} 

For { the Bernoulli-VB} method, we have 
$$p(y_0|x_0,X,y) \approx \prod_{j=1}^{p-1}\sum_{\gamma_j = 0}^1{\rm Ber}(\gamma_j;P_j)\frac{1}{y_0 !}\int \exp\{-\exp\{x_0\Gamma{ \boldsymbol\beta}\}\} \exp\{y_0 x_0 \Gamma { \boldsymbol\beta}\} {\rm N}({ \boldsymbol\beta}; { \boldsymbol\mu}_{{ \boldsymbol\beta}(B)}, { \boldsymbol\Sigma}_{{ \boldsymbol\beta}(B)}) \;d{ \boldsymbol\beta}$$
Using re-paramerization $\lambda_{0\gamma} = \exp\{x_0\Gamma{ \boldsymbol\beta}\}$, we see that 
\begin{align}
p(y_0|x_0,X,y) & \approx\prod_{j=1}^{p-1}\sum_{\gamma_j = 0}^1{\rm Ber}(\gamma_j;P_j) \frac{1}{y_0 !}\int e^{-\lambda_{0\gamma}} \lambda_{0\gamma}^{y_0} {\rm LN}(\lambda_{0\gamma}; x_0\Gamma{ \boldsymbol\mu}_{{ \boldsymbol\beta}(B)}, x_0\Gamma{ \boldsymbol\Sigma}_{{ \boldsymbol\beta}(B)}\Gamma x_0^\top) \;d\lambda_{0\gamma}.\label{Lpd21}
\end{align}
Since, for the Bernoulli-VB method, after using sparsity-enforcing thresholds, the parameters ${ P_j^{(B)}}$ take only 0 and 1 values, { equation \eqref{Lpd21} can be expressed more succinctly as}
\begin{align}
p(y_0|x_0,X,y) &\approx \frac{1}{y_0 !}\int e^{-\lambda_{0}} \lambda_{0}^{y_0} {\rm LN}(\lambda_{0}; x_0 { P^{(B)}} { \boldsymbol\mu}_{{ \boldsymbol\beta}(B)}, x_0 { P^{(B)}} { \boldsymbol\Sigma}_{{ \boldsymbol\beta}(B)} { P^{(B)}} x_0^\top) \;d\lambda_{0}.\label{Lpd2}
\end{align}

\section{Simulation study}

{ To examine} the performance of the proposed models, we have conducted a simulation study as follows. In each of $N=1000$ replications, we have generated samples of size $n$, {from} a sparse Poisson regression model, whose vector of coefficients ${ \boldsymbol\beta}_{p\times 1}$, with $p$, is first generated from { $N(\mu_0,\sigma_0)$} and then { is multiplied by a vector $\mathbf{z}$ of length $p$ with 0 and 1 elements, to set some elements of ${ \boldsymbol\beta}$ to zero.} The rows of the covariate matrix $X$ are generated independently from a multivariate normal distribution with { mean $\mu_X$} and a variance-covariance matrix with elements ${\Sigma}_{ij} = \sigma^2_X \cdot 0.3^{|i-j|}$, to model a slight multicollinearity. { We consider the cases $n=30,100$, and $p=10,200$, 
to {successfully} cover the low- and high-dimensional scenarios. For the cases with $p=10$, we let $\mu_0 = 0.7, \sigma_0 = 0.5, \mu_X = 0.1, \sigma^2_X = 1$, and we set $\mathbf{z} = (1, 0,1,0,0,0,1,0,1,0)^\top$, and for the cases with $p = 200$, we let $\mu_0 = 0.1, \sigma_0 = 0.6, \mu_X = 0.1, \sigma^2_X = 0.05$, and we randomly choose 60 (including the first one) out of 200 values of $\mathbf{z}$ to be equal 1, and the remaining are set to 0. The case $n=100, \; p = 10$, called the low-dimensional scenario, and the case $n=30, \; p=200$, called the high-dimensional scenario, are studied in the following subsections, while the two remaining cases ($n=30,\;p=10$ and $n=100,\;p=200$) are given in the supplementary material for the sake of brevity.}

Three proposed VB models (Laplace, CS, Bernoulli) are fitted to each generated dataset. Furthermore, two frequentist alternatives, LASSO and SCAD-penalized Poisson regression models, are considered for comparison. The sparsity parameters of both LASSO and SCAD models are optimized based on the Corrected Akaike Information Criterion 
$${\rm AICc} = -\log \widehat{L}({ \boldsymbol\beta}|x) + 2 {\rm df} + 2{\rm df}({\rm df}+1)/(n-{\rm df}-1),$$
where $\widehat{L}({ \boldsymbol\beta}|x)$ is the estimated likelihood of the model and ${\rm df}$ is the { degrees of freedom} of the model, that is, the number of nonzero coefficients in the regression model. 


To compare the performance of the 3 proposed models with each other and with  LASSO and SCAD-penalized Poisson regression models, the regression coefficient relative errors of methods $M = L, C, B$ are computed as follows
$${\rm CRE} = \frac{\frac{1}{N}\sum_{t = 1}^N (\widehat{{ \boldsymbol\beta}}^M - { \boldsymbol\beta})^\top (\widehat{{ \boldsymbol\beta}}^M - { \boldsymbol\beta})}{\frac{1}{N}\sum_{t = 1}^N { \boldsymbol\beta}^\top{ \boldsymbol\beta}}.$$
Furthermore, the train-set and test-set relative errors are obtained, respectively, as 
$${\rm TRRE} = \frac{\frac{1}{N}\sum_{t = 1}^N (\widehat{y}_{t}^{\rm train} - {y}_{t}^{\rm train})^\top  (\widehat{y}_{t}^{\rm train} - {y}_{t}^{\rm train})}{\frac{1}{N}\sum_{t = 1}^N ({y}_{t}^{\rm train} - \bar{y}_{t}^{\rm train})^\top  ({y}_{t}^{\rm train} - \bar{y}_{t}^{\rm train})},$$
and 
$${\rm TSRE} = \frac{\frac{1}{N}\sum_{t = 1}^N (\widehat{y}_{t}^{\rm test} - {y}_{t}^{\rm test})^\top  (\widehat{y}_{t}^{\rm test} - {y}_{t}^{\rm test})}{\frac{1}{N}\sum_{t = 1}^N ({y}_{t}^{\rm test} - \bar{y}_{t}^{\rm test})^\top  ({y}_{t}^{\rm test} - \bar{y}_{t}^{\rm test})},$$
where ${y}_{t}^{\rm train} = ({y}_{1t},\ldots,{y}_{nt})^\top$ and ${y}_{t}^{\rm test} = ({y}^\prime_{1t},\ldots,{y}^\prime_{n^\prime t})^\top$, are the train-set (80\% of the full sample) and test-set (20 \% of the full sample) response samples, respectively, { generated during $t$th iteration, {\color{black} $\bar{y}_{t}^{\rm train} = \frac{1}{n}\sum_{i=1}^n {y}_{it}$ and $\bar{y}_{t}^{\rm test} = \frac{1}{n'}\sum_{i=1}^n {y'}_{it}$.} }

The false negative rate and the false positive rate are also defined as follows
$${\rm  FNR}=\frac{\#\{j;\;1\leq j\leq p,\; \beta_j\neq 0,\; \hat{\beta}^M_j=0\}}{\#\{j;\;1\leq j\leq p,\; \beta_j\neq 0\}},$$
$${\rm  FPR}=\frac{\#\{j;\;1\leq j\leq p\;\beta_j= 0,\; \hat{\beta}^M_j\neq 0\}}{\#\{j;\;1\leq j\leq p,\; \beta_j= 0\}},$$
for $M=L,C,B$, where $\#A$ stands for the {cardinality} of the set $A$. These criteria are used to examine the { sparsity} performance of the competitive methods. 

The computation time of { the 3 proposed} models, as well as the LASSO and SCAD-penalized Poisson regression models, is also computed in each iteration. 

For the Laplace-VB method, we set the hyper-priors $\nu = 0.0001$ and $\delta =0.01$, to obtain a mean of 0.01 and variance of 1 for the parameter {$\eta$}, which controls the sparsity of the coefficients. The hyper-parameters ${ \rho}_1$ and ${ \rho}_2$ for the CS-VB and Bernoulli-VB methods are {\color{black} both set to 1.} Also, all $a_j$ and $b_j$, hyperparameters for the Bernoulli-VB method, are set to 0.01. The hyperparameter $A$ is set to 0.01 for the Laplace-VB and CS-VB methods. 

{ \subsection{Low dimensional scenario}

Figures \ref{box1} and \ref{box2} show the box plots of all aforementioned criteria for all competitive methods, for the low-dimensional scenario, with {$p=10$, $n=100$, and $\mathbf{z} = (1,0,1,0,1)^\top$}. In this scenario, all models are also compared to their MCMC versions. In implementing the MCMC methods, we utilized the R packages R2OpenBUGS and rjags within R (version 4.2.1), in conjunction with OpenBUGS (version 3.2.3) and JAGS (version 4.3.1). Each chain comprised 10,000 iterations, with a burn-in of 5,000 and a thinning interval of 10. {\color{black}The Gelman-Rubin diagnostics are computed for each parameter and reported in the supplementary material. These values confirm the convergence of the MCMC algorithm for most parameters.} 

{ Among the VB variants, Bernoulli-VB achieve the lowest median CRE with limited variability, whereas CS-VB and LAPLACE-VB show slightly more dispersion and occasional outliers, suggesting sensitivity to data characteristics or prior assumptions. Comparisons with the corresponding MCMC counterparts (Bernoulli-MCMC, CS-MCMC, and LAPLACE-MCMC) indicate that the VB framework can achieve similar accuracy levels at a markedly reduced computational cost, albeit with modest trade-offs in variance. Overall, the figure underscores the proposed VB methods as effective and computationally efficient alternatives for coefficient estimation, maintaining accuracy close to established methods while offering scalability advantages. 

LASSO achieves lowest median TRRE values with minimal spread, indicating strong and consistent in-sample fitting. 
The VB methods exhibit slightly higher medians with a modestly wider interquartile range, indicating some additional variability yet remaining competitive with state-of-the-art baselines. Among the proposed methods, Bernoulli-VB delivers the lowest TRRE median, demonstrating reliable fitting performance, while CS-VB and LAPLACE-VB show comparable accuracy but with a broader distribution, reflecting model flexibility and responsiveness to diverse data structures. Relative to their MCMC counterparts, the VB approaches achieve similar median errors while offering the advantage of reduced computation times, a key benefit in large-scale or iterative modeling contexts. Overall, the figure demonstrates that the proposed VB methods maintain strong general in-sample accuracy while providing a balance between efficiency and predictive reliability.

All methods achieve relatively low median TSRE values (around 0.05-0.08), suggesting strong predictive performance. The VB methods are competitive with established baselines, with Bernoulli-VB and LAPLACE-VB displaying compact interquartile ranges comparable to LASSO and SCAD, while CS-VB shows a slightly higher median and a few large-value outliers, potentially reflective of sensitivity to specific test scenarios. Compared to their MCMC counterparts, the VB methods maintain similar median accuracy but generally exhibit fewer extremely high errors, underscoring a balance between stability and efficiency. Notably, the sparsity-inducing VB variants (CS-VB and LAPLACE-VB) deliver consistent out-of-sample performance while avoiding the higher computational cost of MCMC sampling, reinforcing their practicality for large-scale predictive modeling. Overall, the figure highlights the ability of the proposed VB framework to retain robust generalization while ensuring computational scalability.}

Note that for the computation of TSRE values for the LASSO and SCAD methods, the prediction of the response is computed as 
$$\hat{y}_i = \exp(X_{\rm test}\hat{{ \boldsymbol\beta}}_{\rm sparse}),$$
in which $\hat{{ \boldsymbol\beta}}_{\rm sparse}$ is the sparse estimator obtained by the LASSO and SCAD methods, while 
the TSRE values for the VB methods are computed using the ppmf functions \eqref{Lpd} for the CS-VB and Laplace-VB and the ppmf function \eqref{Lpd2} for the Bernoulli-VB method. The mode of \eqref{Lpd} is considered as the point predictor of the response value. 

{ The VB methods achieve competitive median FPRs compared with established baselines such as LASSO and SCAD, with CS-VB and LAPLACE-VB showing particularly compact interquartile ranges, suggesting consistent performance across datasets. Bernoulli-VB maintains a low median FPR (at least compared to LASSO) but exhibits a slightly wider spread, pointing to occasional variability under certain conditions. In direct comparison to their MCMC counterparts, the VB methods demonstrate similar FPR control (except the Bernoulli-VB method), while benefiting from the substantially reduced computational demands inherent to the variational framework. Taken together, the figure reinforces that the proposed VB methodologies not only maintain strong predictive accuracy but also effectively limit the false positive rate.

All three VB methods achieve median FNR values at or very close to zero, indicating excellent detection capability.  Comparisons with corresponding MCMC methods show that VB approaches can match or closely approximate the highest attainable FNR without the heavy computational burden. LASSO and SCAD, by contrast, display wider variability and lower medians, highlighting the advantage of VB's probabilistic modeling in maintaining sensitivity. Overall, the figure underscores that the proposed VB methods pair high detection power with scalability.

All three VB variants demonstrate dramatically reduced runtimes, with relative times on the order of 
$10^{-3}$ compared to CS-MCMC, reflecting speed-ups of several hundred to over a thousand times. Among them, CS-VB shows the baseline value for VB implementations, while Bernoulli-VB and LAPLACE-VB perform similarly, confirming that the computational advantage is consistent across different prior structures.}

Although the computation time of the VB methods is a little more than the LASSO and SCAD methods, this comparison is not fair, since the LASSO and SCAD methods use {C++} programming for accelerating the computation, while we have done all computations in the R software. So, it seems that the VB methods are very fast, as expected. The CS-VB method seems to be the fastest VB method. {\color{black} All computations were performed using R version 4.3.2 on a machine with a Core i5-10210U CPU.}

Table \ref{cover} presents the average (standard deviation) of the coverage probabilities of the highest posterior density (HPD) confidence intervals for the regression coefficients for three VB methods at level 0.95. {\color{black} Table \ref{cover} indicates that the average coverage probabilities of nearly all regression coefficients approximate the nominal value. In some cases, the average coverage probabilities for the HPD intervals are occasionally below the nominal 0.95 level for both the VB and MCMC methods. This pattern is observed even for the MCMC benchmarks, which suggests that the under-coverage is not a consequence of the variational approximation but is inherent to the model and prior specification. It reflects the frequentist operating characteristics of the Bayesian procedure with these specific priors in the given simulation setting. The key finding is that the VB methods closely replicate the coverage properties of the corresponding MCMC methods, demonstrating their reliability for uncertainty quantification in this context.}

\begin{table}
\centering\color{black}
{ \caption{The average (standard deviation) of coverage probabilities of the HPD intervals for the regression coefficients for three VB and MCMC methods at level 0.95.}\label{cover}
\begin{tabular}{lcccccc}
\toprule
 & Intercept & ${ \boldsymbol\beta}_1$ & ${ \boldsymbol\beta}_2$ & ${ \boldsymbol\beta}_3$ & ${ \boldsymbol\beta}_4$ \\
\hline\hline
Bernoulli-VB & 0.92 (0.27) & 0.94 (0.24) & 0.94 (0.24) & 0.96 (0.20) & 0.98 (0.14) \\
CS-VB & 0.92 (0.27) & 0.96 (0.20) & 0.95 (0.22) & 0.96 (0.20) & 0.96 (0.20) \\
LAPLACE-VB & 0.92 (0.27) & 0.95 (0.22) & 0.94 (0.24) & 0.96 (0.20) & 0.97 (0.17) \\
Bernoulli-MCMC & 0.96 (0.20) & 0.99 (0.10) & 0.98 (0.14) & 0.96 (0.20) & 0.99 (0.10) \\
CS-MCMC & 0.94 (0.24) & 0.97 (0.17) & 0.94 (0.24) & 0.97 (0.17) & 0.99 (0.10) \\
LAPLACE-MCMC & 0.95 (0.22) & 0.96 (0.20) & 0.96 (0.20) & 0.97 (0.17) & 0.97 (0.17) \\
\hline\hline
\end{tabular}
\vspace{0.5cm}
\begin{tabular}{lcccccc}
 & ${ \boldsymbol\beta}_5$ & ${ \boldsymbol\beta}_6$ & ${ \boldsymbol\beta}_7$ & ${ \boldsymbol\beta}_8$ & ${ \boldsymbol\beta}_9$ \\
\hline\hline
Bernoulli-VB & 0.97 (0.17) & 0.87 (0.34) & 0.98 (0.14) & 0.88 (0.32) & 0.94 (0.24) \\
CS-VB & 0.94 (0.24) & 0.96 (0.20) & 0.98 (0.14) & 0.94 (0.24) & 0.91 (0.29) \\
LAPLACE-VB & 0.94 (0.24) & 0.95 (0.22) & 0.98 (0.14) & 0.94 (0.24) & 0.92 (0.27) \\
Bernoulli-MCMC & 0.98 (0.14) & 0.96 (0.20) & 0.98 (0.14) & 0.94 (0.24) & 0.96 (0.20) \\
CS-MCMC & 0.95 (0.22) & 0.96 (0.20) & 0.98 (0.14) & 0.96 (0.20) & 0.94 (0.24) \\
LAPLACE-MCMC & 0.96 (0.20) & 0.96 (0.20) & 0.99 (0.10) & 0.96 (0.20) & 0.93 (0.26) \\
\hline\hline
\end{tabular}}
\end{table}


\begin{table}
\centering\color{black}
\caption{Average {(standard error)} of accuracy values for the parameters and hyper-parameters of the VB against an MCMC benchmark for different priors.}\label{tab1}
\begin{tabular}{c c c c c c}
\hline\hline
 &   &  & Parameters & &  \\
\hline
Method & $\beta_0$ & $\beta_1$ & $\beta_2$ & $\beta_3$ & $\beta_4$ \\
\hline
{Bernoulli} & 92.19 {(8.55)} & 87.50 {(21.65)} & 85.60 {(12.17)} & 83.50 {(23.51)} & 88.50 {(21.04)} \\
{CS} & 95.26 {(1.44)} & 94.68 {(2.44)} & 90.35 {(4.44)} & 94.59 {(2.41)} & 94.92 {(2.61)} \\
{Laplace} & 95.28 {(1.34)} & 95.57 {(1.09)} & 95.65 {(1.14)} & 95.60 {(1.23)} & 95.48 {(1.35)} \\
\hline
 & $\beta_5$ & $\beta_6$ & $\beta_7$ & $\beta_8$ & $\beta_9$ \\
\hline
{Bernoulli} & 87.00 {(21.93)} & 86.50 {(13.80)} & 86.00 {(22.45)} & 85.63 {(13.79)} & 83.50 {(23.51)} \\
{CS} & 95.01 {(2.67)} & 95.21 {(1.62)} & 94.82 {(2.98)} & 95.19 {(1.85)} & 94.60 {(2.58)} \\
{Laplace} & 95.57 {(1.38)} & 96.01 {(1.29)} & 95.64 {(1.34)} & 95.55 {(1.24)} & 95.59 {(1.24)} \\
\hline
& & $\gamma_1 [Z_1]$ & $\gamma_2 [Z_2]$ & $\gamma_3 [Z_3]$ & $\gamma_4 [Z_4]$  \\
\hline
Bernoulli & & 30.75 {(11.05)} & 24.77 {(42.90)} & 32.03 {(15.10)} & 53.95 {(22.02)}  \\
CS & & 10.90 {(1.95)} & 83.04 {(10.62)} & 11.37 {(2.10)} & 49.27 {(7.81)}  \\
\hline
& $\gamma_5 [Z_5]$ & $\gamma_6 [Z_6]$ & $\gamma_7 [Z_7]$ & $\gamma_8 [Z_8]$ & $\gamma_9 [Z_9]$ \\
\hline
Bernoulli & 62.47 {(19.18)} & 25.86 {(43.64)} & 29.87 {(10.34)} & 28.00 {(44.90)} & 29.97 {(14.14)} \\
CS & 51.38 {(6.51)} & 61.66 {(6.84)} & 10.99 {(2.30)} & 60.12 {(6.11)} & 10.72 {(2.10)} \\
\hline
  & $\sigma$ & $\pi_1$ & $\pi_2$ & $\pi_3$ & $\pi_4$ \\
\hline
Bernoulli & & 77.33 {(6.66)} & 86.68 {(6.78)} & 76.73 {(7.25)} & 77.97 {(6.02)} \\
CS & 84.66 {(5.02)} & 76.07 {(2.85)} & 84.98 {(6.10)} & 75.65 {(3.31)} & 76.36 {(2.96)} \\
\hline
 & $\pi_5$ & $\pi_6$ & $\pi_7$ & $\pi_8$ & $\pi_9$ \\
\hline
Bernoulli &  77.52 {(6.78)} & 88.07 {(1.13)} & 76.65 {(7.06)} & 87.65 {(1.04)} & 75.86 {(8.15)} \\
CS &  75.84 {(3.16)} & 94.69 {(0.99)} & 75.86 {(3.03)} & 94.62 {(0.98)} & 75.97 {(3.27)} \\
\hline
 & $\alpha_0$ & $\alpha_1$ & $\alpha_2$ & $\alpha_3$ & $\alpha_4$ \\
\hline
Bernoulli & 76.94 {(--)} & 47.84 {(16.32)} & 78.87 {(6.93)} & 46.73 {(17.25)} & 47.98 {(16.39)} \\
\hline
 & $\alpha_5$ & $\alpha_6$ & $\alpha_7$ & $\alpha_8$ & $\alpha_9$ \\
\hline
Bernoulli & 50.89 {(17.98)} & 76.74 {(--)} & 47.04 {(16.49)} & 76.64 {(--)} & 45.40 {(17.75)} \\
\hline
 & $\tau_0$ & $\tau_1$ & $\tau_2$ & $\tau_3$ & $\tau_4$ \\
\hline
Laplace & 75.74 {(1.96)} & 87.71 {(1.17)} & 89.82 {(1.12)} & 87.92 {(1.12)} & 87.78 {(1.21)} \\
\hline
 & $\tau_5$ & $\tau_6$ & $\tau_7$ & $\tau_8$ & $\tau_9$ \\
\hline
Laplace & 87.91 {(1.37)} & 88.05 {(1.77)} & 87.69 {(1.29)} & 86.30 {(1.65)} & 87.66 {(1.20)} \\
\hline\hline
\end{tabular}
\end{table}

\begin{sidewaysfigure}
\centerline{\includegraphics[scale=0.6]{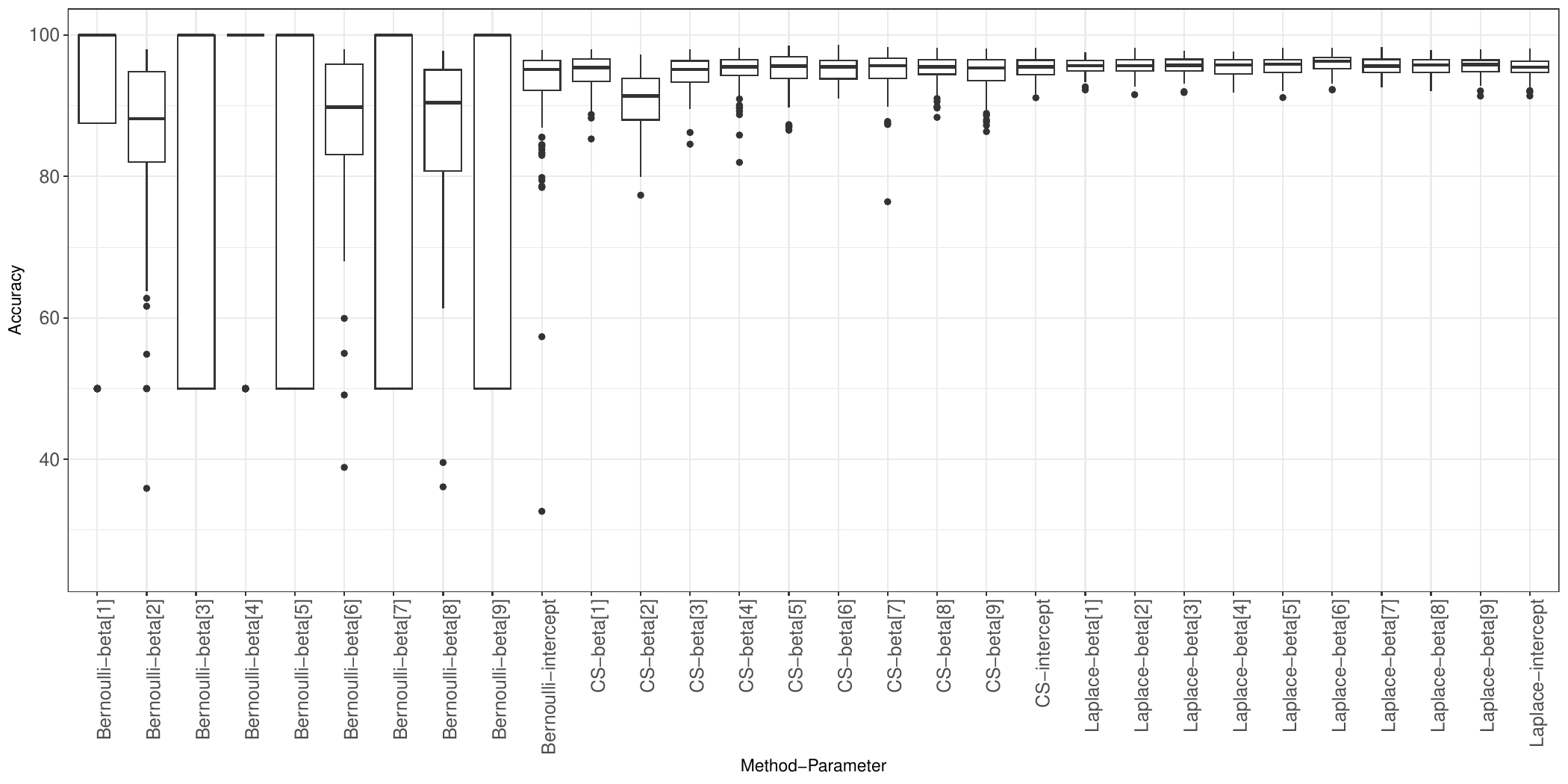}}
\caption{The boxplots of the accuracies for the {\color{black} regression} coefficients for different VB-methods.}\label{boxac1}
\end{sidewaysfigure}

\begin{figure}
\centerline{\includegraphics[scale=0.4]{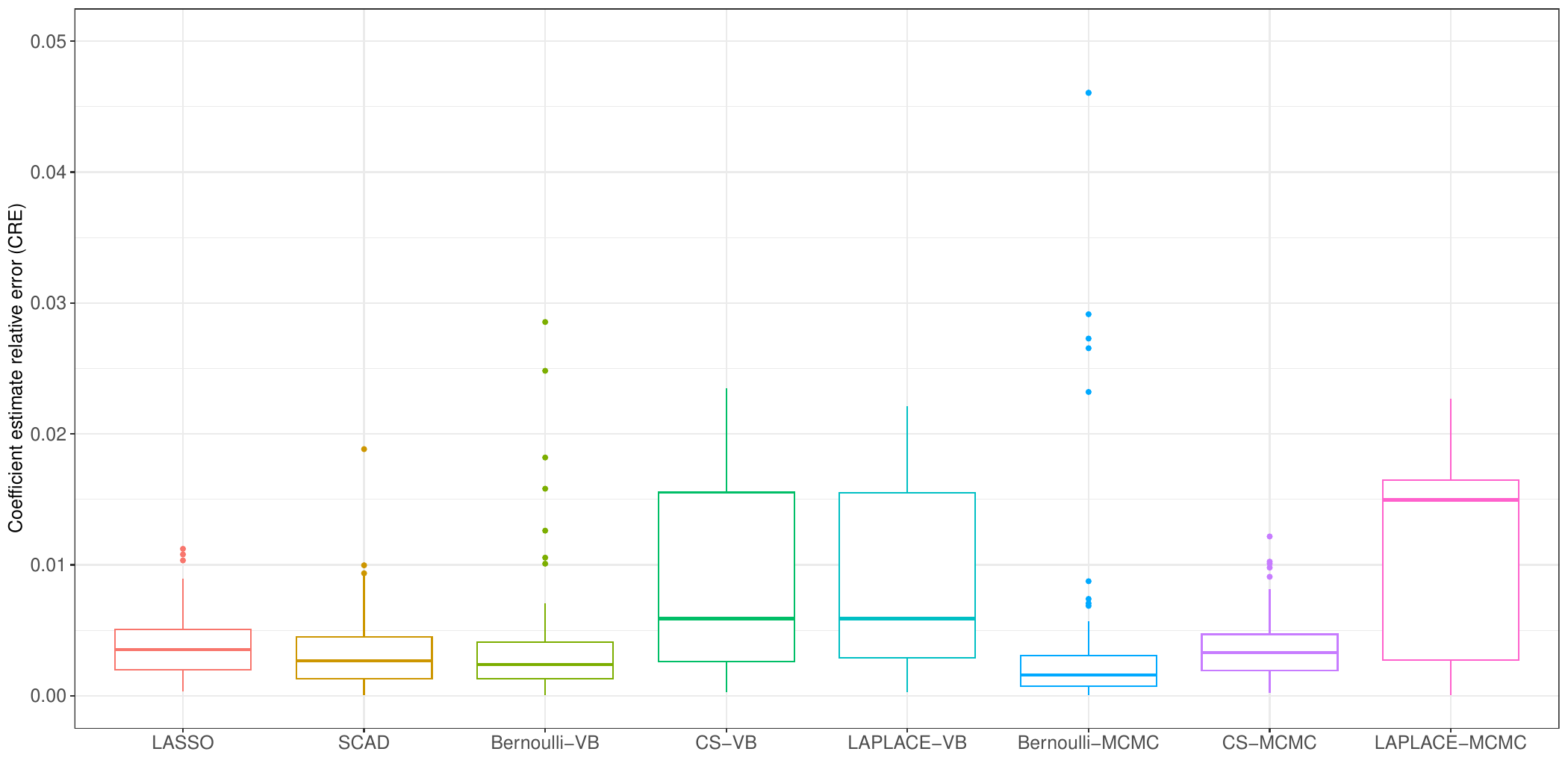}}
\centerline{\includegraphics[scale=0.4]{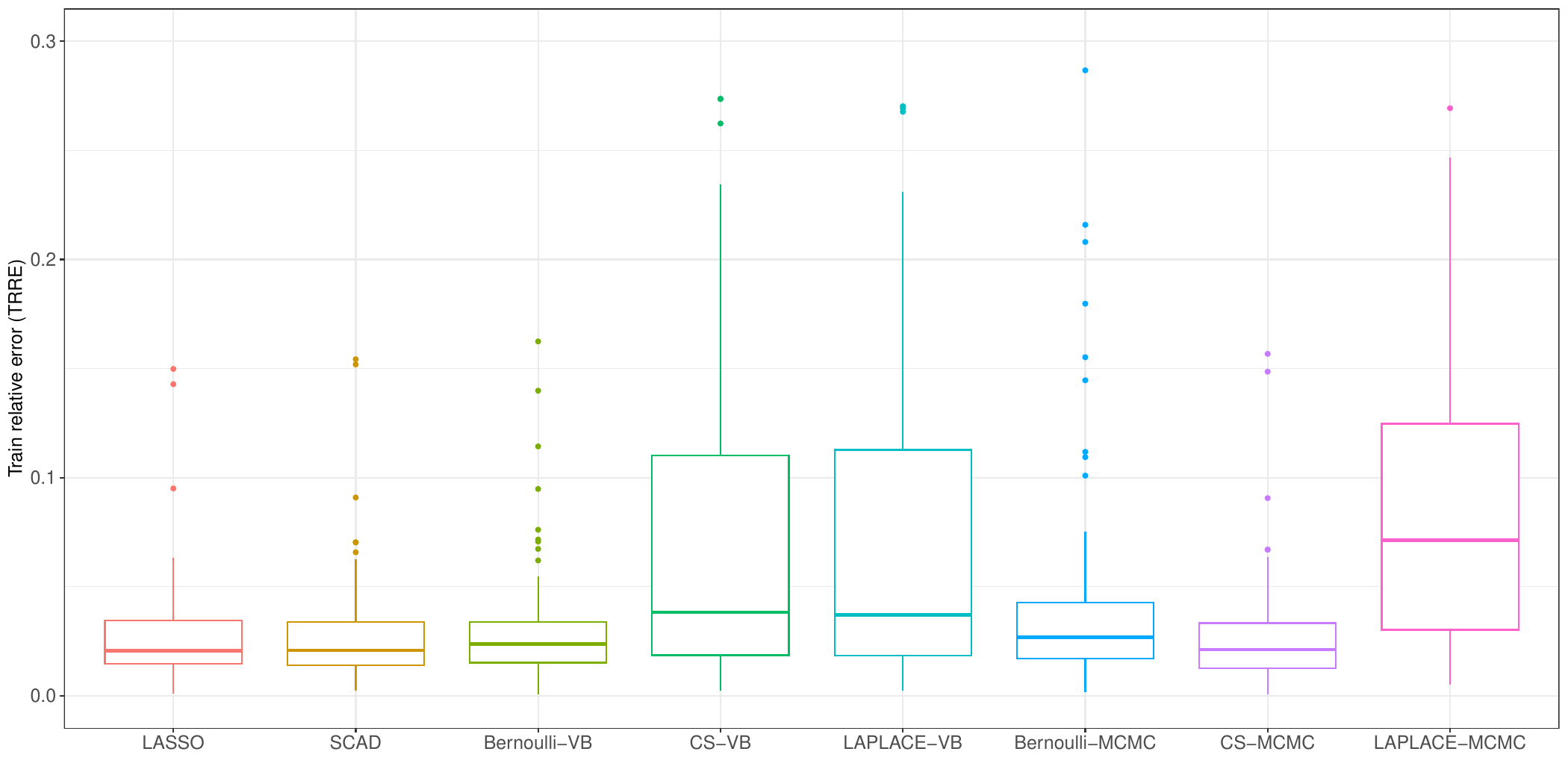}}
\centerline{\includegraphics[scale=0.4]{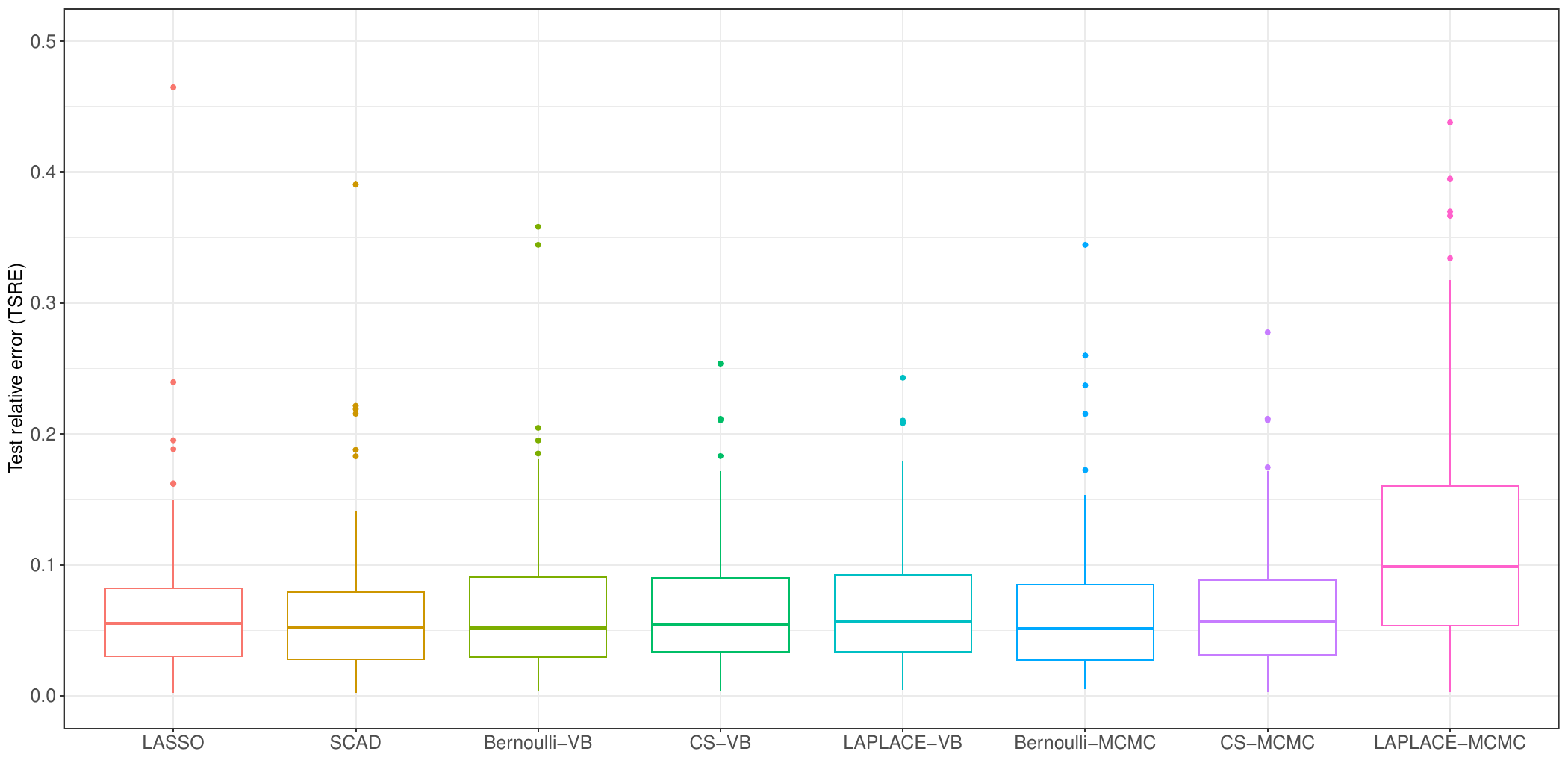}}
\caption{The low dimensional scenario simulation results: the coefficient relative error (top), the train relative error (middle), and the test relative error (bottom) for different methods.}\label{box1}
\end{figure}

\begin{figure}
\centerline{\includegraphics[scale=0.4]{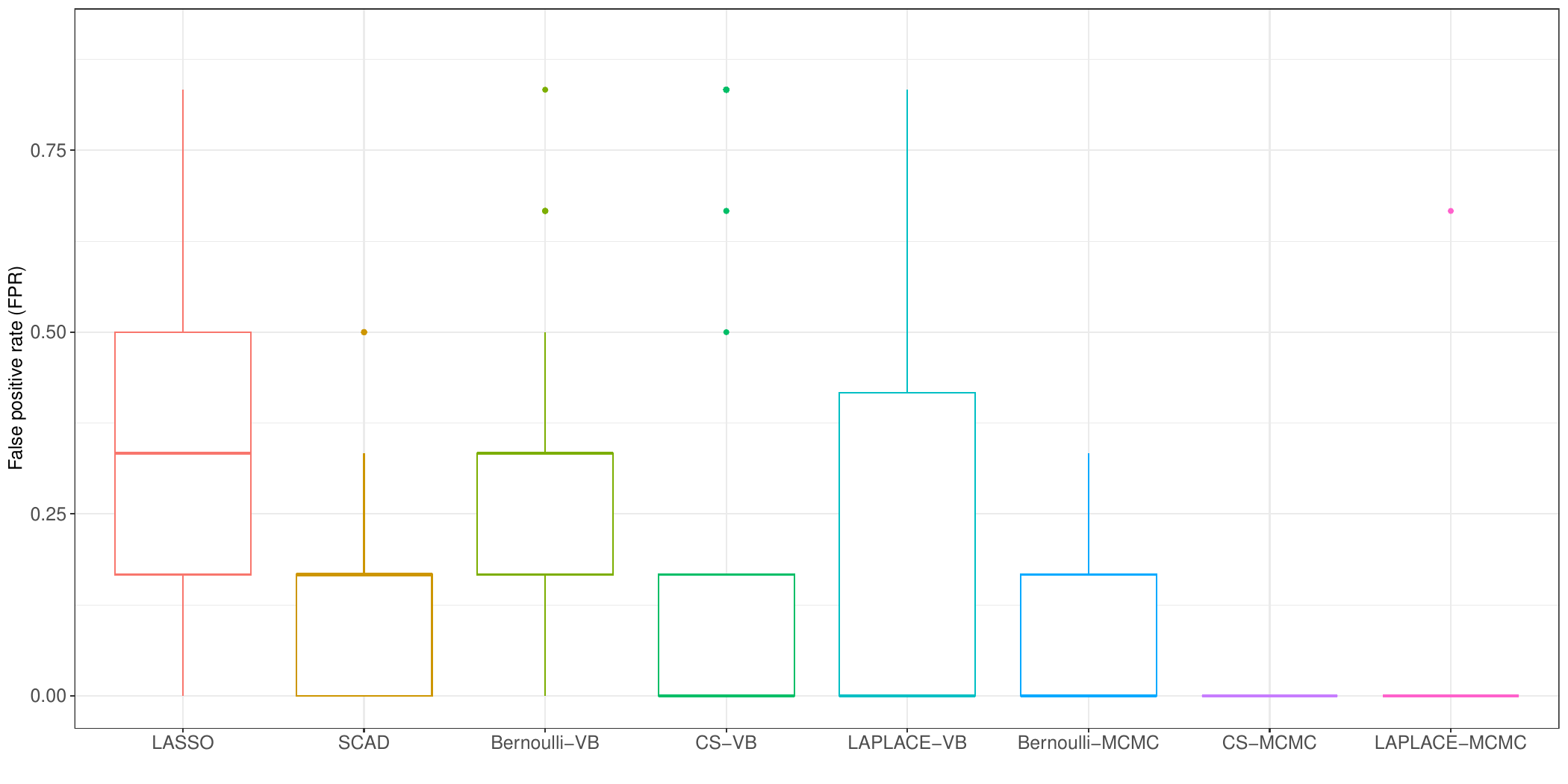}}
\centerline{\includegraphics[scale=0.4]{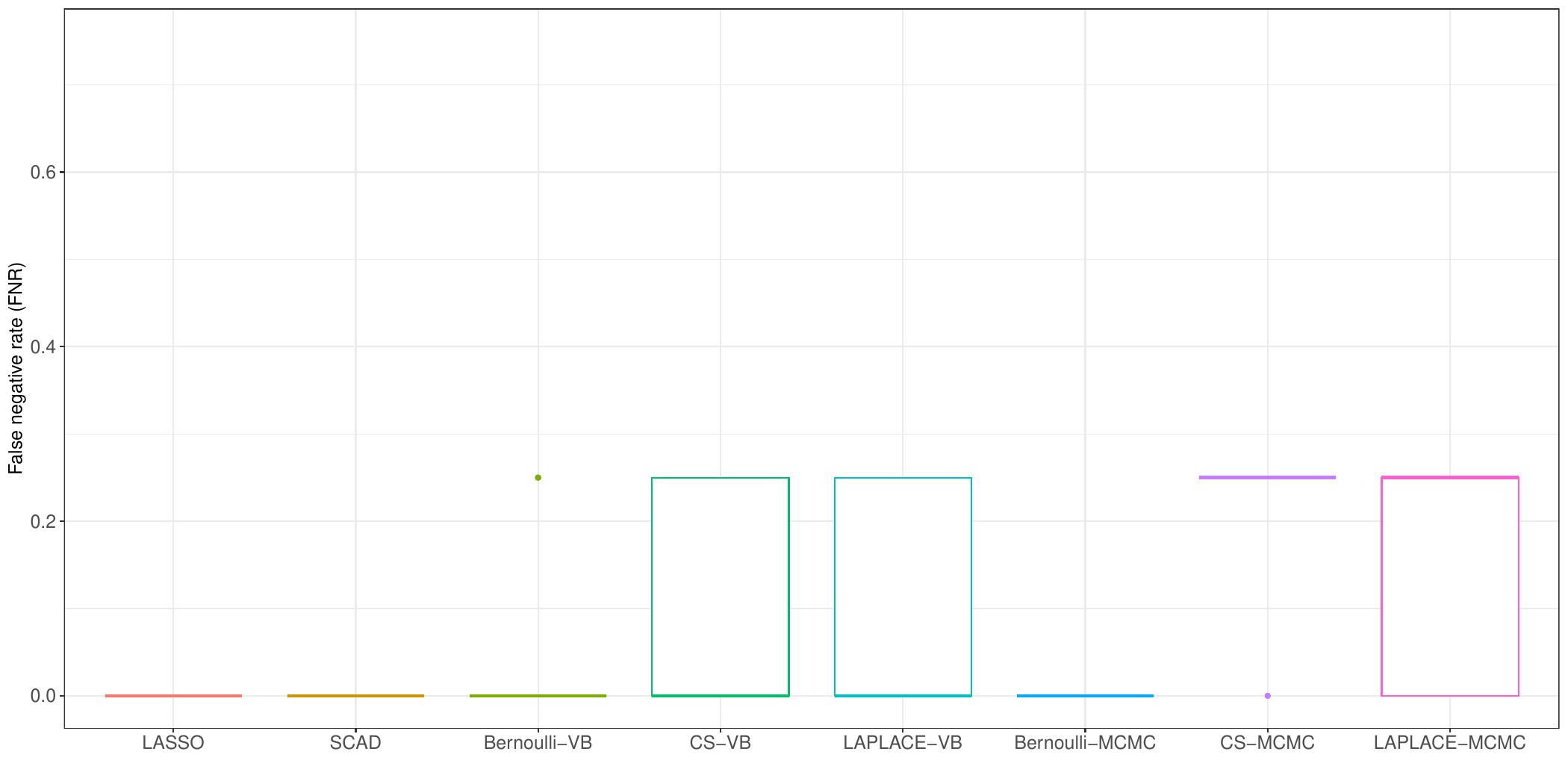}}
\centerline{\includegraphics[scale=0.4]{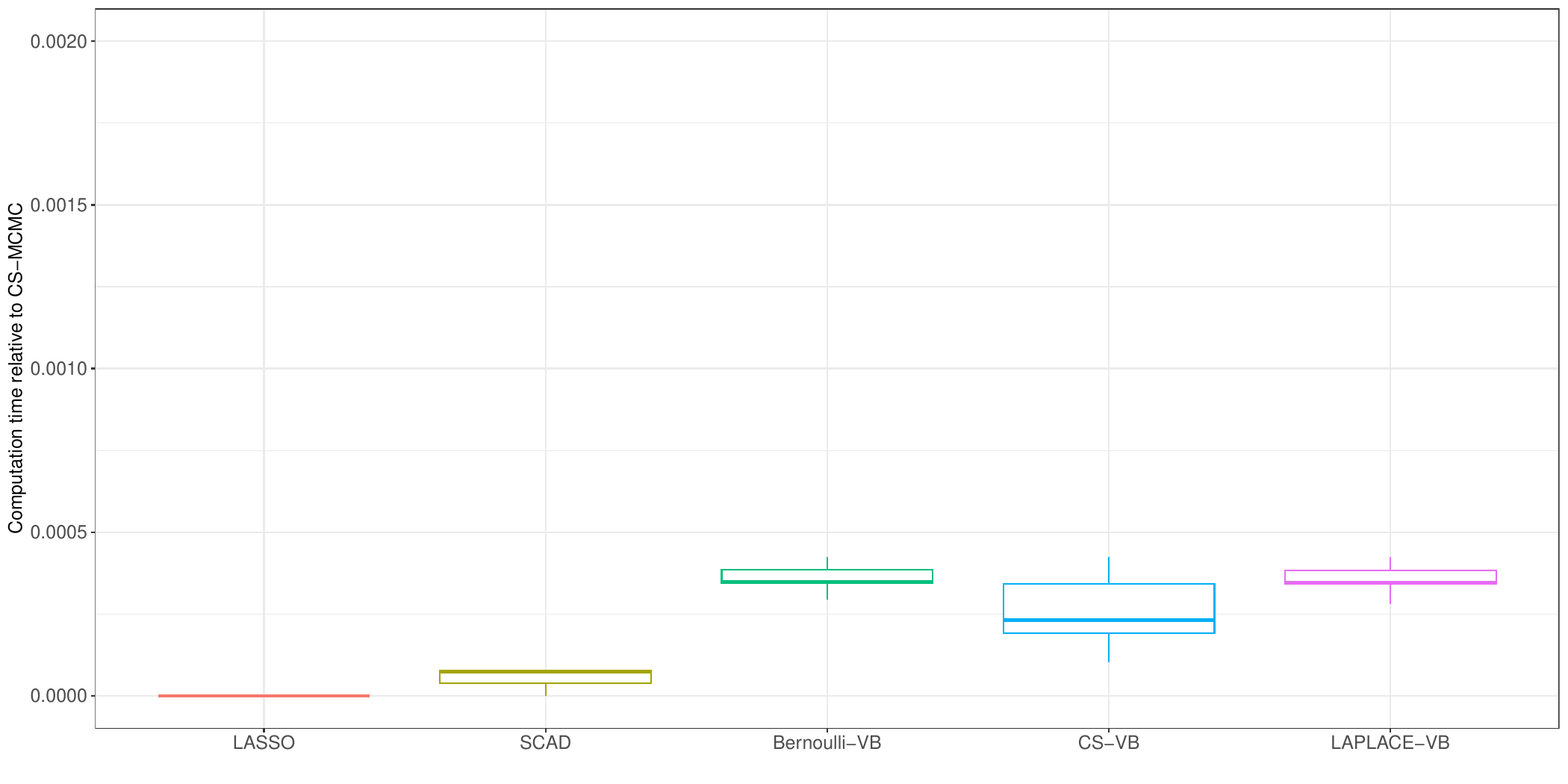}}
\caption{The low dimensional scenario simulation results: the FPR (top), the {\color{black} FNR} (middle), and the relative computation time (bottom) for different methods.}\label{box2}
\end{figure}

\begin{figure}
\centerline{\includegraphics[scale=0.4]{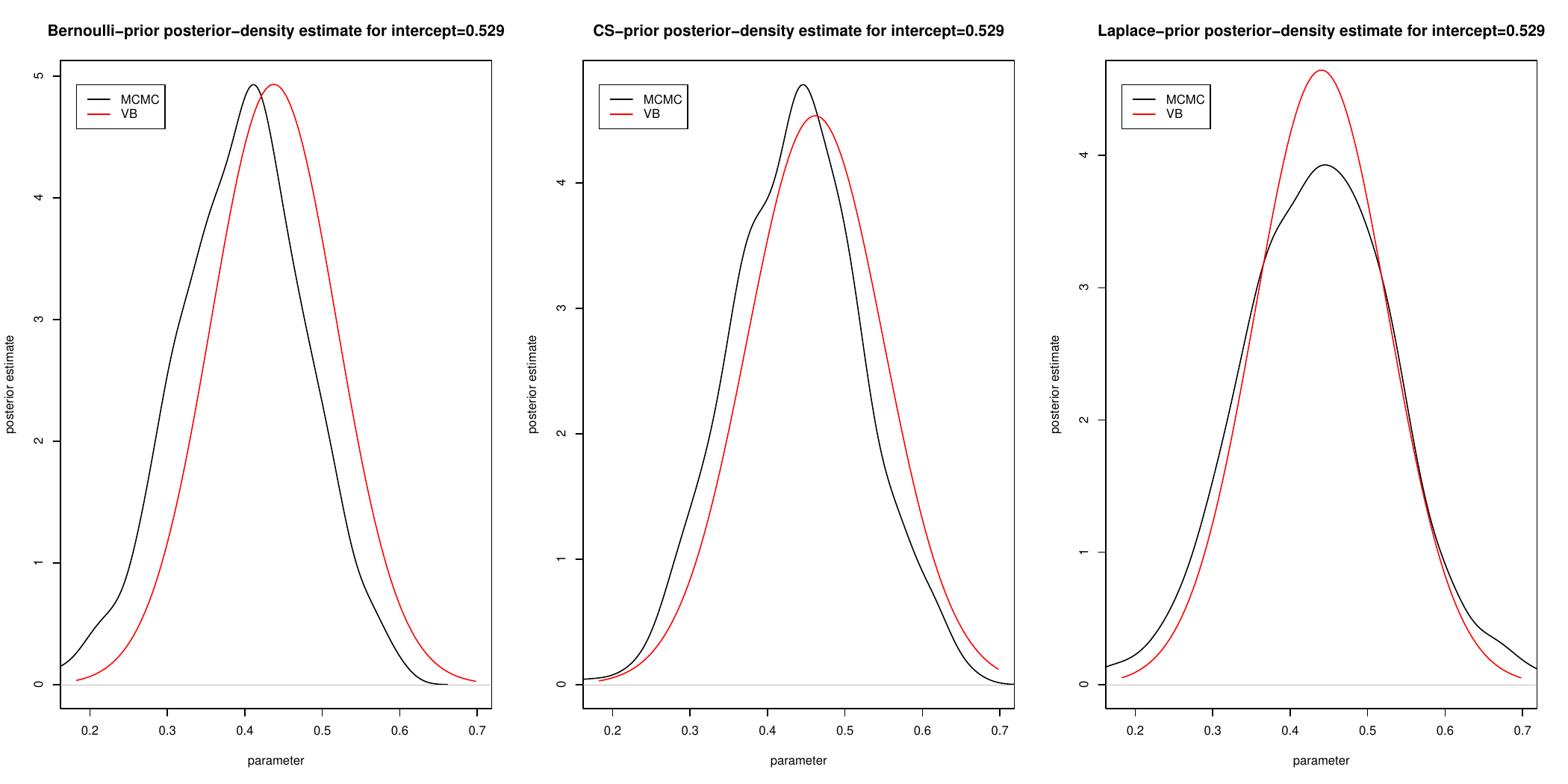}}
\centerline{\includegraphics[scale=0.4]{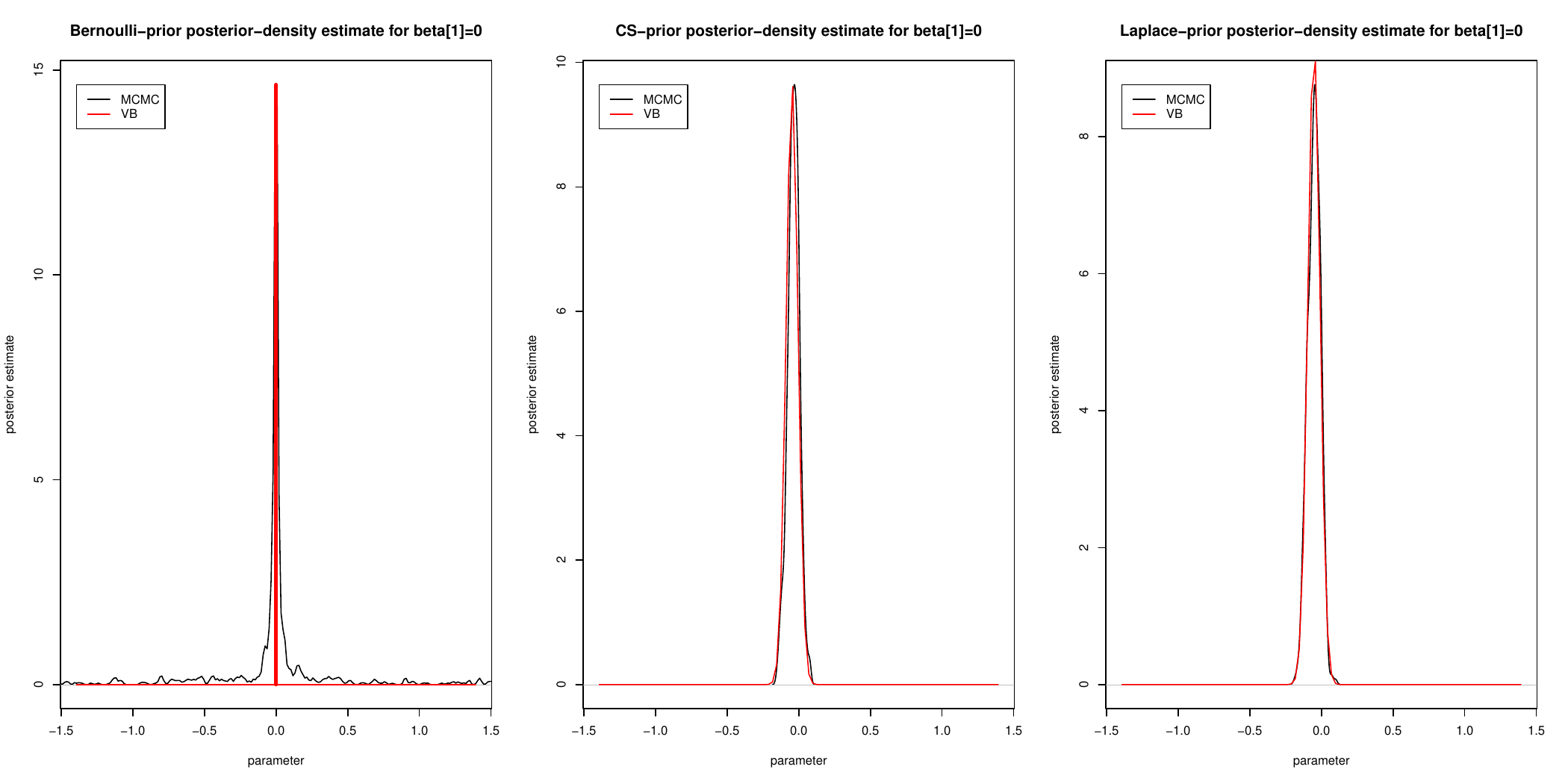}}
\centerline{\includegraphics[scale=0.4]{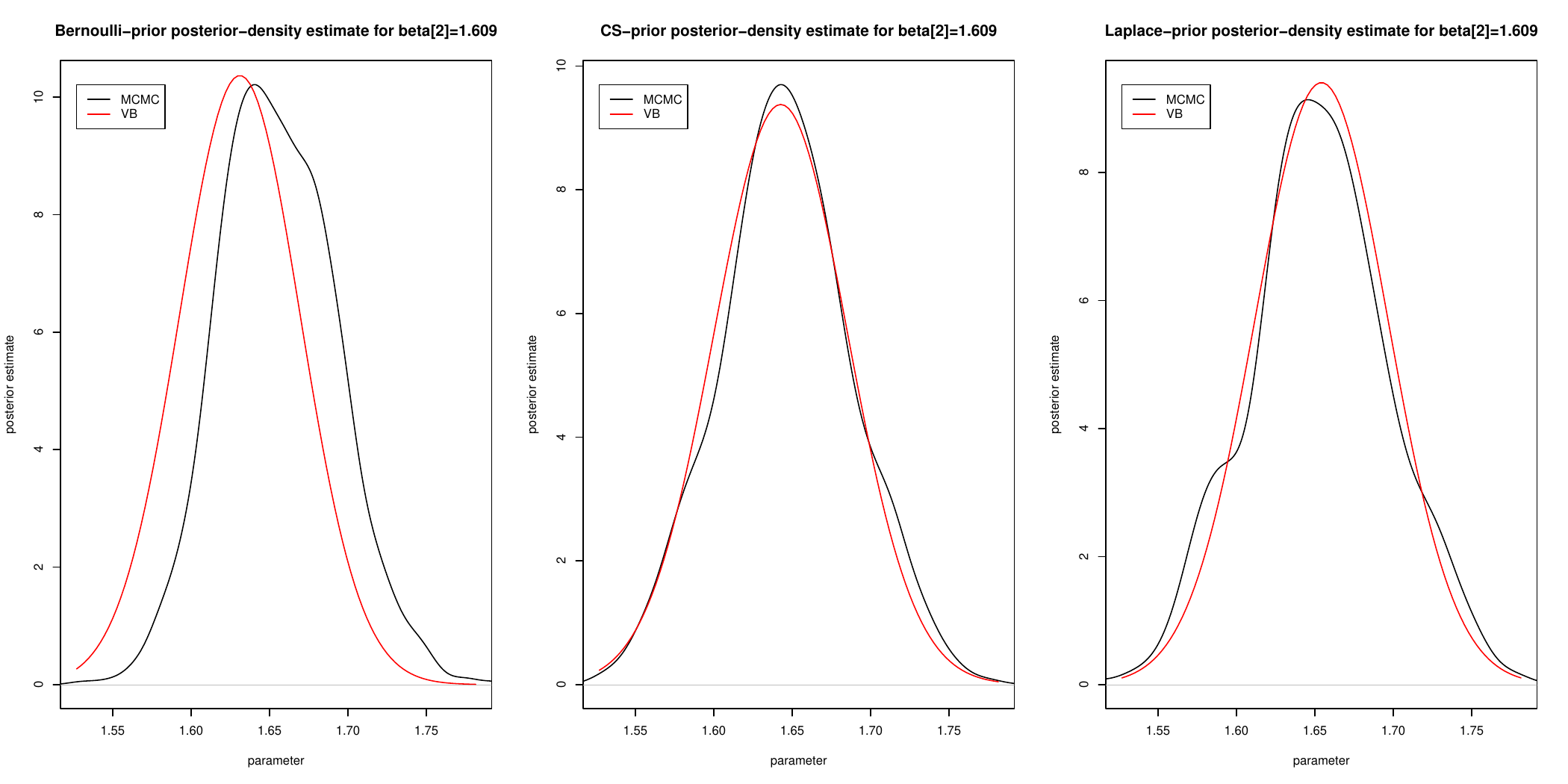}}
\caption{ Approximate posterior density functions of the first three regression coefficients for different models.}\label{dens1}
\end{figure}

\begin{figure}
\centerline{\includegraphics[scale=0.4]{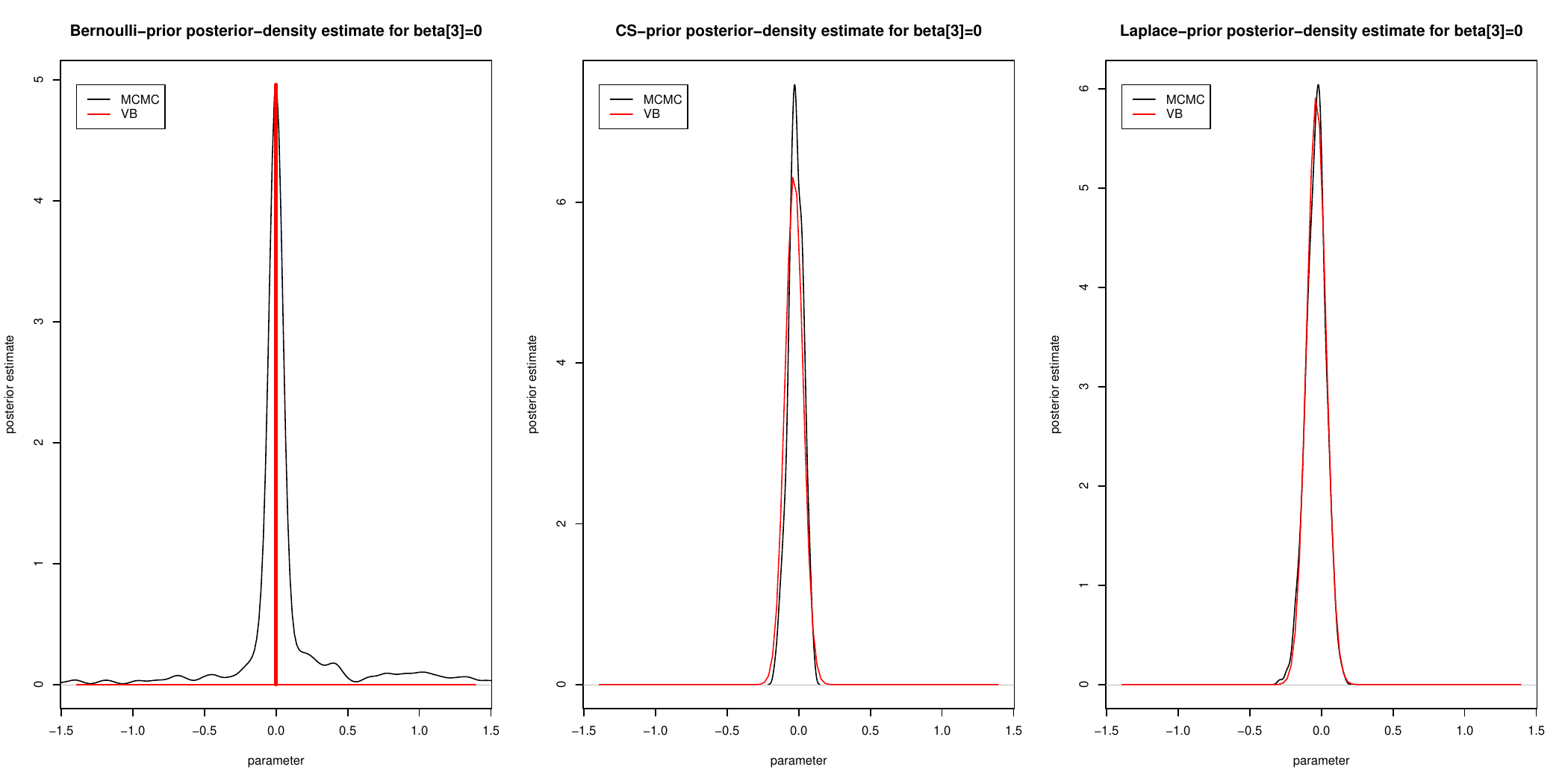}}
\centerline{\includegraphics[scale=0.4]{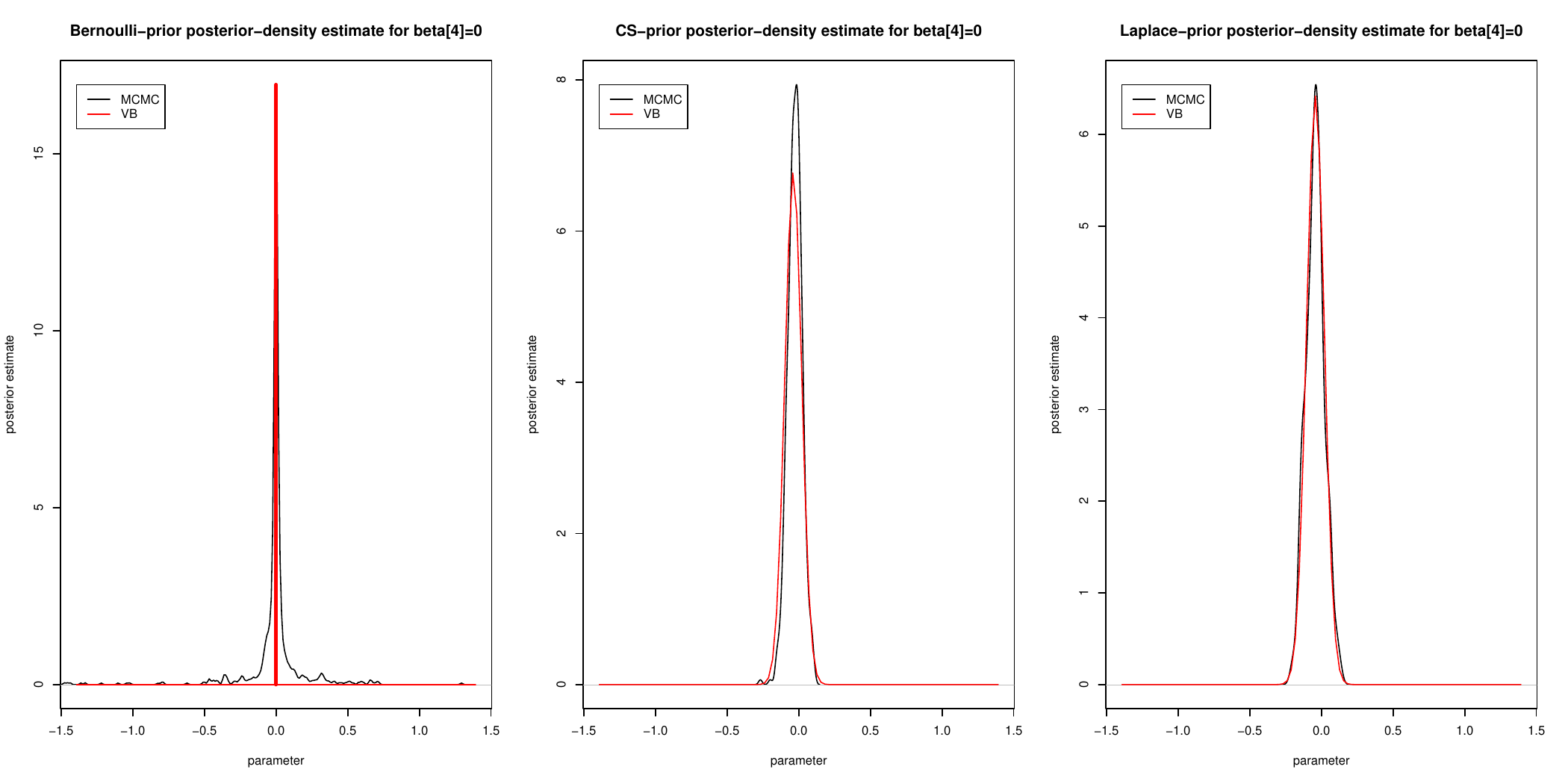}}
\centerline{\includegraphics[scale=0.4]{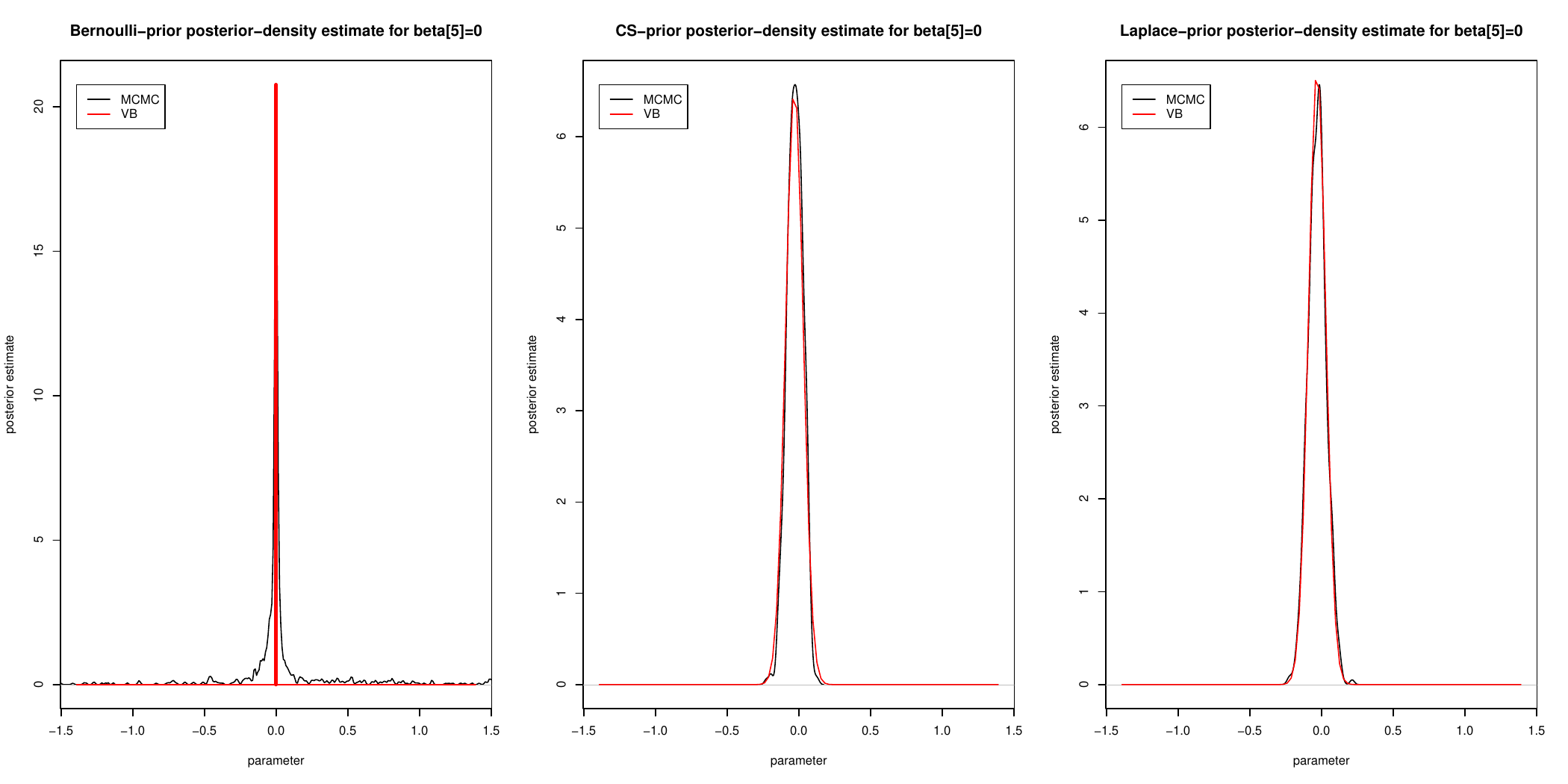}}
\caption{ Approximate posterior density functions of the second three regression coefficients for different models.}\label{dens2}
\end{figure}
\begin{figure}
\centerline{\includegraphics[scale=0.3]{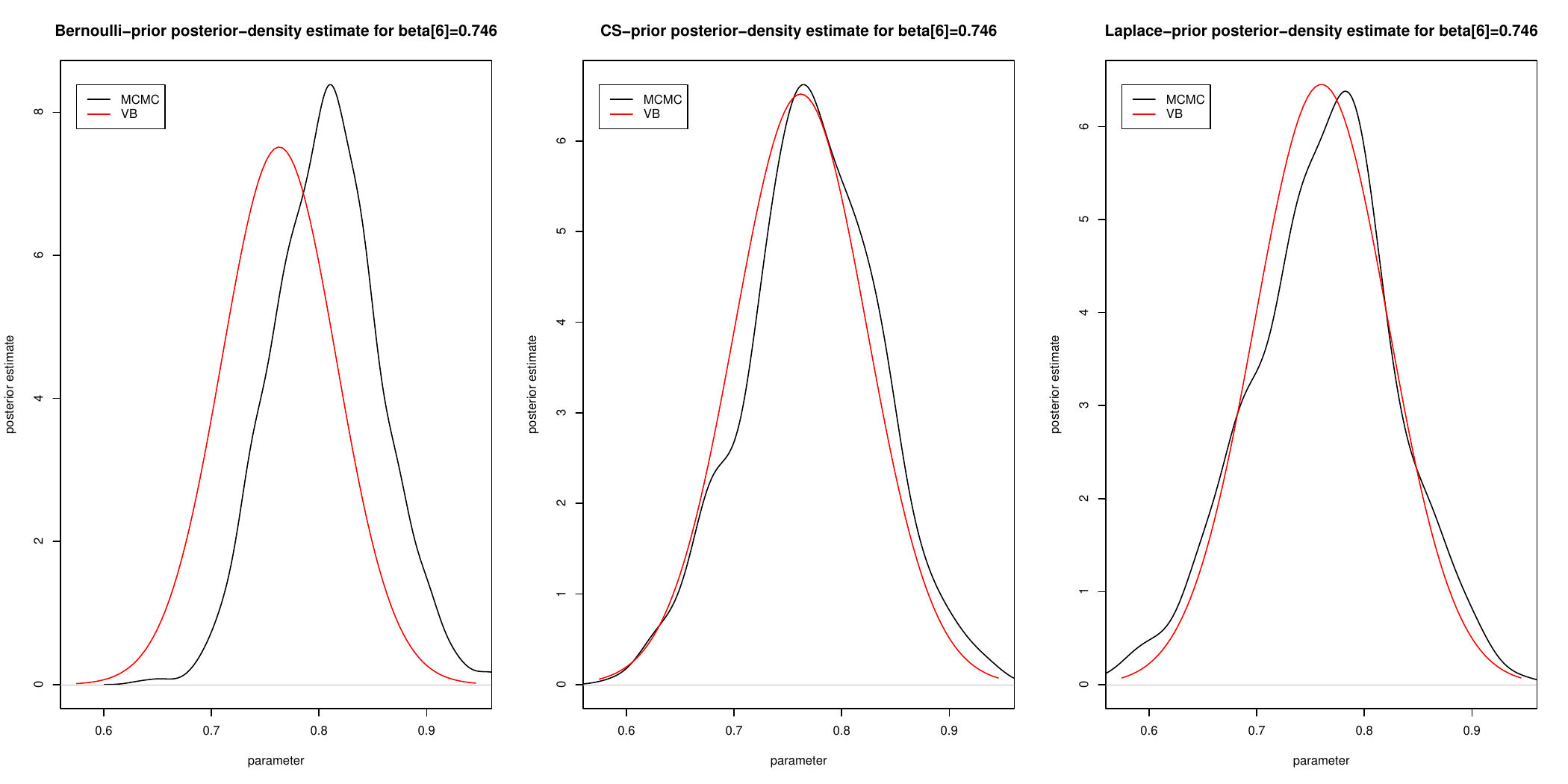}}
\centerline{\includegraphics[scale=0.3]{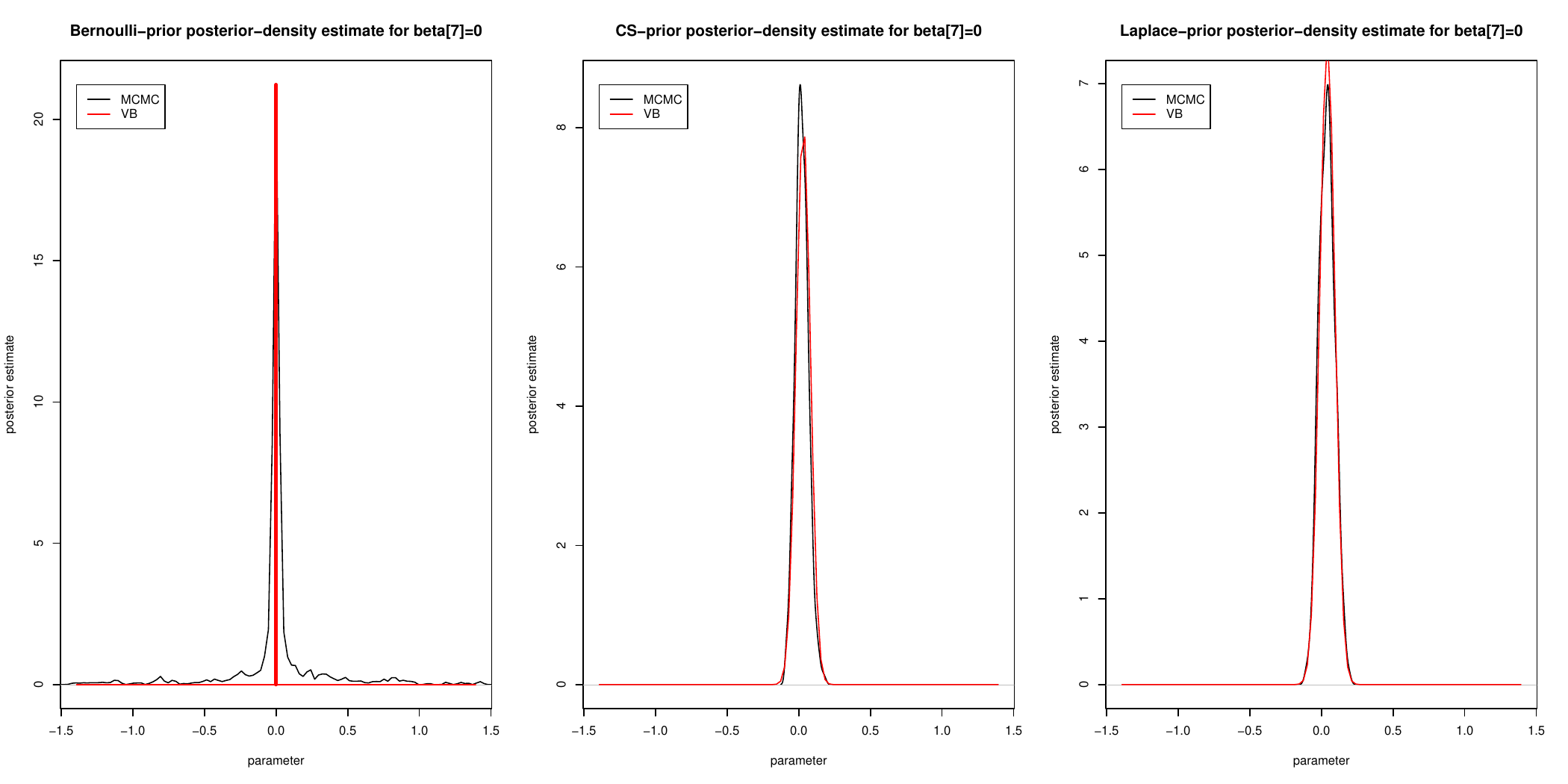}}
\centerline{\includegraphics[scale=0.3]{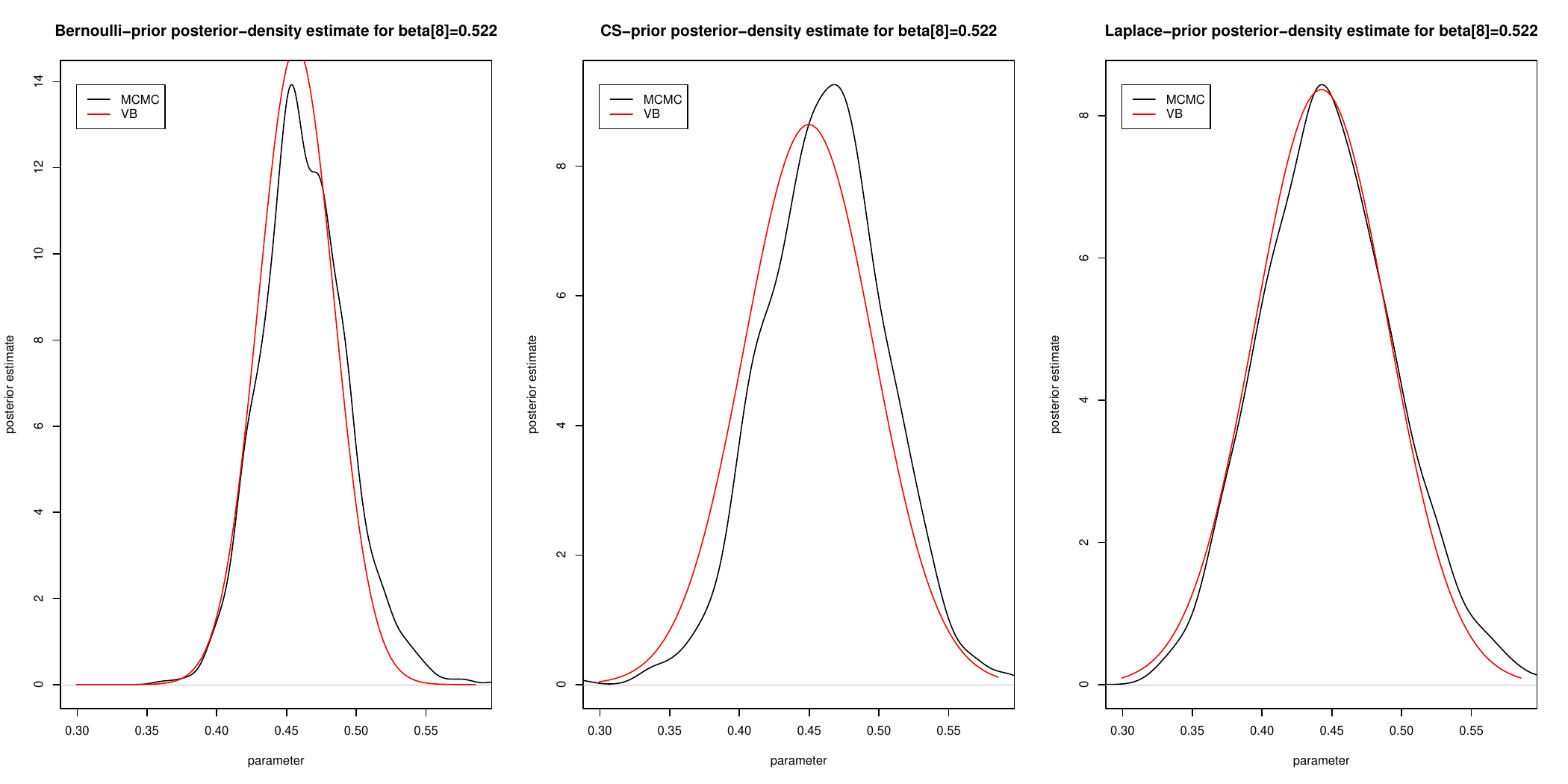}}
\centerline{\includegraphics[scale=0.3]{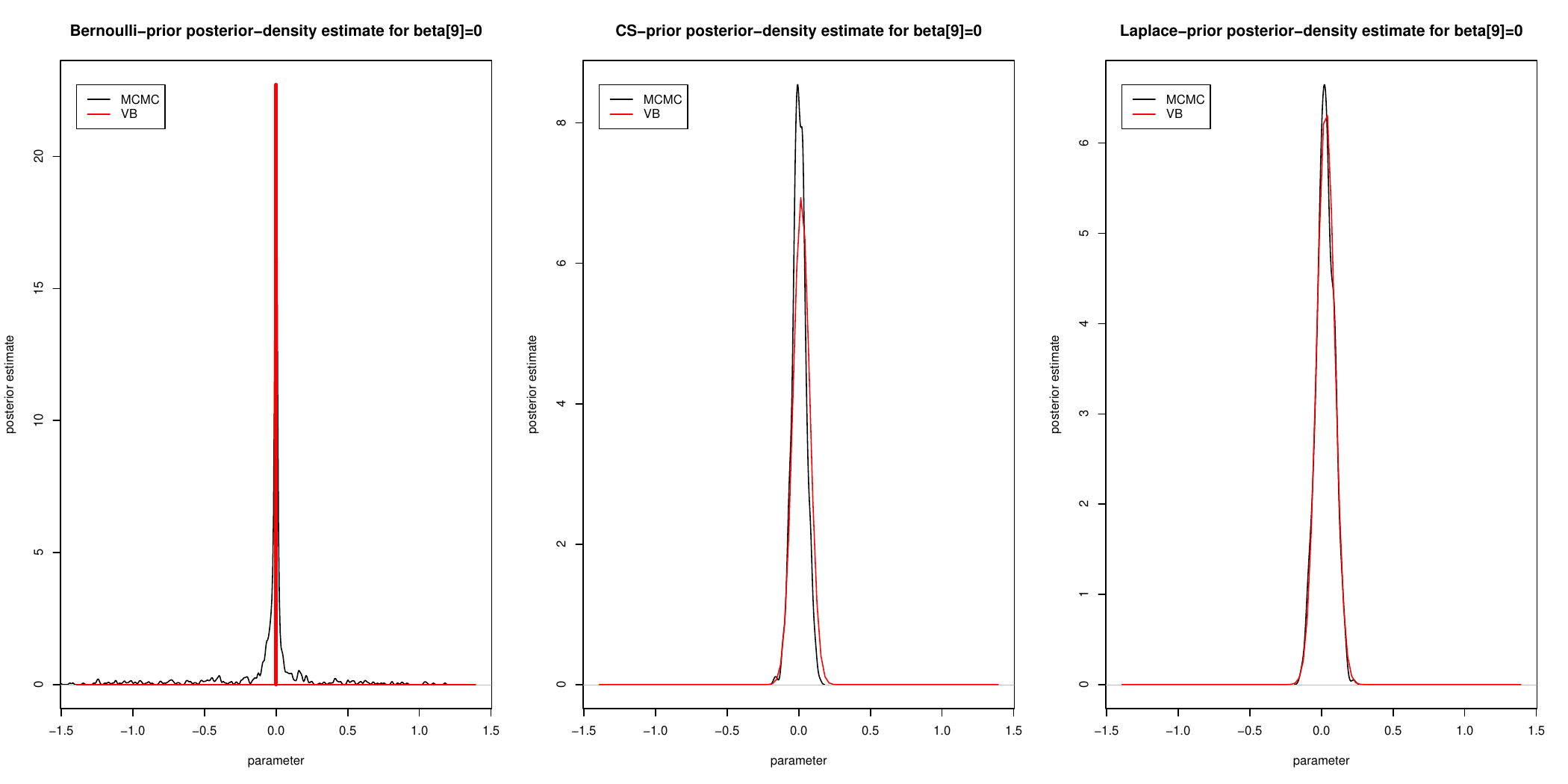}}
\caption{ Approximate posterior density functions of the last four regression coefficients for different models.}\label{dens3}
\end{figure}

{ To examine} the accuracy of the approximation of the posterior distribution by VB methods, the accuracy measure \citep[see][]{lu15} is computed for each parameter component $\theta_j,\; j=1,\ldots,k,$ as follows
\begin{equation}\label{acc}
{\rm accuracy}(q(\theta_j)) = 100\left(1-\frac{1}{2}\int_{-\infty}^\infty |q(\theta_j) - p(\theta_j|y)| d\theta\right)\%,
\end{equation}
where $p(\theta_j|y)$ is the kernel density estimator using the MCMC sample.
{ If $q(\theta) = p(\theta|y)$ for all values of $\theta$, we have ${\rm accuracy}(q(\theta_j)) = 100\%,$ while since both $q(\cdot)$ and $p(\cdot|y)$ are probability density functions, the minimum value of ${\rm accuracy}(q(\theta_j))$ is zero, which occurs when $q(\cdot)$ and $p(\cdot|y)$ have separate supports.}

{\color{black} Note that the variational posteriors of the regression coefficients under the Bernoulli-VB method are not directly comparable to their MCMC estimated densities. In this case, we would need to compute the accuracy of the variational posterior of ${ \boldsymbol\beta}\Gamma$. Thus, the marginal densities are either $q(\beta_j)$ or degenerate at zero (when $\gamma_j=0$). {Based on the criterion in Equation \eqref{acc}, the accuracy of the variational posterior is reduced, while in this scenario the point mass at zero offers a notably more accurate posterior.} Thus, for the simulation study where the true $\beta_j$ is known, we have replace $p(\beta_j|y)$ by $\delta(\beta_j)p(\beta_j|y)$ in \eqref{acc}, for the Bernoulli-VB method.}

For the Bernoulli components $q(\gamma_j)$ and $q(Z_j)$ for Bernoulli-VB and CS-VB methods, the accuracy is computed as follows
$${\rm accuracy}(q(\theta_j)) =100\left(1-\frac{1}{2}\left(\left|q(\theta_j = 1) - \frac{1}{T}\sum_{t=1}^T I(\theta_{jt} = 1)\right| +\left|q(\theta_j = 0) - \frac{1}{T}\sum_{t=1}^T I(\theta_{jt} = 0)\right|\right)\right)\%,$$
where $\theta_{j1},\ldots,\theta_{jT}$ are MCMC samples. { If $q({ \theta}_j = k) = \frac{1}{T}\sum_{t=1}^T I({ \theta}_{jt} = k)$ for $k = 0,1$, we have ${\rm accuracy}(q(\theta_j)) = 100\%,$ and when $q({ \theta}_j = k) = 1$ and $\frac{1}{T}\sum_{t=1}^T I({ \theta}_{jt} = k) = 0$ for some $k \in \{0,1\}$, we have ${\rm accuracy}(q(\theta_j)) = 0\%$.}

{The average (standard error)} of accuracy values are given in Table \ref{tab1}. {The boxplots for the accuracy of the regression coefficients are given in Figure \ref{boxac1}.} Furthermore, the $q$-functions as well as the kernel density estimator of the marginal posteriors of the regression coefficients, for a single iteration of the simulation study, are given in Figures \ref{dens1}, \ref{dens2}, and the corresponding plots for the other parameters are given in {the supplementary material}. 

As one can see from {Table \ref{tab1} and Figure \ref{boxac1}, the accuracy of the posterior approximation for the regression coefficients is high. Table \ref{tab1} also presents the accuracy for the parameters ${ \boldsymbol\gamma}$ and $\mathbf{Z}$ of the VB methods against an MCMC benchmark for the Bernoulli and CS priors. {As can be seen from these values, the CS-VB has higher accuracies than the Bernoulli-VB method}. The accuracy for the $\pi$ hyperparameter is { moderate} for both Bernoulli-VB and CS-VB methods. Regarding the variance hyper-parameters of the Bernoulli-VB and Laplace-VB methods, we can see from Table \ref{tab1} that { the Bernoulli-VB method has lower accuracies for the variance components, which correspond with the zero coefficients compared to the other components, while the Laplace-VB method achieves almost equal accuracies for all components $\tau_0,\ldots,\tau_4$. }

\begin{figure}
\centerline{\includegraphics[scale=0.4]{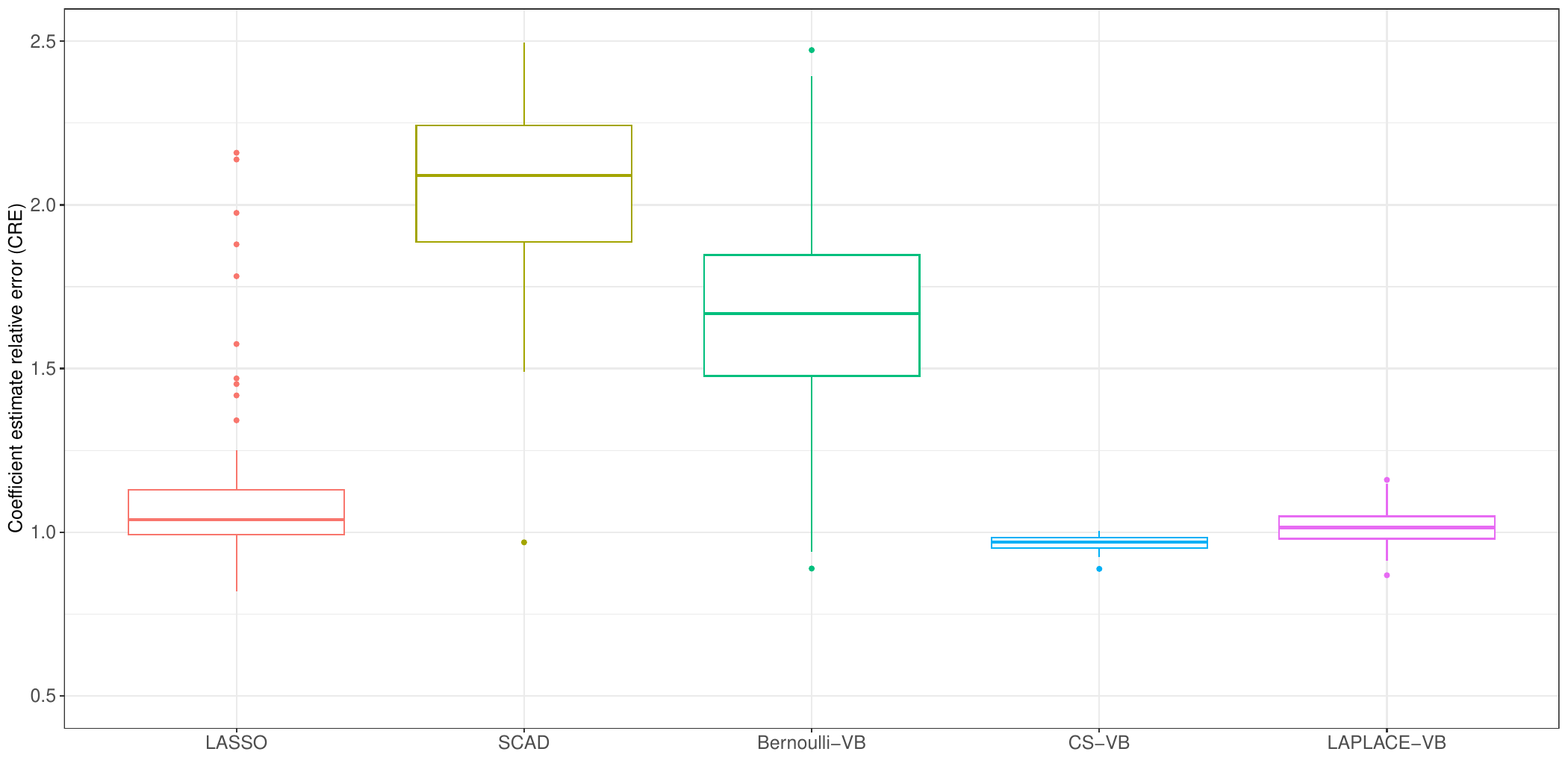}}
\centerline{\includegraphics[scale=0.4]{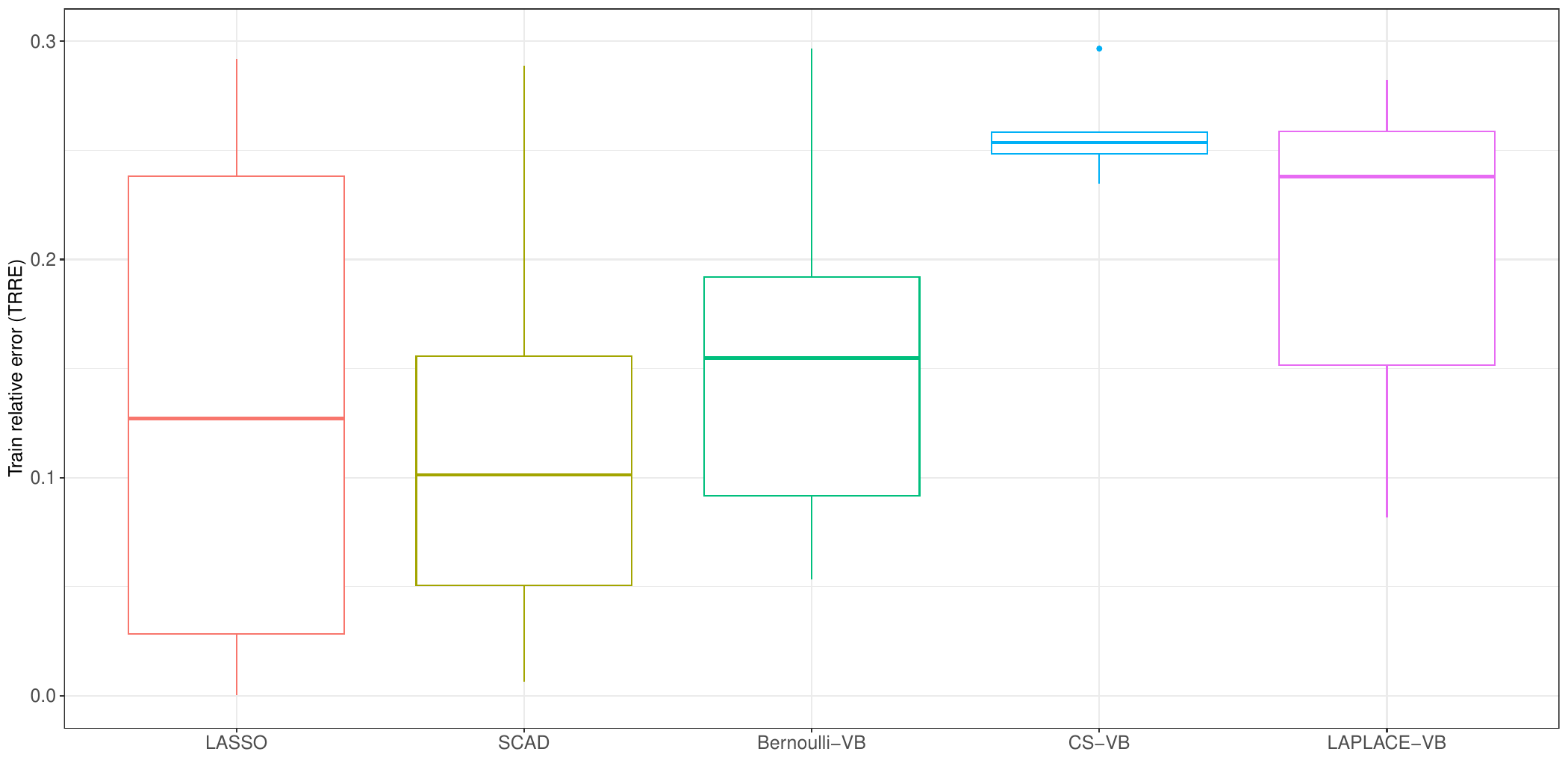}}
\centerline{\includegraphics[scale=0.4]{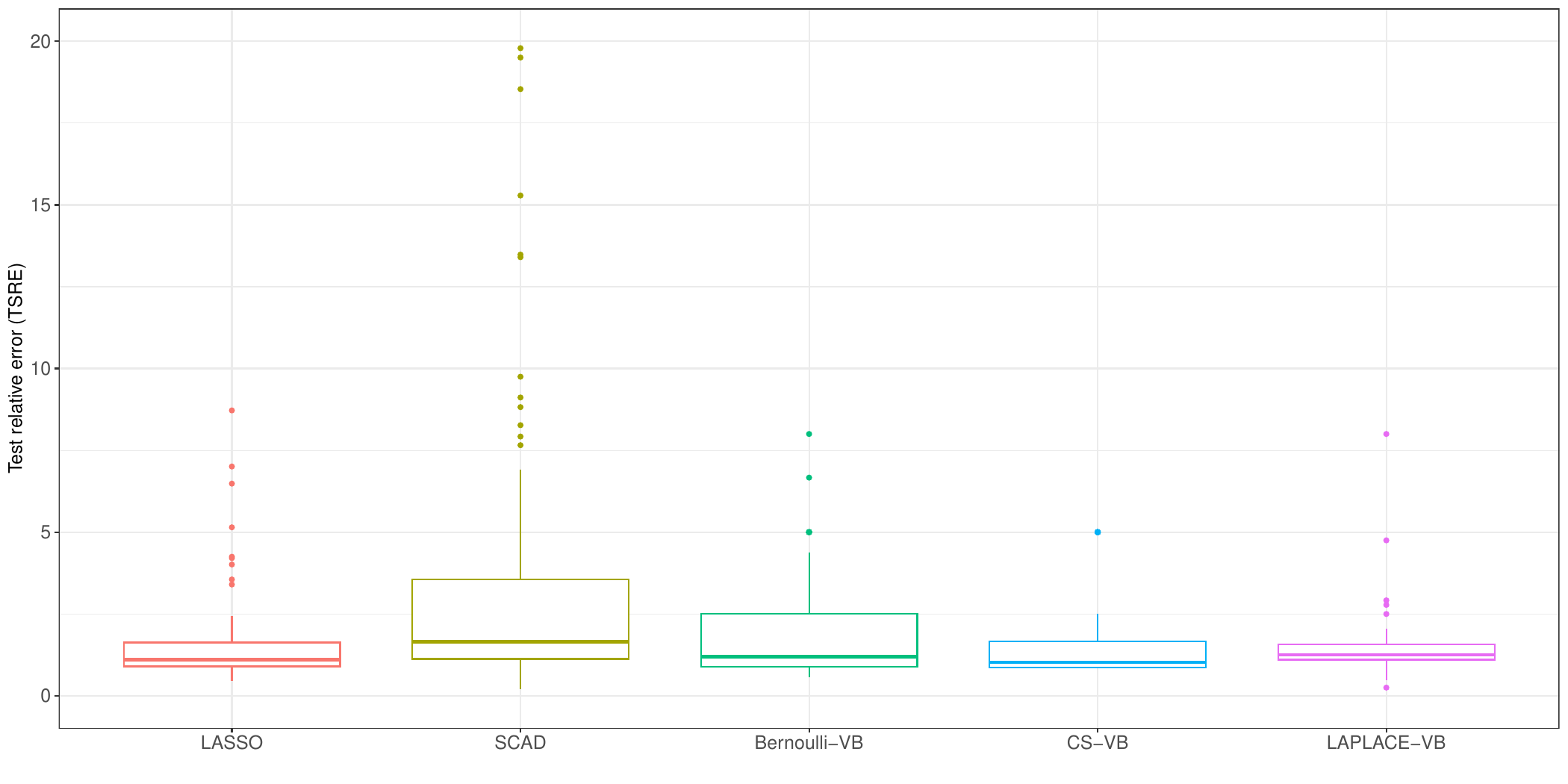}}
\caption{The high-dimensional scenario simulation results: the coefficient relative error (top), the train relative error (middle), and the test relative error (bottom) for different methods.}\label{box3}
\end{figure}

\begin{figure}
\centerline{\includegraphics[scale=0.4]{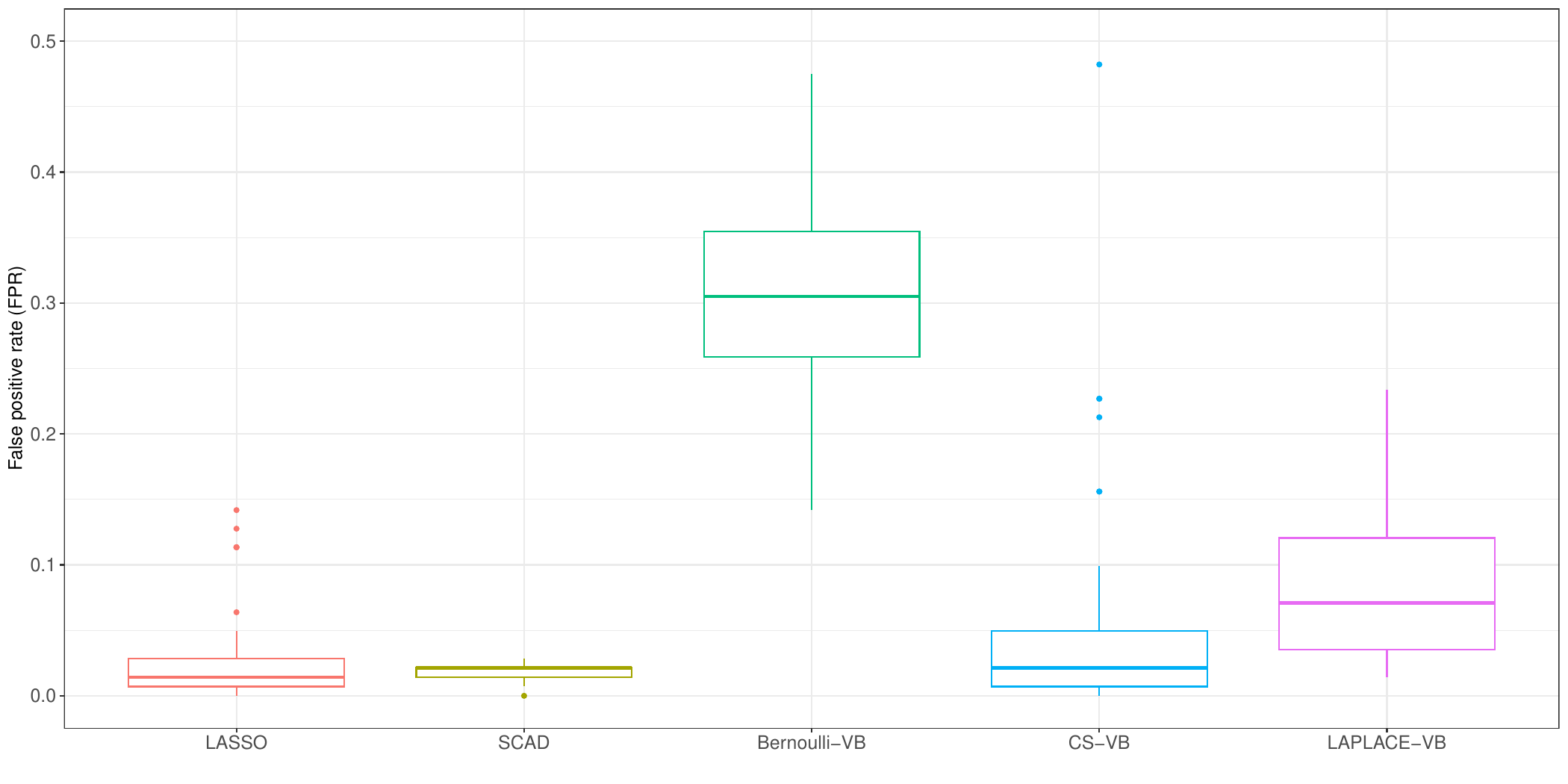}}
\centerline{\includegraphics[scale=0.4]{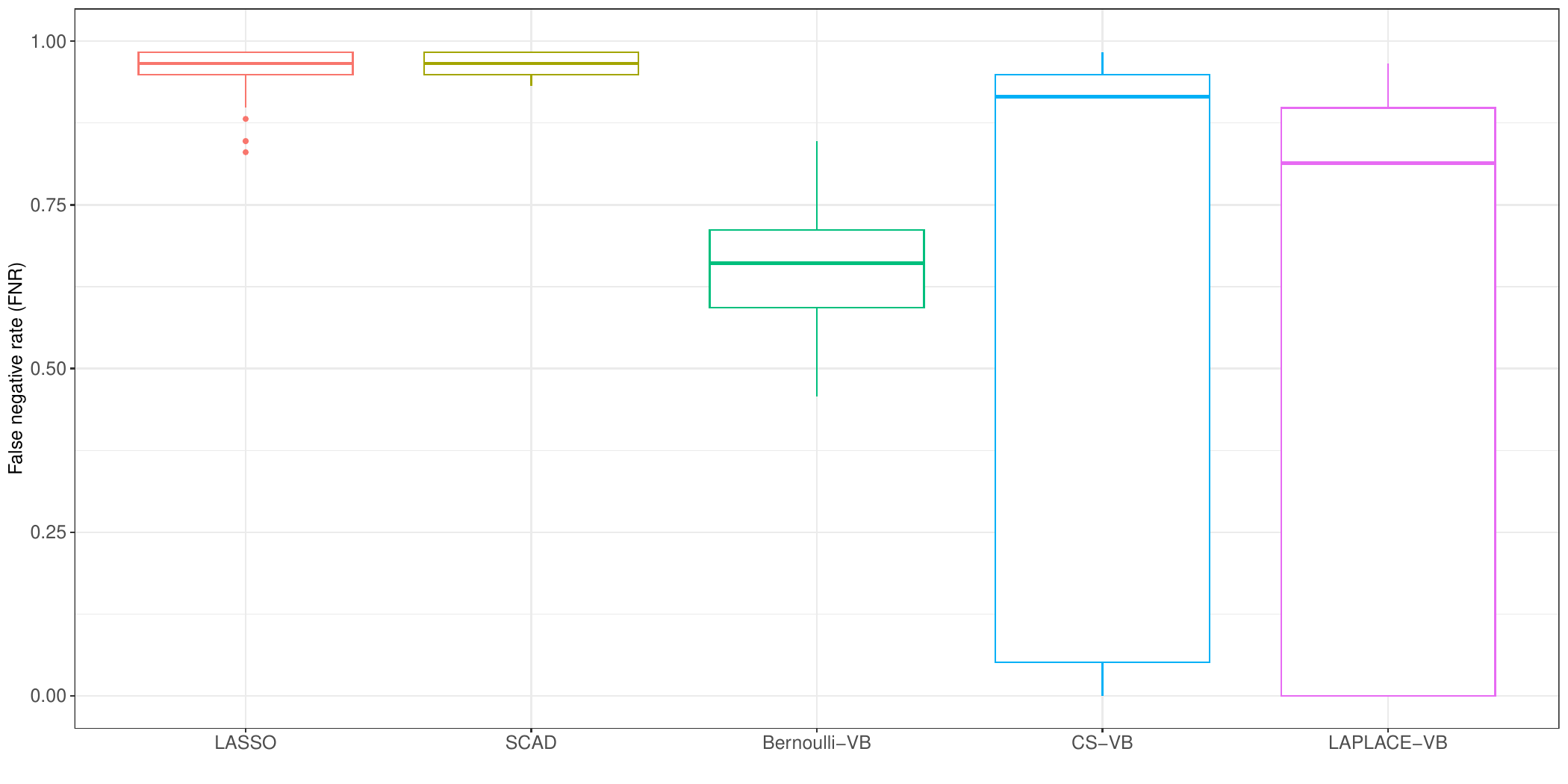}}
\centerline{\includegraphics[scale=0.4]{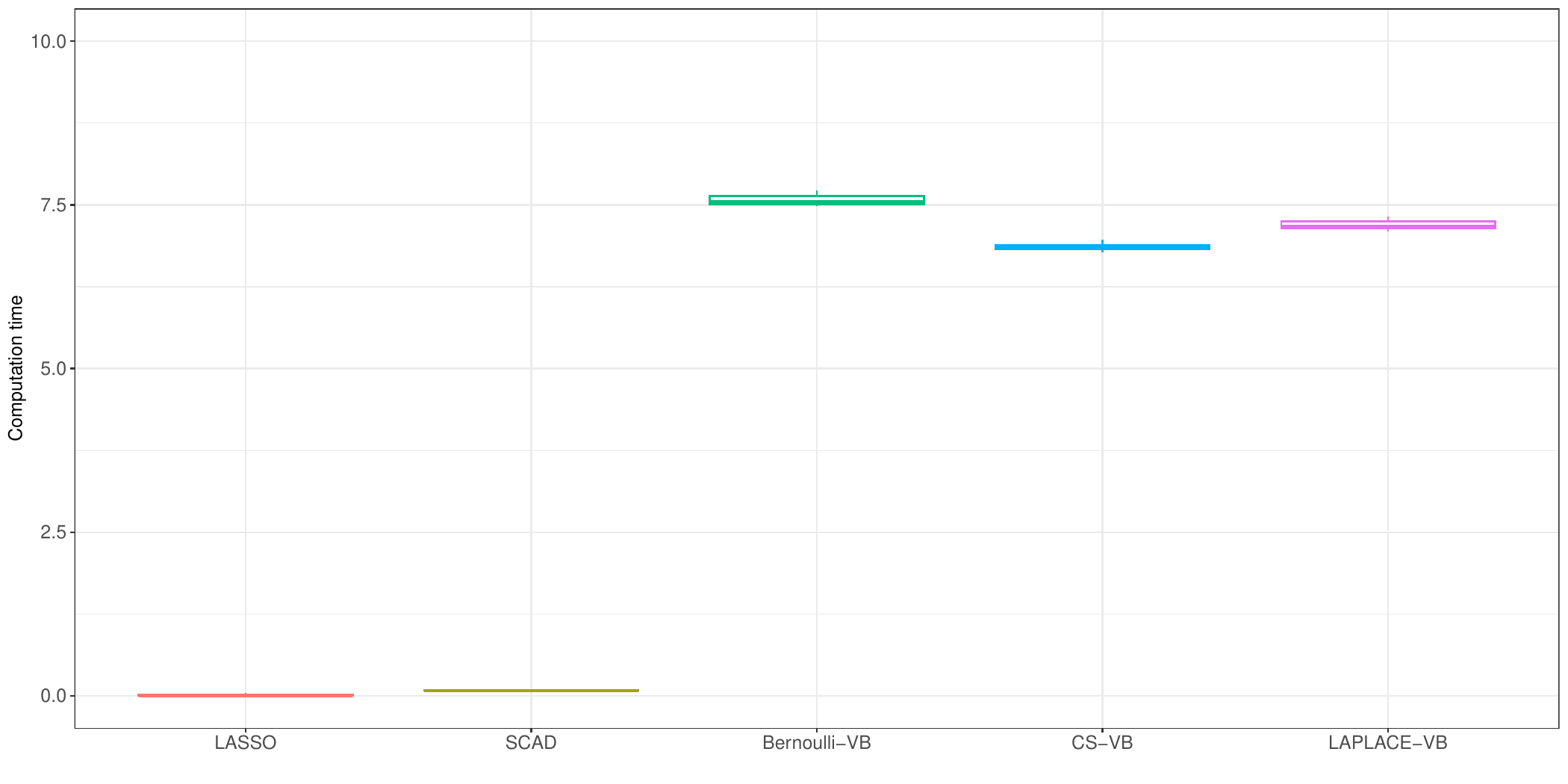}}
\caption{The high-dimensional scenario simulation results: the FPR (top), the {\color{black} FNR} (middle), and the computation time (bottom) for different methods.}\label{box4}
\end{figure}

{\subsection{High {\color{black} dimensional} scenario}

To evaluate the performance of the proposed methods in a high-dimensional setting, we consider the case with
$p = 200$, $n = 30$, and {We randomly selected 60 values of 
$\mathbf{z}$ (including the first) to be set to 1, with the remaining values assigned 0.}. Figures \ref{box3} and \ref{box4} show the boxplots of all aforementioned criteria for all competitive methods.

{ For CRE, the VB approaches achieve competitive median errors close to LASSO and SCAD, with Laplace-VB generally showing the tightest distribution with the lowest median among the proposed methods. The CS-VB method is the second-best VB method concerning the CRE criterion and has achieved the highest TRRE value. In TRRE and TSRE, the VB variants maintain accuracy on par with or close to the best-performing baselines, with Bernoulli-VB yielding the lowest training error and CS-VB and LAPLACE-VB demonstrating stable, consistent generalization. In sparsity performance, all VB methods control FPR effectively, particularly CS-VB and LAPLACE-VB, which show compact variability. It seems that in this case, all methods have high FNR values, while the CS-VB and Laplace-VB methods can achieve lower FNR compared to the LASSO and SCAD methods for some samples. 

Overall, these figures confirm that the proposed VB framework delivers strong and well-balanced performance across estimation, prediction, and classification metrics, while simultaneously offering the dramatic computational advantages inherent to variational inference, reducing runtimes by several orders of magnitude compared to MCMC without sacrificing accuracy or stability.



 On average, the VB methods require approximately 100 times the computation time of the SCAD method, yet their total runtime remains under 10 seconds.}

\section{Benchmark real data analysis}

\begin{figure}
\centerline{\includegraphics[scale=0.33]{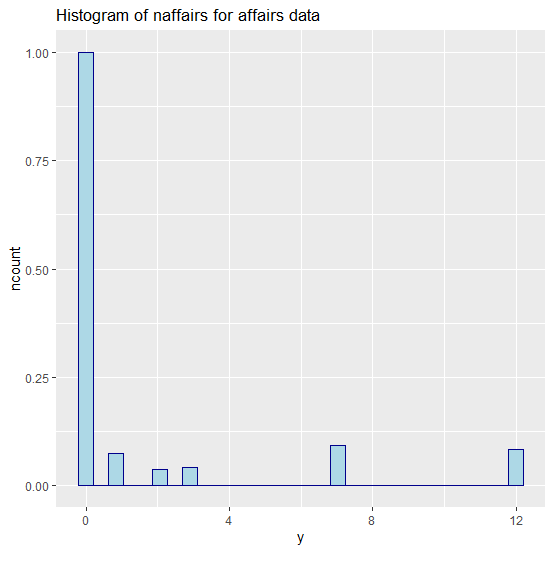}\includegraphics[scale=0.33]{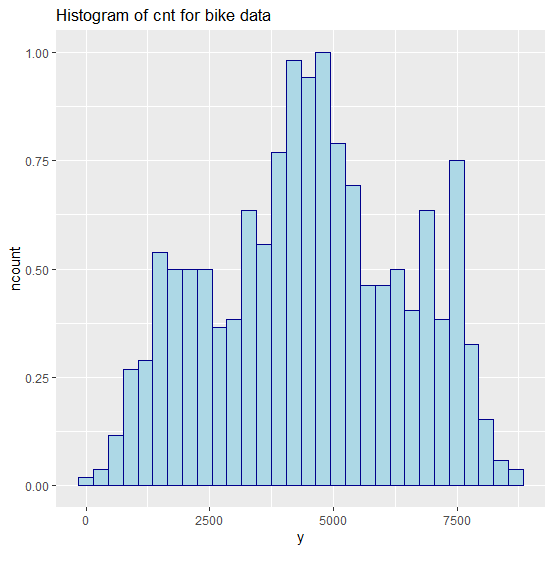}\includegraphics[scale=0.33]{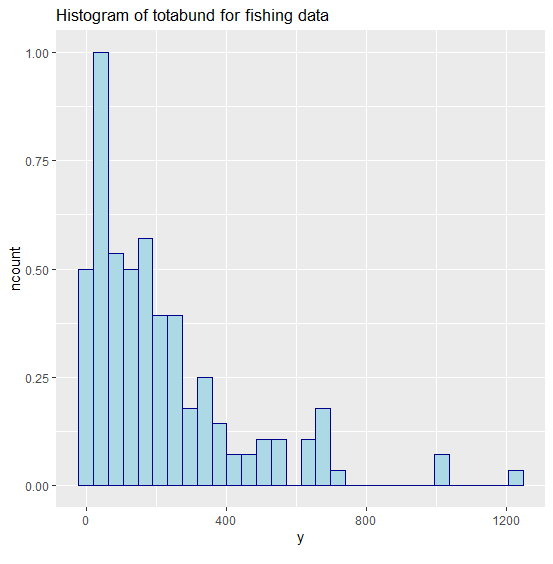}}
\centerline{\includegraphics[scale=0.33]{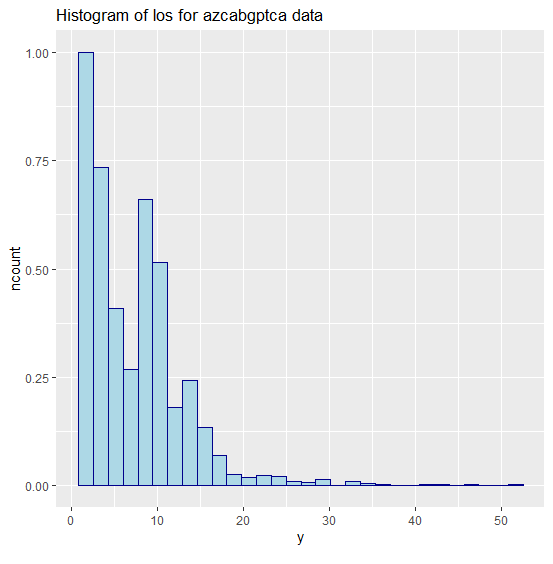}\includegraphics[scale=0.33]{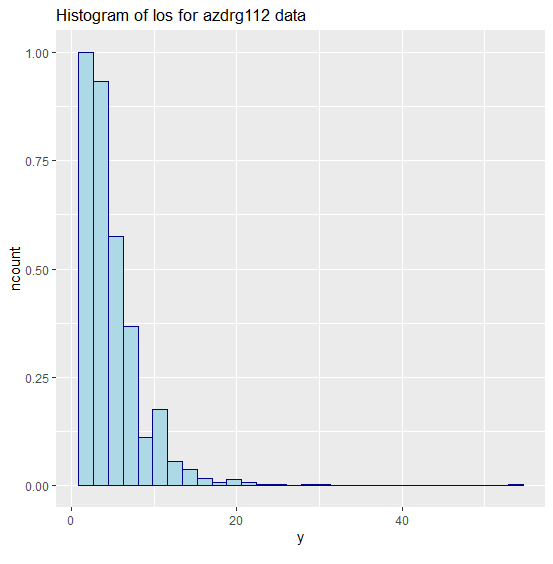}\includegraphics[scale=0.33]{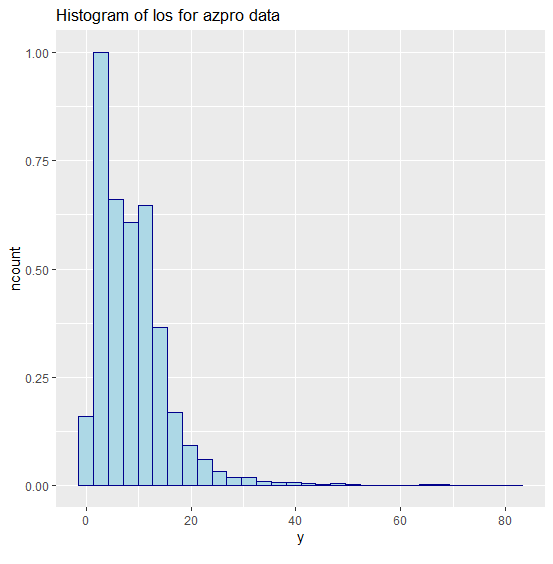}}
\caption{Histograms of the responses for the benchmark real data sets. Top left; affairs, top middle; bike sharing, top right; fishing, bottom left; azcabgptca, bottom middle; azdrg112, bottom right; azpro.}\label{hists}
\end{figure}

To examine the prediction performance of the proposed VB methods, we have considered 5 benchmark real data sets with a count response variable. It is worth noting that we do not claim that these datasets are suitable for Poisson regression. Indeed, all these datasets exhibit overdispersion, making a Negative Binomial model a strong candidate; however, our primary aim is to benchmark the performance of the proposed sparse variational Bayes algorithms within the Poisson framework. Using these challenging, real-world datasets provides a stringent test that demonstrates our method's robustness in accurate variable selection and its computational advantages, even under realistic conditions of model misspecification. {\color{black} Furthermore, we have also fitted the LASSO negative binomial regression from the \texttt{mpath} package in \texttt{R}, and the results show that the Poisson regression models generally perform better than it in terms of test relative error.}

 {The affairs data set \citep{f79}, available in the R package COUNT, contains 601 observations and 18 variables, including naffairs (the number of affairs in the past year) and 17 covariates related to children, marital happiness, religiosity, and age.} \cite{gr03}, modeled this data using Poisson regression, although given the amount of over-dispersion in the data, employing a negative binomial model is an appropriate strategy. The variable naffairs is considered the response variable. The top left graph of Figure \ref{hists} shows the histogram of the response variable naffairs. { The bike-sharing dataset records hourly and daily rental counts from 2011 to 2012 in the Capital Bikeshare system, along with corresponding weather and seasonal information.} This data set includes 731 observations and 14 variables (except date and instant number), including cnt (the response variable), and 13 covariates, including season, year, month, holiday, weekday, working day, 5 weather variables, casual, and registered. The top middle graph in Figure \ref{hists} presents the histogram of the response variable for the bike-sharing data set. { The three remaining data sets, azcabgptca, azdrg112, and azpro, from the R package COUNT pertain to samples from Arizona hospital cardiovascular patient files, collected in 1991 (azcabgptca, azpro) and 1995 (azdrg112), involving patients who received one of two standard procedures: CABG or PTCA.} The data set azcabgptca has 1959 observations on 6 variables, azdrg112 has 1,798 observations on 4 variables, and azpro has 3589 observations on 6 variables. { In all three data sets, the response variable is los (length of hospital stay), with covariates including procedure type, sex, age, and additional factors.} The bottom plots of Figure \ref{hists} show the histograms of the los response variable in azcabgptca, azdrg112, and azpro, respectively, from left to right. 

\begin{table}
\centering\color{black}
\caption{Test relative error means and (standard deviations) for 10 random partitions of 6 benchmark real data sets.}\label{tab4}
\vspace{2mm}
\begin{tabular}{c c c c c c c}
\hline\hline
Data set & LASSO-NB & LASSO-Poiss & SCAD-Poiss & Bernoulli-VB & CS-VB & LAPLACE-VB \\
\hline
affairs & 0.928 & 0.909 & 0.909 & 0.915 & 0.917 & 0.918 \\
 & (0.084) & (0.086) & (0.087) & (0.070) & (0.068) & (0.065) \\
\hline
bike sharing & 0.112 & 0.053 & 0.054 & 0.053 & 0.054 & 0.054 \\
 & (0.024) & (0.005) & (0.005) & (0.004) & (0.005) & (0.005) \\
\hline
azcabgptca & 0.539 & 0.537 & 0.538 & 0.556 & 0.556 & 0.549 \\
 & (0.039) & (0.038) & (0.038) & (0.034) & (0.040) & (0.039) \\
\hline
azdrg112 & 0.851 & 0.851 & 0.851 & 0.877 & 0.850 & 0.850 \\
 & (0.028) & (0.028) & (0.029) & (0.022) & (0.027) & (0.027) \\
\hline
azprocedure & 0.629 & 0.626 & 0.626 & 0.632 & 0.627 & 0.627 \\
 & (0.026) & (0.025) & (0.026) & (0.023) & (0.025) & (0.025) \\
\hline
azpro & 0.622 & 0.619 & 0.619 & 0.625 & 0.619 & 0.620 \\
 & (0.048) & (0.048) & (0.048) & (0.045) & (0.047) & (0.047) \\
\hline\hline
\end{tabular}
\end{table}

We randomly partition the observations of each dataset into a training set (80 \%) and a test set (20 \%), and replicate the random partitioning 10 times to compute the test set relative prediction error and compare it for all considered competitors. Table \ref{tab4} presents the means and standard deviation of test set relative errors for 10 random partitions of all data sets, for three VB methods, obtained from ppmfs in \eqref{Lpd} and \eqref{Lpd2}, as well as those for LASSO and SCAD Poisson regression methods. { Table \ref{tab4} demonstrates that, for all six benchmark data sets, the proposed VB methods perform comparably to frequentist sparse Poisson regression approaches.}
\begin{figure}
\centerline{\includegraphics[scale=0.37]{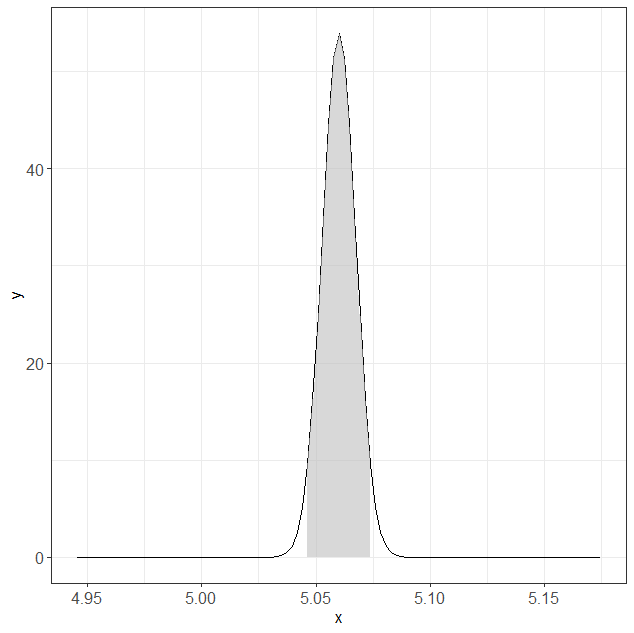}\includegraphics[scale=0.37]{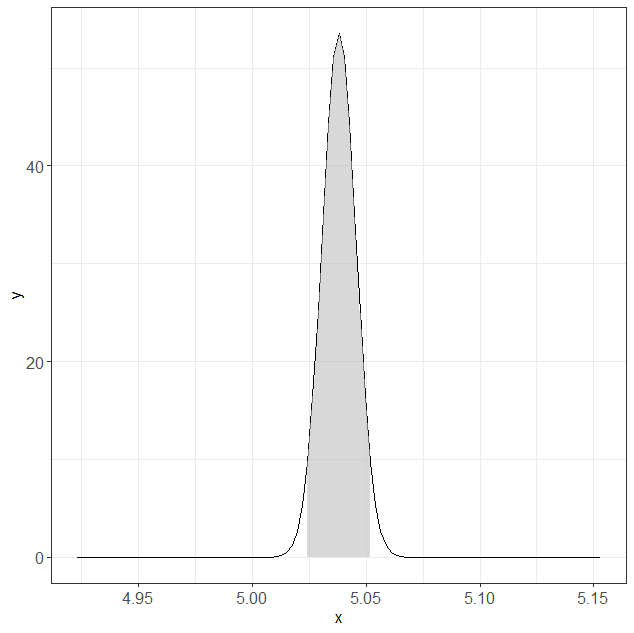}\includegraphics[scale=0.37]{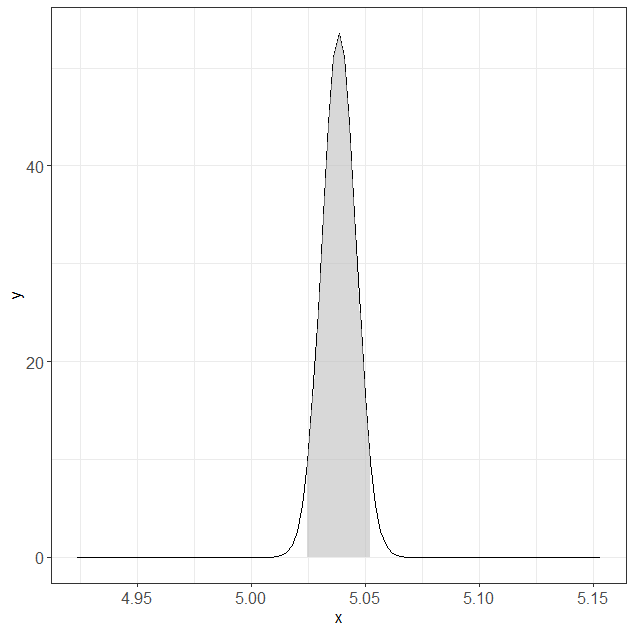}}
\centerline{\includegraphics[scale=0.37]{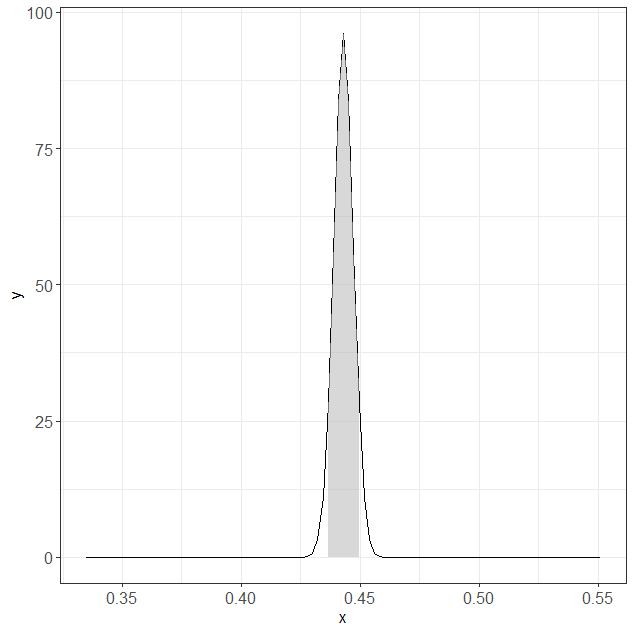}\includegraphics[scale=0.37]{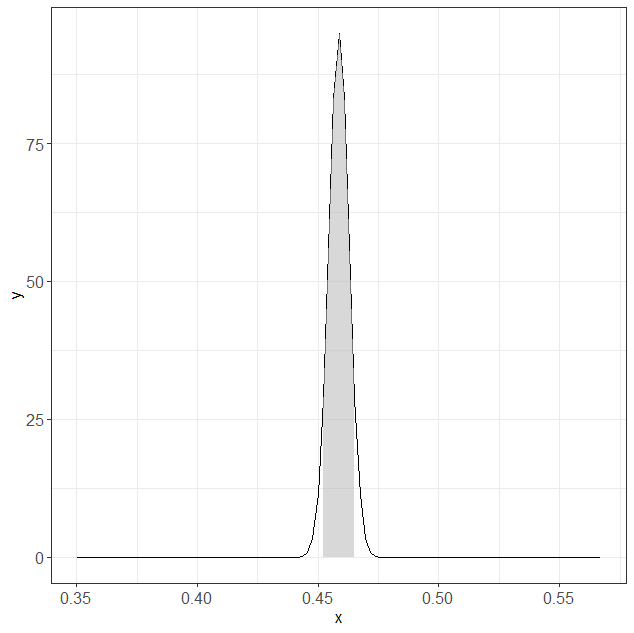}\includegraphics[scale=0.37]{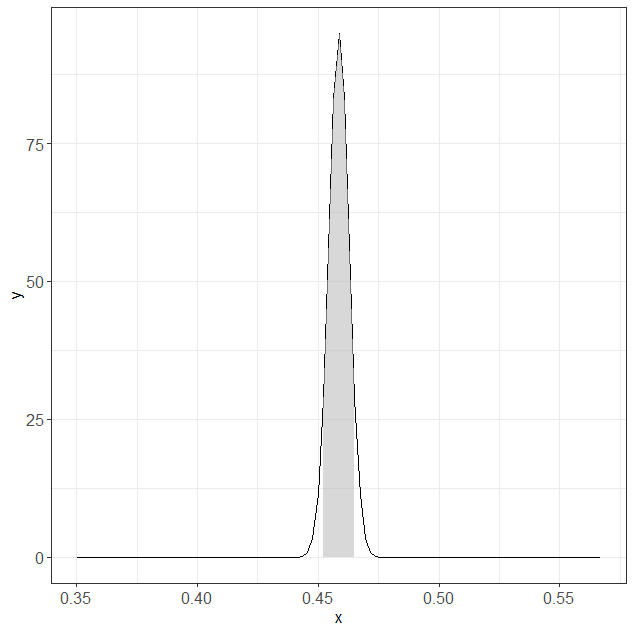}}
\centerline{\includegraphics[scale=0.37]{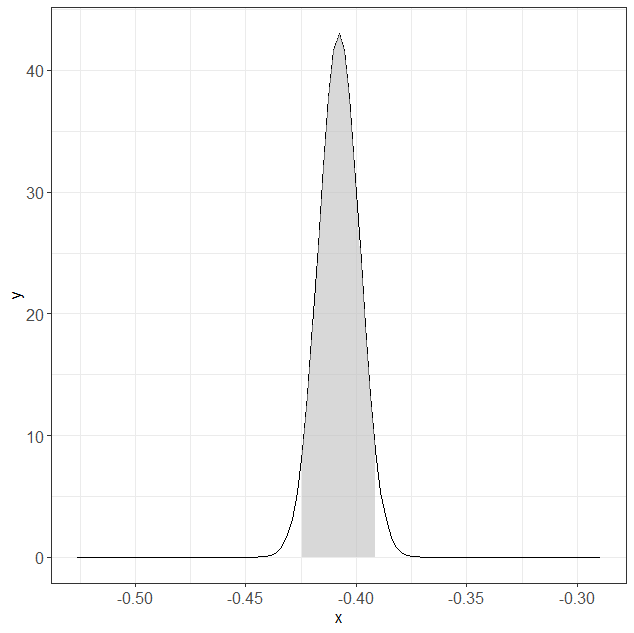}\includegraphics[scale=0.37]{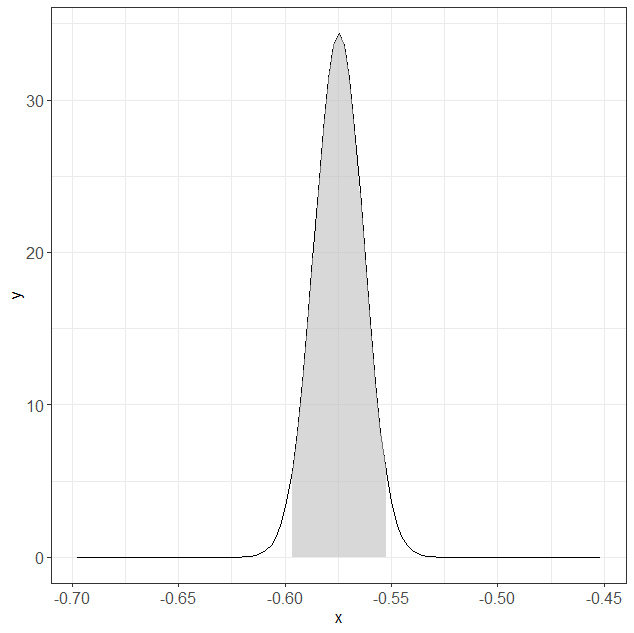}\includegraphics[scale=0.37]{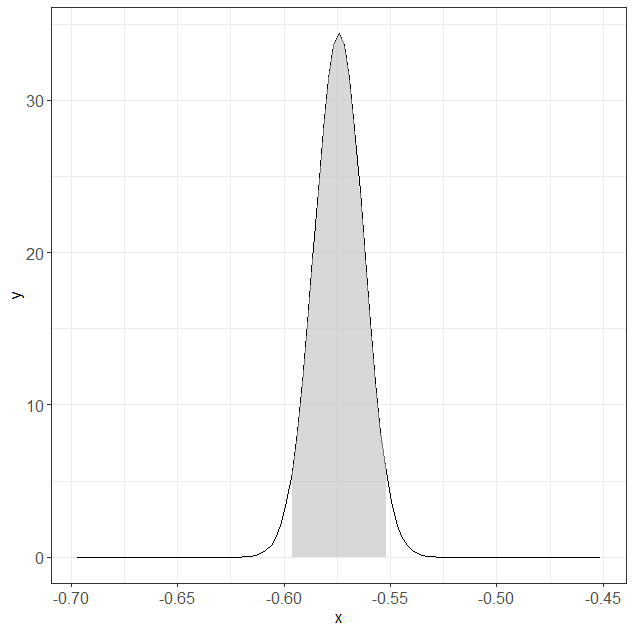}}
\centerline{\includegraphics[scale=0.37]{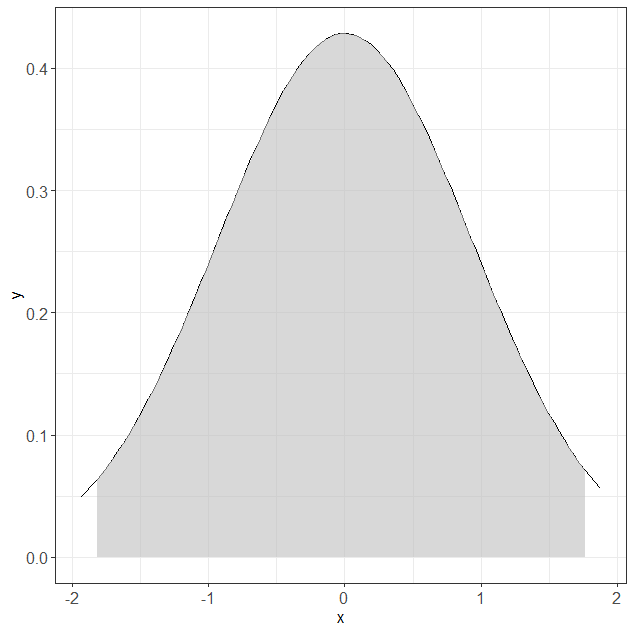}\includegraphics[scale=0.37]{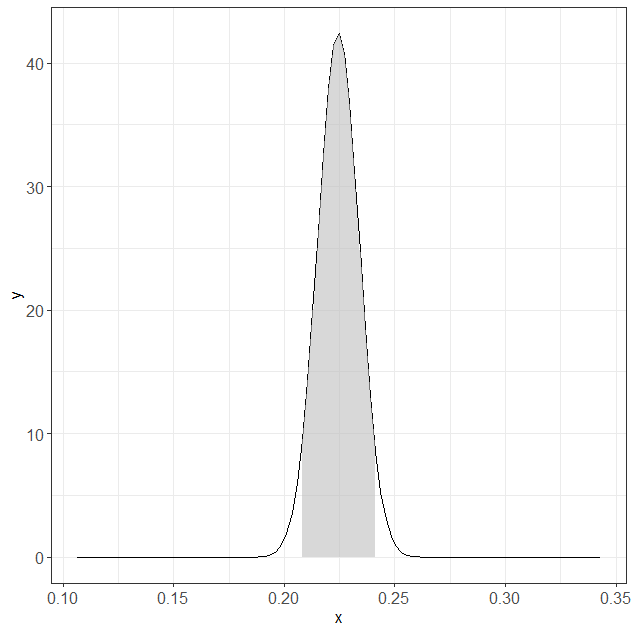}\includegraphics[scale=0.37]{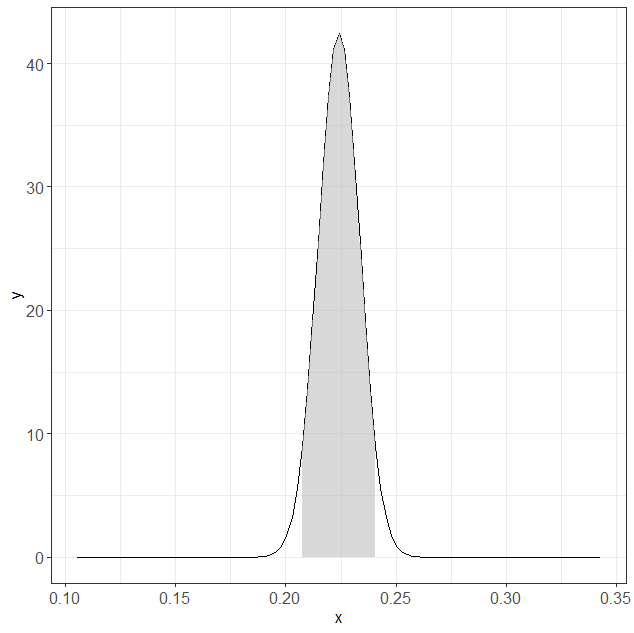}}
\caption{The posterior approximated densities of the regression coefficients, and the corresponding HPD intervals, for the fishing data set. Columns from left to right are associated with Bernoulli-VB, CS-VB, and Laplace-VB, and rows from top to bottom are associated with the intercept and 3 regression coefficients: density, mean-depth, and swept-area.}\label{fishing}
\end{figure}
{ \subsection{Fishing data set}

\begin{sidewaysfigure}
\centerline{\includegraphics[scale=0.45]{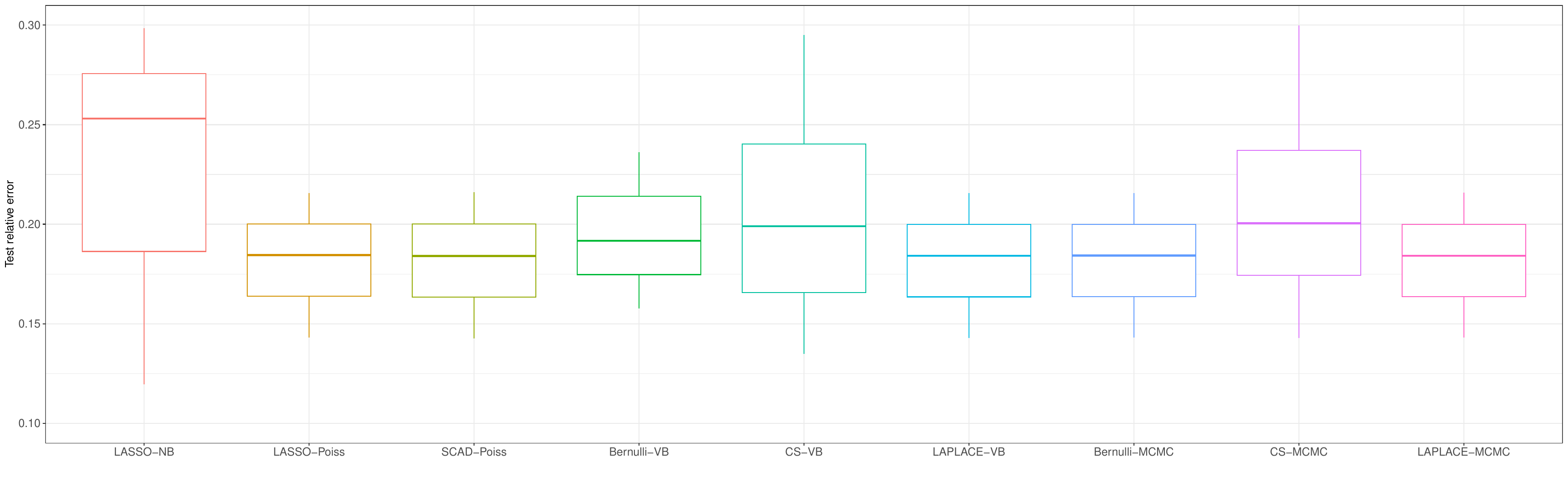}}
\caption{Test relative error boxplots for 10 random partitions of fishing data set.}\label{boxf}
\end{sidewaysfigure}

\begin{figure}
\centerline{\includegraphics[scale=0.3]{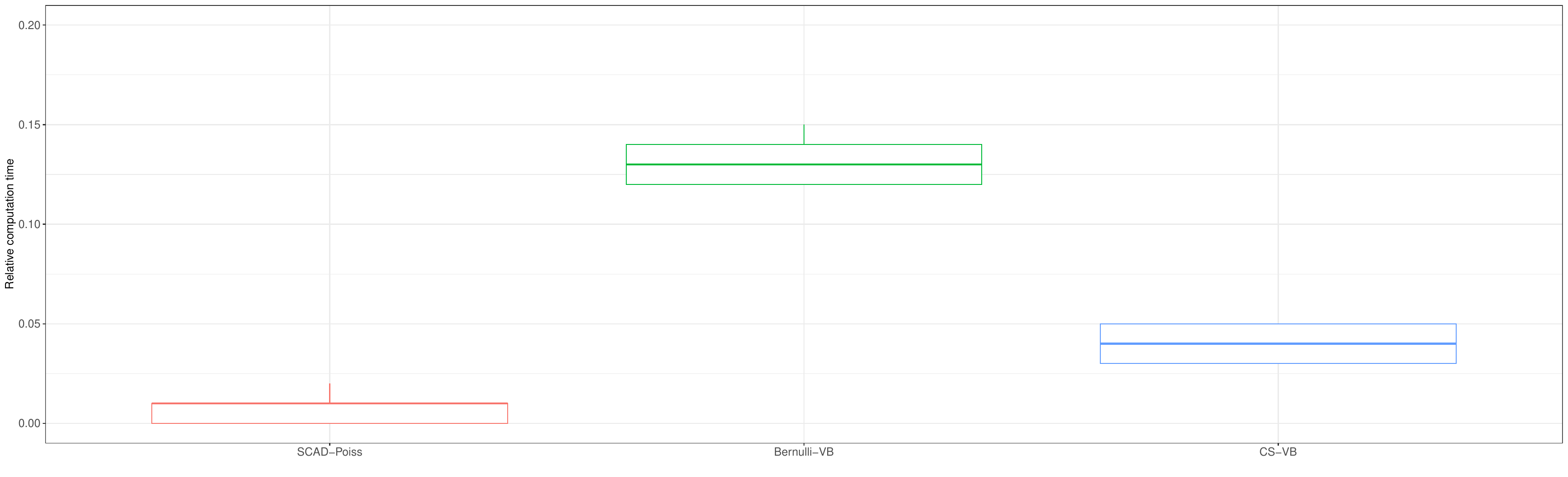}}
\caption{ The boxplots of relative computational time of VB methods compared to the MCMC methods for 10 random partitions of the fishing data set.}\label{boxr}
\end{figure}

We have chosen one of the benchmark data sets, that is the fishing data \citep{zea13}, for a more extensive numerical study. { For this dataset, we further compare the predictive performance of the VB methods against their MCMC counterparts, and present the HPD intervals for the model coefficients.} 

{\color{black} The fishing data, adapted from \cite{bea09}, investigate the effects of commercial fishing on certain deep-sea fish populations when operations expanded into deeper waters than in prior years. Observations from 147 sites include totabund (total fish per site) and six covariates: depth, area, foliage density, catch site, year, and period. In this study, only three covariates, density, mean depth, and swept area, are used.}  The top right plot in Figure \ref{hists} shows the histogram for this data set.

The boxplots for the test relative errors of all competitive methods, including also the MCMC versions of each proposed VB method, are shown in Figure \ref{boxf}. { As depicted in the plot, the VB methods achieve predictive accuracy equivalent to that of their MCMC versions for the fishing data set.} Furthermore, Figure \ref{boxr} presents the boxplots of the relative computational time of VB methods compared to the MCMC methods for 10 random partitions of the fishing data set. { From the figure, it is evident that the VB methods, implemented in R, outperform their MCMC counterparts, implemented in C++, by a factor of at least 28 in computational speed.} The posterior approximated densities of the regression coefficients, and the corresponding HPD intervals, for the fishing data set, are plotted in Figure \ref{fishing}. }

\section{Concluding remarks}

The VB methods proposed in this paper provide a fast Bayesian inference about the parameters of the sparse Poisson regression model. These methods are { substantially} faster than the MCMC methods. The proposed VB methods are used for simultaneous estimation and variable selection in the Poisson regression model. { The findings indicate that the proposed Bayesian inference achieves performance equivalent to the LASSO and SCAD penalized Poisson regression models. Beyond matching predictive accuracy, the Bayesian approach offers further advantages, including direct access to highest posterior density intervals, parameter posterior distributions, and the predictive probability mass function for the response given the data.} It is observed that the proposed VB methods provide a good approximation of the marginal posterior densities. {\color{black} A limitation of the proposed models is their sensitivity to the assumption of the Poisson distribution of the response, which fails when there is over-dispersion and under-dispersion in the response observations.}  Another limitation is due to the quadratic approximation used based on the idea from \cite{ja00}, which results in a Gaussian approximation of $q(\boldsymbol\beta)$, which might not provide a good approximation for the posterior. {\color{black} The authors address all of these issues, {\color{black}including development of the VB model for the sparse negative binomial regression,} in their proposed directions for future research.} A guideline for choosing the prior is given in Appendix B, based on the results of the simulation study. The codes and data for this paper are available online on GitHub at \url{https://github.com/mortamini/VBSparsePoisson}. 

{\section*{Acknowledgments}

The authors would like to thank five anonymous reviewers for their valuable comments and corrections, which significantly improved the results of this paper.}


\appendix
\section*{Appendix (A): Derivation of $q(\cdot)$ density functions}
\setcounter{equation}{0}
 \renewcommand{\theequation}{A\thesection.\arabic{equation}}
 Here, we propose the details of the derivation of the VB  components,  as well as, the computation of the ELBO, for the three proposed VB models.
\subsection*{Laplace prior} 

For model \eqref{lapm}, and using \eqref{vbf}, we have 

\begin{align*}
\log q( { \boldsymbol\beta} ) = & {\rm E}_{-{ \boldsymbol\beta}} \left[ \log \tilde{p}(\mathbf{y} \vert { \boldsymbol\beta}, \boldsymbol\xi) + \log p({ \boldsymbol\beta} \vert {\boldsymbol\tau}) \right] + \text{Const.} \\
= & {\rm E}_{-{ \boldsymbol\beta}} \left[- \sum_{i=1}^{n} M_{\xi_i} \mathbf{X}_i { \boldsymbol\beta} + \sum_{i=1}^{n} y_i \mathbf{X}_i { \boldsymbol\beta} \right. \\
 & - \left. \frac{1}{2} \sum_{i=1}^{n} e^{\xi_i} { \boldsymbol\beta}^\top \mathbf{X}_i \mathbf{X}_i^\top { \boldsymbol\beta} - \frac{1}{2} { \boldsymbol\beta}^\top \text{diag}(\tau_j^{-1}) { \boldsymbol\beta} \right] + \text{Const.} \\
= & (\mathbf{y}-M_{\boldsymbol\xi})^\top \mathbf{X} { \boldsymbol\beta} - \frac{1}{2} { \boldsymbol\beta}^\top \left[ S_X^{\boldsymbol\xi} + \text{diag}({\rm E}_q({\boldsymbol\tau}^{-1})) \right] { \boldsymbol\beta} + \text{Const.,}
\end{align*}
which is the kernel of the Gaussian distribution. Also, we have 
\begin{align*}
\log q({ \eta}) = & {\rm E}_{- { \eta}} \left[ {\sum_{j=1}^{p-1}} \log p(\tau_j \vert { \eta}) + \log p({ \eta}) \right] + \text{Const.} \\
= &  {\rm E}_{- { \eta}}  \left[ {(p-1)} \log { \eta} - \frac{{ \eta}}{2} {\sum_{j=1}^{p-1}} \tau_j + (\nu -1) \log { \eta} - { \eta} \delta \right] + \text{Const.} \\
= & {(p + \nu - 2)} \log { \eta} - {\eta} \left( \frac{1}{2} {\sum_{j=1}^{p-1}} {\rm E}_q(\tau_j) + \delta \right) + \text{Const.},
\end{align*}
which is the kernel of the gamma distribution. For the variance parameters $\tau_j$, for {$j=1,\ldots,p-1$}, we can see that 
\begin{align*}
\log q(\tau_j ) = &  {\rm E}_{- \tau_j} \left[ \log p(\beta_j \vert \tau_j) + \log p(\tau_j\vert{ \eta}) \right] + \text{Const.} \\
 = & {\rm E}_{- \tau_j} \left[ - \frac{1}{2} \log (\tau_j) - \frac{1}{2} \dfrac{\beta_j^2}{\tau_j} - \frac{{ \eta}}{2} \tau_j \right]  + \text{Const.}\\
 = &- \frac{1}{2} \log (\tau_j) - \frac{1}{2} \dfrac{ {\rm E}_q [\beta_j^2]}{\tau_j} - \frac{ {\rm E}_q [{ \eta}]}{2} \tau_j  + \text{Const.}, 
\end{align*}
which is the logarithm of the generalized inverse Gaussian density. For $\tau_0$, we have 
\begin{align*}
\log q (\tau_0) = & {\rm E}_{- \tau_0} \left[ \log p(\beta_0 \vert \tau_0 ) + \log p(\tau_0 \vert a) \right] + \text{Const.} \\
= & {\rm E}_{- \tau_0} \left[-\frac{1}{2} \log \tau_0 - \frac{1}{2} \dfrac{\beta_0^2}{\tau_0} - \frac{3}{2} \log \tau_0 - \dfrac{1}{a \tau_0 } \right] + \text{Const.} \\
= & -2 \log \tau_0 - \frac{1}{\tau_0} \left[ \frac{1}{2} {\rm E}_q(\beta_0^2) + {\rm E}_q(a^{-1}) \right] + \text{Const.}, 
\end{align*}
Finally, we can see that 
\begin{align*}
\log q(a) = & {\rm E}_{-a} \left[  \log p(\tau_0 \vert a) + \log p(a) \right] + \text{Const.}  \\
 = &  {\rm E}_{-a} \left[ - \dfrac{1}{2} \log a - \dfrac{1}{a \tau_0} - \dfrac{3}{2} \log a - \dfrac{1}{Aa} \right] + \text{Const.} \\
 = &- 2 \log a - \dfrac{1}{a} ({\rm E}_q(\tau_0^{-1} )+ A^{-1}) + \text{Const.},
\end{align*}
which is the kernel of the inverse gamma distribution. 

To obtain the ELBO, we compute the following terms 
\begin{align*}
{\rm E}_q (\log \tilde{p} (\mathbf{y}\vert { \boldsymbol\beta} , \boldsymbol\xi) ) = & - M_{\boldsymbol\xi}^T (1+\mathbf{X} { \boldsymbol\mu}_{{ \boldsymbol\beta}(L)} ) - \frac{1}{2}\sum_{i=1}^{n}  {\xi_i^2} e^{\xi_i} \\
& - \frac{1}{2} {\rm tr}(S_{X}^{\boldsymbol\xi} D_{{ \boldsymbol\beta}}^{(L)}) + \mathbf{y} ^\top \mathbf{X} { \boldsymbol\mu}_{{ \boldsymbol\beta}(L)} + \text{Const.},
\end{align*}
where $D_{{ \boldsymbol\beta}}^{(L)} = {\rm E}_q({ \boldsymbol\beta}{ \boldsymbol\beta}^\top) = { \boldsymbol\Sigma}_{{ \boldsymbol\beta}(L)} + { \boldsymbol\mu}_{{ \boldsymbol\beta}(L)}{ \boldsymbol\mu}_{{ \boldsymbol\beta}(L)}^\top, $
\begin{align*}
{\rm E}_q(\log p({ \boldsymbol\beta} \vert \tau)) = & -\frac{1}{2} {\rm E}_q(\log \tau_0) - \frac{1}{2} {\sum_{j=1}^{p-1}} {\rm E}_q( \log \tau_j) \\
& - \frac{1}{2} {\rm E}_q(\tau_0^{-1}) {\rm E}_q(\beta_0^2) - \frac{1}{2} {\sum_{j=1}^{p-1}} {\rm E}_q(\tau_j^{-1} ) {\rm E}_q(\beta_j^2)+ \text{Const.},
\end{align*}
\begin{align*}
{\rm E}_q(\log p({ \eta})) = & (\nu-1 ) {\rm E}_q(\log { \eta}) - \delta {\rm E}_q({ \eta}) + \text{Const.}\\
= & (\nu -1 ) \left[ \psi({ p+ \nu -1}) - \log \left(\delta + \frac{1}{2}  {\sum_{j=1}^{p-1}} {\rm E}_q(\tau_j)\right) \right] \\
- &  \dfrac{\delta({ p+ \nu -1})}{\delta + \frac{1}{2}  {\sum_{j=1}^{p-1}} {\rm E}_q(\tau_j)}+ \text{Const.},
\end{align*} 
\begin{align*}
{\rm E}_q(\log p(\tau_0 \vert a)) = & {\rm E}_q\left(- \frac{1}{2} \log a - \frac{3}{2} \log \tau_0 - \frac{1}{a \tau_0}\right)+ \text{Const.} \\
= & - \frac{1}{2} \log (A^{-1} + {\rm E}_q(\tau_0^{-1}))- \frac{3}{2} \log (d_{00}^{(L)}/2 + {\rm E}_q(a^{-1})) \\
& - {\rm E}_q(a^{-1}) {\rm E}_q(\tau_0^{-1})+ \text{Const.},
\end{align*}
\begin{align*}
{\rm E}_q(\log p(\tau \vert { \eta})) = & {\sum_{j=1}^{p-1}} E \left[ \log { \eta} -  \frac{1}{2} { \eta} \tau_j \right]+ \text{Const.} \\
= &  - {(p-1)} \log \left(\delta + \frac{1}{2} {\sum_{j=1}^{p-1}} {\rm E}_q(\tau_j)\right) - \frac{1}{2} {\rm E}_q({ \eta}) {\sum_{j=1}^{p-1}} {\rm E}_q(\tau_j)+ \text{Const.},
\end{align*}
\begin{align*}
{\rm E}_q (\log p(a)) = & {\rm E}_q \left(-\frac{3}{2} \log a - \dfrac{1}{Aa} \right) + \text{Const.}\\
= & -\frac{3}{2} \log \big( \frac{1}{A} + E (\tau_0^{-1}) \big) - \dfrac{\frac{1}{A}}{\frac{1}{A} + E (\tau_0^{-1})}+ \text{Const.},
\end{align*}
\begin{align*}
-{\rm E}_q(\log q({ \boldsymbol\beta})) = \frac{1}{2} \log \vert { \boldsymbol\Sigma}_{{ \boldsymbol\beta}(L)} \vert+ \text{Const.},
\end{align*}
\begin{align*}
- {\rm E}_q (\log q ({ \eta})) = & (p+ \nu -1 ) \log \left( \delta + \frac{1}{2} {\sum_{j=1}^{p-1}} {\rm E}_q(\tau_j) \right)+ \text{Const.},
\end{align*}
\begin{align*}
- {\rm E}_q(\log q (\tau_0) ) = &\; 2 {\rm E}_q( \log (\tau_0) ) + {\rm E}_q( \tau_0^{-1} ) \big(d_{00}^L/2+ {\rm E}_q(a^{-1}) \big) \\
&- \log (d_{00}^L/2 + {\rm E}_q(a^{-1})) + \text{Const.} \\
= &  \log (d_{00}^L/2 + {\rm E}_q(a^{-1}))+ \text{Const.},
\end{align*}
\begin{align*}
- {\rm E}_q(\log q (\tau) ) = & \;\frac{1}{2} {\sum_{j=1}^{p-1}}{\rm E}_q(\log \tau_j) + \frac{1}{2} {\sum_{j=1}^{p-1}}{\rm E}_q(\tau_j^{-1}) { d_{jj}^{(L)}} \\
&+ \frac{1}{2} {\sum_{j=1}^{p-1}}{\rm E}_q({ \eta}) {\rm E}_q(\tau_j) - \frac{1}{4} {\sum_{j=1}^{p-1}}{\rm E}_q({ \eta})/{ d_{jj}^{(L)}} \\
&+ {\sum_{j=1}^{p-1}}\log K_{1/2}(\sqrt{{\rm E}_q({ \eta}){ d_{jj}^{(L)}}}) + \text{Const.},
\end{align*}
and 
\begin{align*}
- {\rm E}_q(\log q (a)) = & \; 2 {\rm E}_q(\log a) + {\rm E}_q(a^{-1})  \big( {\rm E}_q( \tau_0^{-1} ) + A^{-1} \big) + \text{Const.}\\
= & \; 2 \log \big( {\rm E}_q( \tau_0^{-1} ) + A^{-1} \big) - \psi(1) + {\rm E}_q(a^{-1}) \big( {\rm E}_q( \tau_0^{-1} ) + A^{-1} \big) \\
& 
-\log (A^{-1}+ {\rm E}_q(\tau_0^{-1})) + \text{Const.}
\\
= & \log \big( {\rm E}_q( \tau_0^{-1} ) + A^{-1} \big)+ \text{Const.}
\end{align*}
Adding the above terms, we reach in \eqref{elbo1}. 
\subsection*{Continuous spike and slab prior}
For model \eqref{model2}, we can see that 
\begin{align*}
\log q( { \boldsymbol\beta} ) = &
{\rm E}_{-{ \boldsymbol\beta}} \left[ \log \tilde{p}(\mathbf{y} \vert { \boldsymbol\beta}, \boldsymbol\xi) + \log p({ \boldsymbol\beta} \vert Z,{ \tau^2}) \right] + \text{Const.} \\
& (\mathbf{y} - M_{\boldsymbol\xi})^\top \mathbf{X} { \boldsymbol\beta} - \frac{1}{2} { \boldsymbol\beta}^ \top S_{X}^{\boldsymbol\xi} { \boldsymbol\beta}  - \frac{1}{2} E ({ \tau^{-2}} ) { \boldsymbol\beta}^\top \text{diag} ({ P^{(C)}}) { \boldsymbol\beta} \\
 & - \frac{1}{2} c^{-1} E ({ \tau^{-2}} )  { \boldsymbol\beta}^\top ( \text{I}_p - \text{diag} ({ P^{(C)}})) { \boldsymbol\beta} + \text{Const.},
\end{align*}
which is the kernel of the Gaussian distribution, where ${ P^{(C)}} = (1,P_1^{(C)},\ldots,P_{p-1}^{(C)})^T$, and 
\begin{align*}
\log q({ \tau^2} ) = & {\rm E}_{- { \tau^2}} \left[  \log p({ \boldsymbol\beta} \vert Z,{ \tau^2})+ \log p({ \tau^2} \vert a) \right] + \text{Const.} \\
= & -\frac{p}{2} \log ({ \tau^2}) - \frac{1}{2 { \tau^2}} E \left[ \text{tr(diag}(Z) ) { \boldsymbol\beta} { \boldsymbol\beta}^T \right] \\
& - \dfrac{1}{2 c { \tau^2}} E \left[ \text{tr(diag}(1-Z)) { \boldsymbol\beta} { \boldsymbol\beta}^T \right] - \frac{1}{{ \tau^2}} {\rm E}_q(a^{-1})  \\
& - \dfrac{1}{2} \log { \tau^2} + \text{Const.} \\
= & - \dfrac{p+1}{2} \log { \tau^2} - \dfrac{1}{{ \tau^2}} \left[ \frac{1}{2} \text{tr(diag}({ P^{(C)}}) D_{{ \boldsymbol\beta}}^{(C)}) \right. \\
& + \left.  \dfrac{1}{2c}\text{tr(diag}(1-{ P^{(C)}}) D_{{ \boldsymbol\beta}}^{(C)})  + {\rm E}_q(a^{-1}) \right] + \text{Const.}, 
\end{align*}
which is the logarithm of an inverse gamma density, and 
\begin{align*}
\log q(a) = & {\rm E}_{-a} \left[  \log p({ \tau^2} \vert a) + \log p(a) \right] + \text{Const.}  \\
 = &  {\rm E}_{-a} \left[ \dfrac{1}{2} \log a - \dfrac{1}{a { \tau^2}} + \dfrac{3}{2} \log a - \dfrac{1}{Aa} \right] + \text{Const.} \\
 = & 2 \log a - \dfrac{1}{a} ({\rm E}_q({ \tau^{-2}} )+ A^{-1})+ \text{Const.}, 
\end{align*}
which corresponds with an inverse gamma distribution. Furthermore,
for {$j=1,\ldots,p-1$}, 
\begin{align*}
\log q(Z_j) = &  {\rm E}_{-Z_j} \left[  \log p(\beta_j \vert Z_j,{ \tau^2}) + \log p(Z_j) \right] + \text{Const.} \\
= &  {\rm E}_{-Z_j} \Big [Z_j (- \dfrac{1}{2 { \tau^2}} \beta_j^2 ) + (1 - Z_j) (-  \dfrac{1}{2 c { \tau^2}} \beta_j^2 )   \\
& + Z_j \log \pi_j + (1- Z_j ) \log (1- \pi_j) \Big ] + \text{Const.}\\
& =  Z_j (- {\rm E}_q({ \tau^{-2}}){ d_{jj}^{(C)}}(1-1/c)/{2 }+{\rm E}_q(\log\pi_j) - {\rm E}_q(\log(1-\pi_j)))+\text{Const.},
\end{align*}
which is the kernel of a Bernoulli probability mass function,  
\begin{align*}
\log q(\pi_j) = &  {\rm E}_{-\pi_j} \left[ \log p(Z_j \vert \pi_j) +  \log p(\pi_j )  \right] + \text{Const.} \\
= & E \Big [Z_j \log \pi_j  + (1 -Z_j) (1- \pi_j) + ({ \rho}_1 - 1) \log \pi_j \\
& + ( { \rho}_2 - 1) \log (1-\pi_j) \Big ] + \text{Const.} \\
= & ({ P_j^{(C)}} + { \rho}_1 - 1) \log \pi_j + ( { \rho}_2 - { P_j^{(C)}}) \log(1- \pi_j)+ \text{Const.}, 
\end{align*}
which is the beta distribution. 

Now we compute the following terms to obtain ELBO 
\begin{align*}
{\rm E}_q (\log \tilde{p} (\mathbf{y}\vert { \boldsymbol\beta} , \boldsymbol\xi) ) = & - M_{\boldsymbol\xi}^\top(1+ \mathbf{X} { \boldsymbol\mu}_{{ \boldsymbol\beta}(C)}) - \frac{1}{2}\sum_{i=1}^{n} e^{\xi_i} {\xi_i}^2 \\
& - \frac{1}{2} {\rm tr}(S_{X}^{\boldsymbol\xi} D_{{ \boldsymbol\beta}}^{(C)}) + \mathbf{y}^\top \mathbf{X} { \boldsymbol\mu}_{{ \boldsymbol\beta}(C)}+ \text{Const.},
\end{align*}
\begin{align*}
{\rm E}_q (\log p({ \boldsymbol\beta} \vert \mathbf{Z},{ \tau^2})) = & {\rm E}_q \Big [ {\sum_{j=1}^{p-1}} Z_j  \big(- \frac{1}{2} \log { \tau^2} - \dfrac{1}{2 { \tau^2}} \beta_j^2 \big) \\
& - (1-Z_j) \big( - \frac{1}{2} \log c { \tau^2} - \dfrac{1}{2 c { \tau^2}} \beta_j^2 \big) \Big ] + \text{Const.}\\
= &- \frac{1}{2} {\sum_{j=1}^{p-1}} { P_j^{(C)}} \big( {\rm E}_q(\log{ \tau^2})+{\rm E}_q({ \tau^{-2}}) { d_{jj}^{(C)}} \big) \\
&  - \frac{1}{2} {\sum_{j=1}^{p-1}} (1- { P_j^{(C)}}) \big({\rm E}_q(\log{ \tau^2})+\frac{1}{c} {\rm E}_q({ \tau^{-2}}) { d_{jj}^{(C)}} \big) \\
= & -\frac{1}{2} (1- \frac{1}{c}) {\rm E}_q({ \tau^{-2}}) {\sum_{j=1}^{p-1}} { P_j^{(C)}} { d_{jj}^{(C)}} - \frac{1}{2c} {\rm E}_q({ \tau^{-2}}) {\sum_{j=1}^{p-1}} { d_{jj}^{(C)}} \\
&- \frac{p-1}{2}{\rm E}_q(\log{ \tau^2})+ \text{Const.},
\end{align*}
\begin{align*}
{\rm E}_q(\log p({ \tau^2}\vert a)) = &   -\frac{1}{2} {\rm E}_q(\log a) - \frac{3}{2} {\rm E}_q(\log { \tau^2}) - \frac{1}{2} {\rm E}_q({ \tau^{-2}}) {\rm E}_q(a^{-1}) + \text{Const.},
\end{align*} 
\begin{align*}
{\rm E}_q(\log p(\mathbf{Z}\vert\pi)) = & {\sum_{j=1}^{p-1}} E \left[Z_j \log \pi_j + (1- Z_j) \log (1 - \pi_j) \right] + \text{Const.}\\
= & {\sum_{j=1}^{p-1}} [{ P_j^{(C)}} E\left(\log{\pi_j}\right)+(1-{ P_j^{(C)}}){\rm E}_q(\log(1-\pi_j))] +\text{Const.},
\end{align*}
\begin{align*}
{\rm E}_q(\log p(\pi)) = &  ({ \rho}_1 -1 )  {\sum_{j=1}^{p-1}}  {\rm E}_q(\log\pi_j) +  ({ \rho}_2 -1 )    {\sum_{j=1}^{p-1}} {\rm E}_q(\log (1-\pi_j))+ \text{Const.},
\end{align*}
\begin{align*}
{\rm E}_q (\log p(a)) = &\; {\rm E}_q (-\frac{3}{2} \log a - \dfrac{1}{Aa} ) + \text{Const.}\\
= & -\frac{3}{2} {\rm E}_q(\log a )- A^{-1}{\rm E}_q(a^{-1})+ \text{Const.},
\end{align*}
Furthermore, 
\begin{align*}
-{\rm E}_q(\log q({ \boldsymbol\beta})) = \frac{1}{2} \log \vert { \boldsymbol\Sigma}_{{ \boldsymbol\beta}(C)} \vert+ \text{Const.},
\end{align*}
\begin{align*}
-{\rm E}_q(\log q ({ \tau^2})) = & - \alpha_{{ \tau^2}} \log \beta_{{ \tau^2}} + \log \Gamma (\alpha_{{ \tau^2}}) \\
& + (\alpha_{{ \tau^2}} -1 ) {\rm E}_q(\log { \tau^2}) + \beta_{{ \tau^2}} E ({ \tau^{-2}}) + \text{Const.},
\end{align*}
\begin{align*}
-{\rm E}_q(\log q(\mathbf{Z})) = & -E \left[ {\sum_{j=1}^{p-1}} Z_j \log ({ P_j^{(C)}}) + (1-Z_j) \log(1-{ P_j^{(C)}}) \right] + \text{Const.}\\
= & - {\sum_{j=1}^{p-1}} \left[ { P_j^{(C)}} \log ({ P_j^{(C)}}) + (1-{ P_j^{(C)}}) \log(1-{ P_j^{(C)}}) \right]+ \text{Const.},
\end{align*}
and 
\begin{align*}
-{\rm E}_q (\log q (a)) = & -\log \Big( A^{-1}+ {\rm E}_q({ \tau^{-2}}) \Big) + 2 {\rm E}_q(\log a) +  \Big( A^{-1} + {\rm E}_q({ \tau^{-2}}) \Big)  {\rm E}_q(a^{-1}) + \text{Const.}\\
 = & \log \Big( A^{-1} + {\rm E}_q({ \tau^{-2}}) \Big)+ \text{Const.}
\end{align*}
Summation of the above terms would result in \eqref{elbo2}. 
\subsection*{Bernoulli sparsity enforcing prior}
For model \eqref{model3}, the VB elements are computed as follows. 
\begin{align*}
\log q({ \boldsymbol\beta}) = & {\rm E}_{- { \boldsymbol\beta}} \left[ \log  \tilde{p}(\mathbf{y}; { \boldsymbol\beta}, { \boldsymbol\gamma}, \boldsymbol\xi) + \log p({ \boldsymbol\beta} \vert A) \right] + \text{Const.} \\
= & {\rm E}_{- { \boldsymbol\beta}} \left[ (\mathbf{y}- M_{\boldsymbol\xi})^\top \mathbf{X} \Gamma { \boldsymbol\beta} - \frac{1}{2}  { \boldsymbol\beta}^\top\Gamma S_X^{\boldsymbol\xi}\Gamma{ \boldsymbol\beta} - \frac{1}{2} { \boldsymbol\beta}^\top A { \boldsymbol\beta} \right] + \text{Const.} \\
= & (\mathbf{y}- M_{\boldsymbol\xi})^\top \mathbf{X} \text{diag}({ P^{(B)}}) { \boldsymbol\beta} - \frac{1}{2} { \boldsymbol\beta}^\top \left[ S_X^{\boldsymbol\xi} \odot \Omega + {\rm diag}({\rm E}_q(\alpha)) \right] { \boldsymbol\beta} + \text{Const.}, 
\end{align*}
which results in a Gaussian element, where 
$$\Omega = {\rm E}_q({ \boldsymbol\gamma}{ \boldsymbol\gamma}^T) = ({ P^{(B)}})({ P^{(B)}})^\top + {\rm diag}({ P^{(B)}})({ I_{p}}-{\rm diag}({ P^{(B)}})).$$
 Also, for {$j=1,\ldots,p-1$}, we have 
 \begin{align*}
\log q(\gamma_j) = & {\rm E}_{-\gamma_j} \left[ \log \tilde{p}(\mathbf{y}; { \boldsymbol\beta} , { \boldsymbol\gamma} , \boldsymbol\xi) + \log p(\gamma_j \vert \pi_j) \right] +\text{Const.} \\
=&  {\rm E}_{-\gamma_j}  \Big[ (\mathbf{y} - M_{\boldsymbol\xi})^\top \mathbf{X} \Gamma { \boldsymbol\beta} - \frac{1}{2} { \boldsymbol\beta}^\top \Gamma S_X^{\boldsymbol\xi} \Gamma { \boldsymbol\beta} \\
&+ \gamma_j \log \big( \dfrac{\pi_j}{1-\pi_j} \big) \Big] +\text{Const.} \\
= & {\rm E}_{-\gamma_j}  \Big[  (\mathbf{y} - M_{\boldsymbol\xi})^\top X_j \gamma_j \beta_j - \frac{1}{2} \gamma_j S_{X_{jj}}^{\boldsymbol\xi} \beta_j^2 \\
& - \frac{1}{2} \sum_{i\neq j} \gamma_i \gamma_j S_{X_{ij}}^{\boldsymbol\xi} \beta_i \beta_j + \gamma_j \log\big( \dfrac{\pi_j}{1-\pi_j} \big) \Big] +\text{Const.} \\
= & \gamma_j \Big[ (y - M_{\boldsymbol\xi})^T X_j  { \boldsymbol\mu}_{{ \boldsymbol\beta}(B) j} - \frac{1}{2}  S_{X_{jj}}^{\xi} { { d_{jj}^{(B)}}} \\
& - \frac{1}{2} \sum_{i\neq j}  P_i^{(B)} S_{X_{ij}}^{\boldsymbol\xi} d_{ij}^{(B)} + {\rm E}_q\big(\log\big( \dfrac{\pi_j}{1-\pi_j} \big)\big) \big]+\text{Const.}, 
\end{align*}
which is the kernel of a Bernoulli distribution. Furthermore, for $j=0,1,\ldots,p$, 
\begin{align*}
\log q(\alpha_j) = & {\rm E}_{-\alpha_j} \left[ \log p( { \boldsymbol\beta} \vert A) + \log p(\alpha_j) \right] + \text{Const.} \\
= & (a_j -1/2) \log \alpha_j - (b_j + {d_{jj}}/{2}) \alpha_j+ \text{Const.},
\end{align*}
which corresponds with a gamma density. Also, 
\begin{align*}
\log q( \pi_j) = & {\rm E}_{-\pi_j} \left[ \log p( \gamma_j \vert \pi_j) + \log p(\pi_j) \right] + \text{Const.} \\
= & ({ P_j^{(B)}} + { \rho}_1 -1) \log \pi_j + ( { \rho}_2 - { P_j^{(B)}} ) \log (1- \pi_j )+ \text{Const.},
\end{align*}
which is the logarithm of a beta density. 

The ELBO in \eqref{elbo3} is obtained by computation of the following terms:
 \begin{align*}
{\rm E}_q(\log \tilde{p}(\mathbf{y} \vert { \boldsymbol\beta}, { \boldsymbol\gamma}, \boldsymbol\xi)) =  & {\rm E}_q \Big[ -M_{\boldsymbol\xi}^\top \mathbf{X} \Gamma { \boldsymbol\beta} - \frac{1}{2} { \boldsymbol\beta}^\top \Gamma S_X^{\boldsymbol\xi} \Gamma { \boldsymbol\beta} \\
 & + \mathbf{y}^\top \mathbf{X} \Gamma { \boldsymbol\beta} - \sum_{i=1}^{n} (e^{\xi_i} + \xi_i - \frac{\xi_i^2}{2}) \Big] + \text{Const.}\\
 = & - M_{\boldsymbol\xi} \mathbf{X} { P^{(B)}} { \boldsymbol\mu}_{{ \boldsymbol\beta}(B)} - \frac{1}{2} \text{tr} [D_{{ \boldsymbol\beta}}^{(B)} (S_X^{\boldsymbol\xi} \odot \Omega)] \\
 &+ \mathbf{y}^\top \mathbf{X} { P^{(B)}}{ \boldsymbol\mu}_{ \boldsymbol\beta(B)} - \sum_{i=1}^{n} \big( e^{\xi_i} + \xi_i - \frac{\xi_i^2}{2} \big)+ \text{Const.},
\end{align*}
\begin{align*}
{\rm E}_q(\log p({ \boldsymbol\beta} \vert \boldsymbol\alpha)) = & \frac{1}{2} {\rm E}_q \Big[{\sum_{j=1}^{p-1}} \log \alpha_j - { \boldsymbol\beta}^\top {\rm diag}( \boldsymbol\alpha) { \boldsymbol\beta} \Big]+ \text{Const.} \\
= & \frac{1}{2} {\sum_{j=1}^{p-1}} {\rm E}_q(\log\alpha_j) - \frac{1}{2}\text{tr} \big[ D_{{ \boldsymbol\beta}}^{(B)} \text{diag} ({\rm E}_q(\alpha))\big] \Big]+ \text{Const.},
\end{align*}
\begin{align*}
 {\rm E}_q (\log p( \boldsymbol\alpha)) = & {\sum_{j=1}^{p-1}} {\rm E}_q \left[ (a_j -1 ) \log \alpha_j - b_j \alpha_j \right] + \text{Const.}\\
 =& {\sum_{j=1}^{p-1}}  \big[ (a_j - 1) {\rm E}_q(\log\alpha_j)- b_j \frac{a_j+1/2}{b_j+\frac{d_{jj}}{2}} \big]+ \text{Const.},
\end{align*}
\begin{align*}
{\rm E}_q(\log p({ \boldsymbol\gamma} \vert \pi)) = & {\sum_{j=1}^{p-1}} {\rm E}_q \left[ \gamma_j \log \pi_{j} + (1- \gamma_j) \log (1-\pi_{j} ) \right]+ \text{Const.} \\
= & {\sum_{j=1}^{p-1}} \left[ { P_j^{(B)}} {\rm E}_q(\log\pi_j) + (1- { P_j^{(B)}}) {\rm E}_q(\log(1-\pi_j))\right]+ \text{Const.},
\end{align*}
\begin{align*}
{\rm E}_q(\log p(\pi) ) = &({ \rho}_1 - 1)  {\sum_{j=1}^{p-1}}  {\rm E}_q(\log\pi_j) +({ \rho}_2-1) {\sum_{j=1}^{p-1}} {\rm E}_q(\log(1-\pi_j))+ \text{Const.},
\end{align*}
Also, 
\begin{align*}
-{\rm E}_q(\log q({ \boldsymbol\beta})) = &\frac{1}{2} \log  \vert { \boldsymbol\Sigma}_{{ \boldsymbol\beta}(B)}\vert+ \text{Const.},
\end{align*}
\begin{align*}
-{\rm E}_q(\log q({ \boldsymbol\gamma})) = & -{\sum_{j=1}^{p-1}} {\rm E}_q \left[ \gamma_j \log { P_j^{(B)}} + (1-\gamma_j) \log (1- { P_j^{(B)}}) \right] + \text{Const.}\\
= & -{\sum_{j=1}^{p-1}} [{ P_j^{(B)}} \log { P_j^{(B)}} + (1 - { P_j^{(B)}}) \log (1- { P_j^{(B)}})]+ \text{Const.}, 
\end{align*}
\begin{align*}
-{\rm E}_q(\log q (\pi) ) = & {\sum_{j=1}^{p-1}} [ \log \Gamma ({ P_j^{(B)}}+{ \rho}_1) + \log \Gamma ({ \rho}_2-{ P_j^{(B)}} +1) ]\\
&- {\sum_{j=1}^{p-1}}({ P_j^{(B)}}+{ \rho}_1 - 1) {\rm E}_q(\log\pi_j)\\
&- {\sum_{j=1}^{p-1}}({ \rho}_2-{ P_j^{(B)}}) {\rm E}_q(\log(1-\pi_j))+ \text{Const.},
\end{align*}
and
\begin{align*}
-{\rm E}_q(\log q( \boldsymbol\alpha)) = & -{\sum_{j=1}^{p-1}} (a_j -1/2) {\rm E}_q(\log\alpha_j)-{\sum_{j=1}^{p-1}}(a_j+1/2)\log(b_j+d_{jj}/2)+ \text{Const.}
\end{align*}
Summing up, we obtain \eqref{elbo3}. 

{\section*{Appendix (B): Prior choice guideline}

Based on the results of the simulation study, we present a guideline for choosing a suitable prior in Table \ref{guid}. The best model is suggested in each of the low- and high-dimensional cases, based on the sparsity, estimation, and prediction evaluation criteria. 

{\color{black} For low-dimensional settings, the recommended approach is consistent across objectives: the Bernoulli-VB or Laplace-VB methods are suggested for enforcing sparsity, performing parameter estimation, and generating predictions, {\color{black} because they have achieved the lowest median FPR, FNR, CRE, and TSRE in Figures \ref{box1} and \ref{box2}}. In contrast, for high-dimensional problems, {\color{black}since these criteria achieve their lowest medians for the CS-VB and Laplace-VB models,} the recommendation shifts to favor these two methods {\color{black}(see also the additional results in the supplementary material).}

In general, the choice of prior depends critically on the dimensionality of the data. The Laplace-VB method stands as a good choice for all dimensions. 
The Bernoulli-VB emerges as a robust and simpler default for low-dimensional problems where explicit sparsity modeling is tractable. For high-dimensional inference, priors incorporating the CS mechanism become essential. Therefore, practitioners should first assess the dimension of their covariate space before applying the corresponding prior pairs from this guideline to ensure optimal model performance in sparsity recovery, estimation accuracy, and predictive quality.}

\begin{table}[H]
\centering
\caption{A guidline for choosing the suitable prior for sparse poisson VB regression model.}\label{guid}
\begin{tabular}{c c c}
\hline
& Low dimension & High dimension \\
\hline
Sparsity    &  Laplace \&  Bernoulli  & CS \& Laplace \\
Estimation & Bernoulli \& Laplace  & Laplace \& CS \\
Prediction & Bernoulli \& Laplace & CS \& Laplace\\
\hline
\end{tabular}
\end{table}

}

\end{document}